\documentclass[aps,prx,twocolumn,longbibliography,superscriptaddress,showpacs]{revtex4-2}
\usepackage{times}
\usepackage{graphicx}
\usepackage{latexsym}
\usepackage{amssymb}
\usepackage{amsmath}
\usepackage{amsfonts}
\usepackage{mathtools}
\usepackage{dsfont}
\usepackage{bbm}
\usepackage{bm}
\usepackage{multirow}
\usepackage{array}
\usepackage{color}
\usepackage{tikz}
\usetikzlibrary{shapes}
\usepackage[normalem]{ulem}
\usepackage{comment}
\usepackage{footmisc}
\usepackage[dvipsnames]{xcolor}
\newcommand{\ket}[1]{\vert#1\rangle}
\newcommand{\bra}[1]{\langle#1\vert}
\usepackage[colorlinks=true, citecolor={blue!80!black}, urlcolor={blue!50!black}, linkcolor = {blue!80!black}]{hyperref}
\allowdisplaybreaks[1]
\usepackage[percent]{overpic}
 \usepackage[utf8]{inputenc}

\newcommand{\higher}{\textit{higher} }

\DeclareMathAlphabet{\mathbbold}{U}{bbold}{m}{n}

\begin{document}

\title{Learning Potts Models and $\mathbb{Z}_3$ Toric Codes: Higher and Ordinary  Nishimori Criticality}

 \author{Rushikesh A. Patil}
 \affiliation{Department of Physics, University of California, Santa Barbara, CA 93106, USA}

 \author{Malte P\"utz}
 \thanks{These two authors contributed equally.}
 \affiliation{Institute for Theoretical Physics, University of Cologne, Z\"ulpicher Straße 77, 50937 Cologne, Germany}

 \author{Rohit Mukherjee}
 \thanks{These two authors contributed equally.}
 \affiliation{Institute for Theoretical Physics, University of Cologne, Z\"ulpicher Straße 77, 50937 Cologne, Germany}

 \author{Guo-Yi Zhu}
 \affiliation{The Hong Kong University of Science and Technology (Guangzhou), Nansha, Guangzhou, 511400, Guangdong, China}

 \author{Simon Trebst}
 \affiliation{Institute for Theoretical Physics, University of Cologne, Z\"ulpicher Straße 77, 50937 Cologne, Germany}

 \author{Andreas W. W. Ludwig}
 \affiliation{Department of Physics, University of California, Santa Barbara, CA 93106, USA}

\begin{abstract}
Motivated by the existence of a \higher  Nishimori line in the Bayesian inference/learning phase diagram of the classical Ising model 
under bond-energy measurements, we identify an analogous line in the corresponding phase diagram of the two-dimensional ($2D$) $q$-state Potts model ($2 < q\leq 4$).
Tuning the Potts model to its critical temperature $\beta_c$,  
this \higher Nishimori line meets the critical line (at $\beta_c$) 
 in a distinct \higher Nishimori critical point 
-- a {\sl tricritical} point at finite inference strength that separates a paramagnetic, a ferromagnetic and a `spin-glass' phase.
 With analytical tools, we discuss the general structure of the
rich phase diagram, which contains two unstable and three stable
fixed points, and obtain a number of {\sl exact} results for 
universal quantities, including the decay exponent of the Edwards-Anderson correlator, 
using a Gaussian measurement protocol which allows for exact calculations.
Using extensive  numerical tools, we confirm these statements for a 
generic, discrete $q$-state measurement protocol and determine precise numerical
estimates for the location of higher and ordinary Nishimori critical points 
as well as RG flows between the various fixed points.
We also discuss the Casimir effective central charges of the critical points in the learning phase diagram,
and their monotonic {\sl decrease} along measurement-induced RG flows, as established non-perturbatively by  
the $c$-effective theorem and its extensions,
and contrast it to the monotonic {\sl increase} along the corresponding RG flows in the random-bond Potts model.
Finally, we discuss a general argument based on Elitzur's theorem that establishes stability of
the ordinary Nishimori critical points in their respective learning phase diagrams.
Our results can be recast in the language of Born measurements on a deformed $\mathbb{Z}_q$ toric code where the tricritical \higher
Nishimori point is an `information' critical point that separates phases with strong, weak, and broken $\mathbb{Z}_q$
symmetry, which correspond to stable quantum, classical, and no memory phases, respectively.
\end{abstract}

\date{\today}
\maketitle

\section{Introduction\label{SecIntro}}

\begin{figure*}[t]
    \includegraphics[width=\linewidth,keepaspectratio]{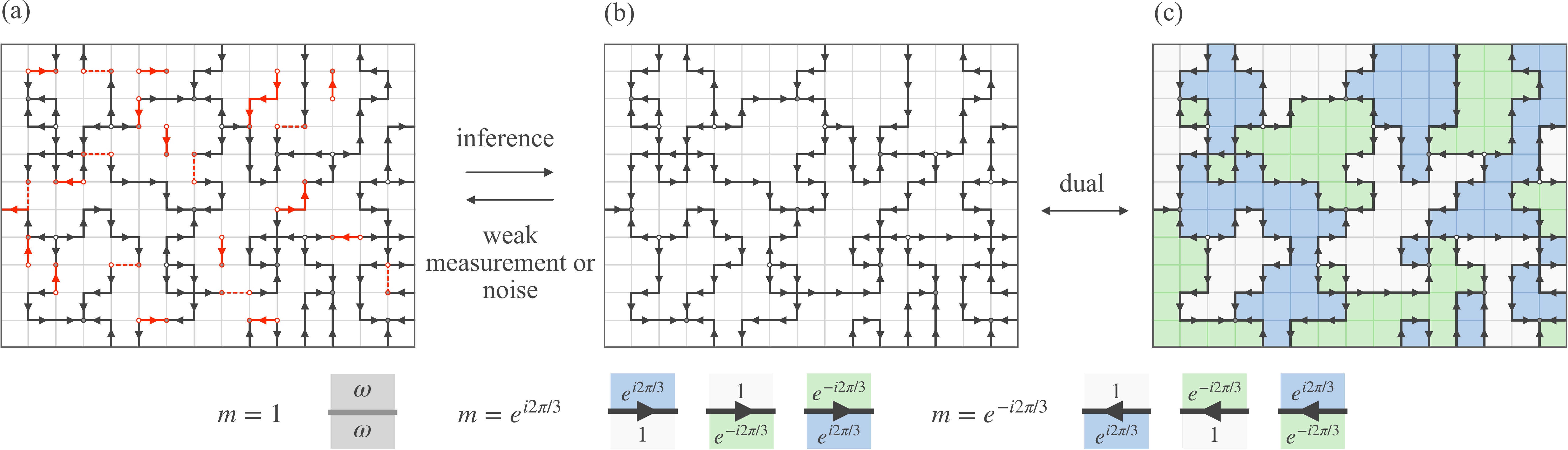}
\caption{
	{\bf Learning the Potts model.}
	Schematically illustrated is the process of learning a configuration of the $3$-state Potts model (with the Potts spins residing on the faces of the squares) 
	from an initial configuration (a) of domain walls	obtained from performing a sequence of bond-energy measurements indicated on the links of the lattice. 
	For the $3$-state Potts model there are three bond types -- no domain wall (gray line), as well as two types of domain walls separating different domains (up/down black arrows).
	For a finite measurement/inference strength, some bonds will not be recognized correctly, which we illustrate by the domain walls marked red.
	The correct domain wall configuration (b) can be reconstructed (``inferred'' or ``learned'') from the erroneous one (a), e.g.\ by reducing it to a domain wall
	configuration with no open-ended segments and enforcing that every vertex is either `divergence-free' (black dot) or has a $\pm 3$ `monopole defect' (white/gray-filled black dots).
	Once a consistent domain wall configuration (b) is found, the corresponding Potts spins (labeled by the white, blue, and green colors of the faces inside the domain walls) 
	can readily be assigned, up to a global $\mathbb{Z}_3$ cyclic permutation of the domain colors, see panel (c). 
	}
	\label{FigSchematic}
\end{figure*}

The toric code~\cite{Kitaev2003} is the archetypal example of how measurements can be used
to prepare a long-range entangled quantum state~\cite{Tantivasadakarn2024} and protect it from decoherence~\cite{DennisKitaevLandahlPreskill}.
Crucially, the outcomes of these measurements, which typically probe only a few qubits directly,
allow one to infer some understanding of the macroscopic quantum state --
the best known example might be the ability to perform quantum error correction
based on a sequence of 4-qubit syndrome measurements.
To read out the quantum memory, one might carry out a sequence of single-qubit measurements 
-- notably this works even 
away from the projective measurement limit (for weak measurements)~\cite{Eckstein2024,Eckstein2025},
where one obtains
only partial information about the qubits (which are fully collapsed only in the projective limit). 
In this quantum setting, each measurement outcome is a probabilistic variable governed by 
Born's rule and whose randomness depends on the correlations in the unmeasured system.
Such inference problems typically exhibit threshold behavior (i.e.\ an 
amount of randomness 
that is 
tolerable), with the underlying phase
transition falling into the universality class of Nishimori transitions (for a Bayes-optimal  setup).

Nishimori criticality is well-known from a classical statistical mechanics context where it
describes the paramagnet-to-ferromagnet transition along a fine-tuned line in the phase diagram
of the random-bond Ising model (RBIM)~\cite{Nishimori_1980,Nishimori1981}. Its first appearance in the context of inference problems
has also been in the classical realm~\cite{Iba_1999,Sourlas_1994}, e.g.\ by employing a Bayesian 
inference protocol to learn a given spin configuration of an (infinite-temperature) Ising model 
through a sequence of bond-energy measurements. 
Every such measurement allows one to condition an initial 
Boltzmann-weight ensemble (the ``prior'' probability distribution)
up to a threshold where it suddenly sharpens and the underlying state has been learned.
The phase diagrams of these purely classical problems turn out to be equivalent to the phase diagrams of the corresponding quantum Rokhsar-Kivelson wavefunctions~\cite{CLHenley_2004} under Born-rule quantum measurements. For example, by mapping first to the Rokhsar-Kivelson problem and then applying a Kramers-Wannier/Wegner duality, one finds that the Bayesian inference phase diagram of the infinite-temperature $2D$ classical Ising model~\cite{Iba_1999} is exactly equivalent to the phase diagram of the toric code ground state under weak Pauli-$Z$ measurements~\cite{Eckstein2024}.

The {\sl randomness} in measurement outcomes is, of course, an essential ingredient of these many-body phenomena. 
However, it is crucial to distinguish 
the randomness in measurement outcomes -- which depends on the correlations in the unmeasured system --
from the uncorrelated randomness in traditional quenched disorder systems. The latter distinction becomes especially apparent when we move to finite temperatures and approach the critical point in the unmeasured statistical system, where degrees of freedom have long-ranged algebraic correlations.
In a slightly more technical language, this distinction between the measurement-induced randomness and uncorrelated quenched impurity-type disorder manifests itself via the distinct replica limits $R\rightarrow0$ and $R\rightarrow1$ that are required to study the respective impurity-type quenched disorder and measurement-induced randomness
problems upon using the replica trick~\cite{JianYouVasseurLudwig2019,BaoChoiAltman2019}.

In a recent work~\cite{PatilPutzTrebstZhuLudwig}, some of us revisited the Bayesian inference/learning phase diagram of the $2D$ Ising model~\cite{PutzGarrattNishimoriTrebstZhu,NahumJacobsen} at finite temperatures, which is exactly dual to the phase diagram of the deformed toric code wavefunction~\cite{CastelnovoChamon,ArdonneFedleyFradkin,PapanikolaouRamanFradkin,IsakovFendleyLudwigTrebstTroyer,Zhu19deform,Verresen25deform} under quantum Born-rule Pauli-$Z$ measurements.
We~\cite{PatilPutzTrebstZhuLudwig} and independent work~\cite{WieseDasNahum}
demonstrated the existence of a novel, distinct \higher  Nishimori line in the learning phase diagram of the Ising model.
This \higher  Nishimori line passes through the previously identified learning tricritical point~\cite{PutzGarrattNishimoriTrebstZhu,NahumJacobsen}, i.e.\ it was demonstrated that the learning tricritical point is a \higher  Nishimori critical point and that it has
an emergent gauge-invariant formulation as a replica-symmetric theory in the $R\rightarrow2$ replica limit, which is larger than the replica symmetry of the gauge-invariant formulation of the ordinary Nishimori line corresponding to the $R\rightarrow1$ replica limit~\cite{LeDoussalHarrisII,GeorgesHanselLeDoussalMaillard,GRL2001}.
This identification allowed Refs.~\cite{PatilPutzTrebstZhuLudwig,WieseDasNahum} to obtain a number of exact results for the universal properties at the learning tricritical point, e.g.\ the power-law exponent of the measurement-averaged second moment of the spin-spin correlator, i.e.\ the Edwards-Anderson correlator, 
was shown to be exactly equal to that of the spin-spin correlation function at the unmeasured Ising critical point. 
Extensive numerical simulations confirmed these statements for the natural binary (instead of Gaussian) bond-energy measurement protocol 
by demonstrating that the Edwards-Anderson correlator 
indeed exhibits the predicted power-law decay at the \higher Nishimori critical point, determining its 
Casimir effective central charge, and mapping out 
an emergent \higher Nishimori line within the high-temperature paramagnetic phase.

In this manuscript, we ask whether this \higher Nishimori physics carries over to other models. In particular, we are asking whether the learning phase diagram
of the deformed $\mathbb{Z}_3$ toric code and its classical analogue, the Bayesian inference problem of the $3$-state Potts model, allow for a similar identification
of enhanced replica symmetry and a \higher Nishimori critical point. 
The latter setting is visualized in Fig.~\ref{FigSchematic} above: If one successively probes the domain wall
structure of a given  Gibbs state of a classical $3$-state Potts model 
(at some finite temperature), 
is there a  sharp threshold (for noisy measurement outcomes) 
above which one is able to readily reconstruct the complete domain
wall structure and therefore infer the actual state? What is the nature of the critical theory at this threshold, and how does it depend on the temperature at which the Gibbs
 state is sampled? What happens at the critical temperature of the Potts model where the underlying 
Gibbs state exhibits algebraic correlations? The answers to these questions
simultaneously address our ability to read out a (deformed) $\mathbb{Z}_3$ toric code memory, 
via single-qutrit Pauli-$Z$ measurements.

\subsubsection*{Summary of Results}

\begin{figure*}[t]
 	   \includegraphics[width=\linewidth]{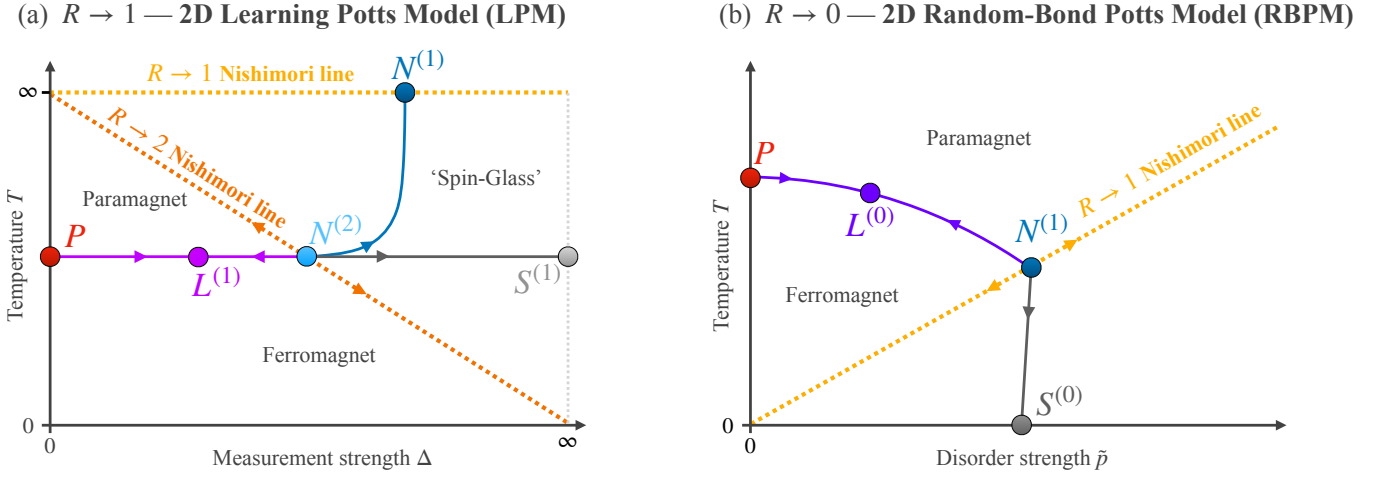}
\caption{{\bf (a) Schematic learning phase diagram of the $2D$ $q$-state Potts model for $2<q\leq 4$} 
with Gaussian measurements defined in Eq.~\eqref{eq:PmsigmaGaussianPotts}. With this specialized measurement protocol both the \higher Nishimori line (dotted dark orange line) $\beta=\Delta$ and the ordinary Nishimori line $\beta=0$ (dotted light orange line) exist in the learning phase diagram at the microscopic level.
The critical points $N^{(1)}$ and $N^{(2)}$ are respectively the ordinary and \higher  Nishimori critical points in the Potts learning phase diagram. $P$ is the unmeasured Potts critical point and ${L}^{(1)}$ is the weak-measurement fixed point discussed in Ref.~\cite{NahumJacobsen}. The RG flows on the $\beta=\beta_c$ line were discussed in Ref.~\cite{NahumJacobsen}, whereas here we explicitly demonstrate that $N^{(2)}$ is a \higher  Nishimori critical point.
 We expect the universal properties of the critical points and the RG flows in the learning phase diagram to be largely independent of the microscopic measurement protocol, given the same global symmetries; 
 see, e.g., the phase diagram in Fig.~\ref{fig:numerical_phase_diagram_withGaussian} for a discrete $q$-state learning protocol.
(b) Schematic phase diagram for the $2D$ random-bond $q$-state Potts model ($2<q\leq 4$) from Ref.~\cite{JacobsenPicco}, obtained with
$q$-state bond-disorder.
}
\label{FigPottsLearningPhaseDiagram}
\end{figure*}

In this work, we obtain the learning phase diagram of the two-dimensional ($2D$) classical $q$-state Potts model with $2<q\leq 4$ 
under bond-energy measurements. For a particularly 
designed  Gaussian bond-energy measurement protocol,
we demonstrate 
that the phase diagram contains, in addition to the {\sl ordinary} Nishimori line lying at the infinite temperature $\beta=0$,
an exact {\sl higher} Nishimori line,
\begin{equation}
    \beta=\Delta,
\end{equation}
where $\Delta$ is the measurement strength in our measurement protocol, 
as illustrated in Fig.~\ref{FigPottsLearningPhaseDiagram}(a). 
With the critical temperature $\beta_c=\frac{1}{q}\ln (1+\sqrt{q})$ of the unmeasured model (see our convention of Potts Hamiltonian in Eq.~\eqref{EqPottsHamiltonian})
determining the phase boundary between the paramagnetic and ferromagnetic phases in our learning phase diagram, we demonstrate the existence of a {\sl tricritical} point at the intersection of the $\beta=\beta_c$ line  and the {\sl higher} Nishimori line $\beta=\Delta$
\begin{equation}
    \beta=\Delta=\beta_c=\frac{1}{q}\ln(1+\sqrt{q}),
\end{equation}
i.e. the tricritical point is a {\sl higher} Nishimori critical point.

Analogous to the Ising learning phase diagram~\cite{PatilPutzTrebstZhuLudwig,WieseDasNahum}, the identification of the tricritical point as a \higher Nishimori critical point implies a number of {\sl exact} results. The most important of these is that we can demonstrate that the Edwards-Anderson (EA) correlator at the tricritical point decays with the power-law exponent of the Potts spin correlator at the critical point in the unmeasured model. For example, for the tricritical point in the $3$-state Potts learning phase diagram, this means that
\begin{equation}  
\overline{\left|
  \langle\omega_i\omega_j^{-1}\rangle_{\vec m}\right|^2}
  \sim \frac{1}{|i-j|^{4/15}} \label{EqEACorrelatorIntro}
\end{equation}
where $\omega_i=1, e^{2\pi i/3},e^{4\pi i/3}$ are Potts spins and $2/15$ is the scaling dimension of the Potts spin at the $3$-state Potts critical point~\cite{DOTSENKO1984Potts}. In general, we can calculate the absolute value squared of any $n$-point correlation function at the higher Nishimori critical point.

We then turn to the discrete, $q$-state
bond-energy measurement protocol which is schematically illustrated in Fig.~\ref{FigSchematic} (for $q=3$) and might be considered the most natural choice for a classical inference setting.
For such a protocol, we cannot make an {\sl exact} statement for the existence of a \higher Nishimori line in the phase diagram, but using extensive numerical simulations we can unambiguously demonstrate the existence of an {\sl emergent} \higher Nishimori line.
Specifically, in our extensive tensor network simulations, we find that the power-law decay of the EA correlator at the tricritical point of the learning phase diagram with 
such a discrete  
protocol is consistent with Eq.~\eqref{EqEACorrelatorIntro}, which is the smoking gun evidence of \higher Nishimori physics.
Going beyond the critical line (at $\beta_c$) in the learning phase diagram,
we numerically identify the emergent \higher Nishimori line by matching the 
finite correlation length of the EA correlator with that of the unmeasured spin-spin correlator in the paramagnetic phase.
The resulting learning phase diagram  
is shown in Fig.~\ref{fig:numerical_phase_diagram_withGaussian}, where the emergent \higher Nishimori line of the discrete protocol is found in close proximity to that for the Gaussian protocol.

At the critical point of the unmeasured $2D$ Potts model, the bond-energy measurements are {\sl relevant} in a 
renormalization group (RG) sense,
and as discussed in Ref.~\cite{NahumJacobsen} (see also~\cite{LUDWIGCardy,Ludwig-Potts-OPE-RG-NPB285-1987-97}), 
measurements drive an RG flow to an intermediate {\sl attractive} fixed point  -- in contrast to the $2D$ Ising
learning phase diagram~\cite{PutzGarrattNishimoriTrebstZhu,NahumJacobsen}
where no such intermediate fixed point exists.
In our phase diagram [Fig.~\ref{FigPottsLearningPhaseDiagram}(a)], we have denoted this attractive fixed point as ${L}^{(1)}$.
As demonstrated in Ref.~\cite{NahumJacobsen}, for the $q$-state Potts model (with $2<q\leq 4$) the universal properties of this
fixed point can be studied perturbatively in $(q-2)$ by using the results of an epsilon expansion originally developed (by one of us) 
for the attractive `ferromagnetic' fixed point $L^{(0)}$ in
the $2D$ random-bond Potts model~\cite{Ludwig-Potts-OPE-RG-NPB285-1987-97,LUDWIGCardy}.
This is possible because both the phase diagram of the random-bond Potts model (RBPM) and the phase diagram of the Potts model under bond-energy measurements are governed by the 
{\sl same} replica theory but with 
{\sl different} required replica limits, $R\rightarrow0$ and $R\rightarrow1$, respectively.

As illustrated in Fig.~\ref{FigPottsLearningPhaseDiagram}, the learning phase diagram we obtain 
exhibits instructive similarities and differences when compared to the phase diagram of the $2D$ RBPM~\cite{JacobsenPicco} 
shown in the right-hand panel (b).
For example, in addition to the distinction of $L^{(0)}$ and $L^{(1)}$ critical points, we should also distinguish 
the multicritical points in the RBPM and the learning phase diagram, respectively: $N^{(1)}$, the
ordinary Nishimori critical point and $N^{(2)}$, the \higher  Nishimori critical point. 
The universal properties of these critical points are again expected to be governed by the 
{\sl same} replica field theory but in 
{\sl different} replica limits;
$R\to 0$ for the RBPM and $R\to1$ for the learning phase diagram reflecting the underlying {\sl quenched} versus {\sl Bayesian/Born} disorder, respectively.

A major distinction between the phase diagrams in Fig.~\ref{FigPottsLearningPhaseDiagram}, is that the learning phase diagram
[Fig.~\ref{FigPottsLearningPhaseDiagram}(a)]
contains both the ordinary and the \higher Nishimori critical point, while the RBPM phase diagram
[Fig.~\ref{FigPottsLearningPhaseDiagram}(b)]
has only the ordinary Nishimori critical point.
The ordinary Nishimori critical point appears in its `gauge-invariant' formulation as a replica theory with a local symmetry in the $R\rightarrow1$ replica limit on the $\beta=0$ line in the learning phase diagram, while the ordinary Nishimori critical point in the RBPM is governed by a `gauge-fixed' version of this
replica theory, where the `gauge-fixing' leads to a replica theory in the $R\rightarrow0$ replica limit. [See the discussion for the $q=2$ (Ising) case in Refs.~\cite{ZhuTantivasadakarnVishwanthTrebstVerresen,NahumJacobsen,PatilPutzTrebstZhuLudwig} (also~\cite{LeDoussalHarrisII}), which straightforwardly generalizes for arbitrary $q$.]
We emphasize that the \higher and ordinary Nishimori critical points are distinct universality classes, and in particular, using our numerical calculations, we establish an RG flow from the higher Nishimori critical point to the ordinary Nishimori critical point in the Potts learning phase diagram [Fig.~\ref{FigPottsLearningPhaseDiagram}(a)].

We use the local symmetry of the $R\rightarrow1$ replica theory for the ordinary Nishimori critical point on the $\beta=0$ line in the learning phase diagram to explain the stability of the critical point 
to the perturbation in (inverse) temperature [Fig.~\ref{FigPottsLearningPhaseDiagram}(a)].
In particular, we apply {\sl Elitzur's theorem}~\cite{Elitzur1975} to the thermal (globally $\mathbb{Z}_q$ symmetric) perturbation of the 
above
replica theory for the ordinary Nishimori transition that exhibits a local 
(`gauge') symmetry~\cite{LeDoussalHarrisII,GRL2001,ZhuTantivasadakarnVishwanthTrebstVerresen,NahumJacobsen}.
The argument we give is a variation of
the seminal argument for stability of `pure-gauge' transitions~\cite{Wegner1971,FradkinShenker1979}, which can be understood as a consequence of Elitzur's theorem, extended to the replica theory.
Our argument is 
rather 
general, and it explains the occurrence of ordinary Nishimori transitions as stable critical universality classes in other monitored
systems, e.g. in the learning phase diagram of the wavefunction-deformed toric code
considered in Ref.~\cite{PutzGarrattNishimoriTrebstZhu}.

Lastly, we discuss the Casimir effective central charges of critical points in the learning phase diagram. 
In particular, using the $c$-effective theorem of Ref.~\cite{PatilLudwig20251} and its extensions~\cite{PatilPutzTrebstZhuLudwig,PatilLudwig20262}, 
it can be non-perturbatively demonstrated that the Casimir effective central charge monotonically {\sl decreases} under the 
measurement-induced RG flow from the unmeasured Potts critical point (central charge $=4/5$) to the attractive fixed point $L^{(1)}$,
and also from the \higher  Nishimori critical point $N^{(2)}$  to
the attractive fixed point $L^{(1)}$ [Fig.~\ref{FigPottsLearningPhaseDiagram}(a)].
Moreover, as noted in Ref.~\cite{PatilLudwig20251}, by combining the result of the $c$-effective theorem in the $R\rightarrow1$ replica limit with
a physically motivated assumption about the central charge
of the replica theory, a simple argument can be made for the {\sl increase} of the Casimir effective central charge 
under the RG flow
in the $R\rightarrow0$ replica limit, which is relevant for systems with quenched impurity-type disorder.
For the random-bond Potts model phase diagram {in} Fig.~\ref{FigPottsLearningPhaseDiagram} (b),
the argument implies that the Casimir effective central charge {\sl increases} under the RG 
flow from the clean Potts critical point to the attractive `ferromagnetic' fixed point $L^{(0)}$ and 
from the ordinary Nishimori critical point
$N^{(1)}$
to the attractive ferromagnetic fixed point $L^{(0)}$.
This is consistent with the known numerical~\cite{JacobsenPicco} and epsilon expansion results~\cite{Ludwig-Potts-OPE-RG-NPB285-1987-97,LUDWIGCardy,DotsenkoJacobsenLewisPicco} for the 
Casimir effective central charge of the attractive `ferromagnetic' fixed point $L^{(0)}$ and numerical results~\cite{JacobsenPicco} for the
ordinary Nishimori critical point $N^{(1)}$
in the 
random-bond $3$-state Potts model.
We summarize the known values of Casimir effective central charges for critical points in both the random-bond $3$-state Potts model 
and the $3$-state Potts learning phase diagram in Fig.~\ref{fig:CentralChargeLadder} towards the end of the paper.

\setcounter{tocdepth}{0}
\tableofcontents

\section{Learning Setup}

\subsection{Bayesian inference protocol for the \texorpdfstring{$q$}{Lg}-state Potts model}

The classical $q$-state Potts model with
a
$q$-fold degree of freedom $\omega_i$ at each site is given by the following Hamiltonian
\begin{equation}
    H[\{\omega_i\}]=-q\sum_{\langle ij\rangle}\delta_{\omega_i, \omega_j} \,,    
    \label{EqPottsHamiltonian}
\end{equation}
where $\langle ij \rangle$ denotes a pair of nearest-neighbor sites, and
where we have introduced a factor of $q$ in the Hamiltonian for later notational convenience. 
Clearly, the energy only depends on whether the states $\omega_i$ on nearest-neighbor sites are equal and not on the explicit values of the variables $\omega_i$.
It will be convenient for us to choose the state  $\omega_i$ to be $q^{\text{th}}$ roots of unity
\begin{equation}
    \omega_i=1,\, e^{2\pi \dot{\iota}/q},\,e^{4\pi \dot{\iota}/q},\,\ldots,\, e^{2\pi(q-1)\dot{\iota}/q}.
\end{equation}
In terms of these roots of unity and using $\delta_{\omega_i,\omega_j}=\frac{1}{q}\sum_{r=0}^{q-1}(\omega_i\omega_j^{-1})^r$, the above Potts Hamiltonian can be written as~\cite{MittagStephen}
\begin{equation}
   -H=\frac{1}{2}\sum_{\langle ij\rangle}\sum_{r=1}^{q-1}[(\omega_i\omega^{-1}_j)^r+(\omega^{-1}_i\omega_j)^r]+\text{const.}
\end{equation}
In the unmeasured model, the probability of a configuration $\{\omega_i\}$ is given by the Boltzmann weight
\begin{equation}
P(\{\omega_i\})=\frac{e^{-\beta H[\{\omega_i\}]}}{Z},\label{EqBoltzmann}
\end{equation}
the ``prior probability distribution''.
Let us consider performing classical measurements on 
a bond $\langle ij\rangle$
of the system with the following local {discrete 
$q$-state
measurement} protocol
\begin{equation}
     P(m_{ij}|\{\omega_{i}\})=\frac{\exp{\{q\tilde{\gamma} \delta_{m_{ij},\omega_i\omega_j^{-1}}\}}}{(q-1)+e^{q\tilde{\gamma}}}, \label{EqMeasurementProtocolA}
\end{equation}
where the measurement outcomes $m_{ij}$ take the following  $q$ possible
values for the bond variable $\omega_i\omega_j^{-1}$
\begin{equation}
    m_{ij}=1,\, e^{2\pi \dot{\iota}/q},\,e^{4\pi \dot{\iota}/q},\,\ldots,\, e^{2\pi(q-1)\dot{\iota}/q},
    \label{LabelEqThreeOutcomeMeasurements}
\end{equation}
and $\tilde{\gamma}$ characterizes the strength of measurements. 
This protocol is a $q$-state symmetric channel, in which the
outcome is correct with probability $[1+(q-1)\gamma]/q$, where
$e^{q\tilde{\gamma}}=[1+(q-1)\gamma]/(1-\gamma)$~\footnote{Equivalently, ${\gamma}=(e^{q\tilde{\gamma}}-1)/(e^{q\tilde{\gamma}}+(q-1))$} and $\gamma\in[0,1]$
interpolates between an uninformative ($\gamma=0$) and a perfect
($\gamma=1$) measurement. 
We use $\gamma$ to indicate the strength of measurement in all figures below.

For performing measurements on all the bonds, we can simply take the product of the
conditional probability distributions 
 \eqref{EqMeasurementProtocolA}
on individual bonds
\begin{equation}
     P(\{m_{ij}\}|\{\omega_{i}\})=\frac{\exp{\{q\tilde{\gamma}\sum_{\langle ij\rangle} \delta_{m_{ij},\omega_i\omega_j^{-1}}\}}}{[(q-1)+e^{q\tilde{\gamma}}]^{N_B}},\label{EqMeasurementProtocolAGlobal}
\end{equation}
where $N_B$ is the number of bonds.
Upon obtaining a particular set of measurement outcomes 
on all bonds
$\{m_{ij}\}$, the state of our knowledge of the configuration $\{\omega_i\}$ is 
encoded by Bayes' theorem in the following probability distribution for  $\{\omega_i\}$, the ``posterior distribution'',
\begin{eqnarray} \nonumber
&&
P(\{\omega_{i}\}|\{m_{ij}\})=\frac{P(\{m_{ij}\}|\{\omega_{i}\})P(\{\omega_i\})}{P(\{m_{ij}\})}
\\  \label{EqBayesRule}
&&
=\frac{\exp\{-\beta H[\{\omega_i\}]+q\tilde{\gamma}\sum_{\langle ij\rangle} \delta_{m_{ij},\omega_i\omega_j^{-1}}\}}{Z\times [(q-1)+e^{q\tilde{\gamma}}]^{N_B}\times P(\{m_{ij}\})}, 
\end{eqnarray}
where $Z$ is the unmeasured Potts partition function. Here,
the probability $P(\{m_{ij}\})$ for obtaining measurement outcomes $\{m_{ij}\}$ is given by
\begin{align}
   P(\{m_{ij}\})&=\sum_{\{\omega_{i}\}}P(\{m_{ij}\}|\{\omega_{i}\})P(\{\omega_i\})\nonumber\\
   &=\frac{ \sum\limits_{\{\omega_i\}}\exp\{-\beta H[\{\omega_i\}]+q\tilde{\gamma}\sum_{\langle ij\rangle} \delta_{m_{ij},\omega_i\omega_j^{-1}}\}}{Z\times [(q-1)+e^{q\tilde{\gamma}}]^{N_B}}
   \label{EqClassicalProbability}
\end{align}
To characterize the effects of learning on our knowledge of the Potts model configurations, we will be interested in calculating the following measurement-averaged moments of various expectation values in the model
\begin{align}
\label{EqDefOfMeasurementAveragedMoments}
&\overline{
\langle { {\cal O}}_1\rangle_{\vec m}
\langle { {\cal O}}_2\rangle_{\vec m}
\cdots
\langle {{\cal O}}_N\rangle_{\vec m}
}=
\sum_{\vec{m}} P(\vec{m})\langle { {\cal O}}_1\rangle_{\vec m}
\langle { {\cal O}}_2\rangle_{\vec m}
\cdots
\langle {{\cal O}}_N\rangle_{\vec m},
\end{align}
where, for fixed measurement outcomes on all links, denoted by ${\vec m}$, the expectation value $\langle {\cal O}_k\rangle_{\vec m}$ ($k=1, \ldots,N$)
of any observable ${\cal O}_k$ in the classical Potts model, which is a function 
of the classical degrees of freedom $\omega_i$,
is evaluated with the posterior distribution in Eq.~\eqref{EqBayesRule}.

\subsection{Quantum measurement protocol for the $\mathbb{Z}_q$ toric code}
\label{SecQuantumProtocol}

The learning problem defined above is exactly equivalent to a
Born-rule measurement problem on a quantum wavefunction. To see this,
consider the Rokhsar--Kivelson (RK) state associated with the Potts
Boltzmann weight of Eq.~\eqref{EqBoltzmann},
\begin{equation}
    |\psi(\beta)\rangle=\frac{1}{\sqrt{Z}}\sum_{\{\omega_i\}}
    e^{-\beta H[\{\omega_i\}]/2}\,|\{\omega_i\}\rangle \,,
    \label{EqRKState}
\end{equation}
where $\beta$ is the inverse temperature of the associated classical Potts model, 
 but here used as a wavefunction `deformation parameter' of a family of RK states as we will now discuss.
At infinite temperature $\beta=0$ this is the uniform product state
$|+\rangle^{\otimes N}$ with
$|+\rangle=\tfrac{1}{\sqrt q}\sum_{\omega\in{\mathbb{Z}_q}}|\omega\rangle$.
Gauging the normal subgroup $\mathbb{Z}_q\subset S_q$ of the symmetry group $S_q$ of the classical Potts
model~\cite{Haegeman_2015} maps this state to a ground state of the
$\mathbb{Z}_q$ toric code~\cite{Kitaev2003}, which can also be obtained by Kramers-Wannier/Wegner duality.
(We focus here on the case $2 < q \leq 4$ and leave the discussion of the $\mathbb{Z}_q$ toric code with $q>4$ to future work.)
At finite temperature $\beta>0$ the same gauging maps Eq.~\eqref{EqRKState} to a
{\sl deformed} $\mathbb{Z}_q$ toric-code
wavefunction, in which oriented domain walls form closed but branching configurations [Fig.~\ref{FigSchematic}]
that carry a Boltzmann line tension $e^{-q\beta}$ per wall segment (equivalently,
an RK-amplitude tension $e^{-q\beta/2}$),
and the thermal transition
of the 2D Potts model becomes the topological transition of this
wavefunction.
For an elementary review of these arguments, 
see footnote~\footnote{
The undeformed $\mathbb{Z}_q$ toric code state is dual to the RK state in Eq.~\eqref{EqRKState} at infinite temperature $\beta=0$, 
which is the product state $|+\rangle^{\otimes N}$ 
of eigenstates of unit eigenvalue of the ${\hat X}_{i}$ 
operators at each site $i$ lying on the face of the square lattice [Fig.~\ref{FigSchematic}].
The RK state at $\beta=0$ is an equal-weight superposition of eigenstates of operators 
${\hat Z}_{i}$ located on the faces of the square lattice.
Eigenstates of operators ${\hat Z}_{ij}$ 
with non-unit eigenvalues $\omega_{ij}$ correspond to domain wall configurations on the links of the lattice: 
Each configuration of eigenvalues $\omega_i$ on the faces of the square lattice uniquely  determines a domain wall configuration (up to boundary terms) on links via $\omega_{ij}=$
$\omega_i {\omega_j}^{-1}$ (going from left to right and top to bottom). Compare Fig.~\ref{FigSchematic}.
The equal-weight superposition of the latter domain-wall configurations, which satisfy a local Gauss-law constraint,
is the ground state of the (undeformed)
$\mathbb{Z}_q$ toric code $\ket{TC}$ with a $q$-dimensional Hilbert space and operators
$\hat{Z}_{ij}$ defined on the links of the square lattice.
When acting on the undeformed $\mathbb{Z}_q$ toric code state,
the factor
\vskip -3mm
\[
\qquad \exp\{
{\beta\over 2} \sum_{\langle i j \rangle}
\sum_{r=1}^{q-1} 
[ 
({\hat Z}_{ij})^r 
+ ({{\hat Z}_{ij}}^{-1})^r 
]
\} \propto
\exp\{\beta q \sum_{\langle i j \rangle} \delta_{\omega_{ij}, 1} \} \,,
\] 
\vskip -1mm
where $\omega_{ij}$ is the eigenvalue of the operator ${\hat Z}_{ij}$,
assigns a factor $\exp\{ - q\beta \times ({\rm total \ number \ domain \ wall \ segments})\}$ to each domain wall configuration. 
Thus, the RK state \eqref{EqRKState} for the finite-temperature $q$-state Potts model is exactly dual to the deformed $Z_q$ toric code state, which, following the above logic, reads
$\exp\{
{1\over 2} {\beta\over 2} \sum_{\langle i j \rangle}
\sum_{r=1}^{q-1} 
[ 
({\hat Z}_{ij})^r 
+ ({{\hat Z}_{ij}}^{-1})^r 
]
\} \ \ket{TC}
$
}.
In this quantum setting, the Bayesian (classical) measurement protocol in
Eq.~\eqref{EqMeasurementProtocolA} 
corresponds to a quantum measurement on the above-defined deformed toric code state.
The latter can be written as a linear combination of simultaneous eigenstates of operators ${\hat Z}_{ij}$  with eigenvalues $g_{ij}\in \mathbb{Z}_q$ on all links $\langle i j \rangle$,
and the corresponding Kraus operators, 
diagonal~\cite{PatilPutzTrebstZhuLudwig,PatilLudwig20251} in this basis, 
read~\footnote{That is, 
on each link $\langle i j \rangle$, $g=g_{ij}\in \mathbb{Z}_q$ is the state variable on this link, and $m = m_{ij}\in\mathbb{Z}_q$ is the discrete measurement outcome on the same link.}
\begin{equation}
    \hat K_{m}
    =\frac{1}{\sqrt{(q-1)+e^{q\tilde\gamma}}}\sum_{g\in \mathbb{Z}_q} e^{\,q\tilde\gamma\,\delta_{m,g}/2}\ket{g}\bra{g} \,,
    \quad m\in\mathbb{Z}_q \,,
    \label{EqKraus}
\end{equation}
which satisfy the POVM condition $\sum_m \hat K_m^\dagger \hat K_m=\mathds{1}$. 
Applied to
every link, the Born probability of a measurement record $\vec m$ on all links
reproduces Eq.~\eqref{EqClassicalProbability}, and the 
post-measurement quantum state is again of RK form, now
with the posterior distribution
Eq.~\eqref{EqBayesRule} as its 
Gibbs-Boltzmann weight -- 
a measurement with outcomes ${\vec m}$ maps 
the RK state of the ``prior'' Boltzmann distribution
\eqref{EqBoltzmann} for configurations $(\{\omega_i\})$ 
to the post-measurement RK state of the
``posterior'' distribution \eqref{EqBayesRule} for configurations of $(\{\omega_i\})$.
Let us note on the side that if the measurement record ${\vec m}$ is
traced out, the above measurement protocol 
is converted to a {\sl dephasing noise} channel, with details shown in Appendix~\ref{app:noise}.

Equal-time correlators of quantum observables diagonal in the Potts basis [used in \eqref{EqRKState} or \eqref{EqKraus}]
in a post-measurement state of the deformed toric code coincide with classical averages
of these observables under the posterior distribution \eqref{EqBayesRule},
and the (classical) measurement-averaged moments in Eq.~\eqref{EqDefOfMeasurementAveragedMoments} acquire the meaning of 
Born-rule averages over quantum trajectories.
This protocol is a direct
$\mathbb{Z}_q$ generalization 
of the weak Pauli-$Z$ measurements that
drive the Nishimori transition of the $\mathbb{Z}_2$ toric
code~\cite{Eckstein2024,Eckstein2025}; the specialized protocol
to be discussed in Sec.~\ref{SecSpecializedMeasurementProtocol} corresponds to replacing the
$q$-state 
measurement defined in \eqref{LabelEqThreeOutcomeMeasurements}
by a weak Gaussian measurement with continuous complex measurement outcomes~\footnote{a continuous complex Gaussian pointer} $m_{ij} \in {\bf C}$.

Specifically, under this dictionary the learning phase diagram of
Fig.~\ref{FigPottsLearningPhaseDiagram}(a) is the measurement phase diagram of the
deformed $\mathbb{Z}_q$ toric code ($2<q\leq 4$) 
with the above-mentioned continuous complex-valued weak Gaussian measurement protocol, and the numerical learning phase diagram with the $q$-state measurement protocol [Eq.~\eqref{EqKraus}] is shown in Fig.~\ref{fig:numerical_phase_diagram_withGaussian}. 
The ordinary Nishimori point
$N^{(1)}$ at $\beta=0$ is the measurement-induced transition of the
undeformed code, upper-bounding~\cite{Putz25nishimori,SchumacherNielsen1996,WildeQuantumInformation}
the optimal error-correction threshold
of the $\mathbb{Z}_q$
toric code under the 
dephasing noise channel~\cite{PhysRevA.91.042331,mukherjee2026nishimorithresholdestimationbayesian}, see App.~\ref{app:noise} for details. 
Meanwhile, the higher Nishimori point $N^{(2)}$
is a tricritical point of the {\sl critically} deformed code under
weak measurements. Finally, the global $\mathbb{Z}_q$  
transformation that bond measurements cannot resolve [Fig.~\ref{FigSchematic}(c)] is
mapped, after gauging, to a corresponding component of the code's
logical information. The coherent information used in
Sec.~\ref{SecNumericalResults} therefore diagnoses the survival of this
encoded global information under the measurements.

\section{Replica Theory and \higher  Nishimori Line}
\label{SecReplicaTheoryAndHigherNishimoriLine}

Let us start our technical analysis of the inference problem with a short overview of its formulation in terms of replica theory,
the natural starting point for Nishimori physics as outlined above.

As a first step, it can be easily verified that the measurement-averaged expectation values in Eq.~\eqref{EqDefOfMeasurementAveragedMoments}
(see, e.g., App.\ A of Ref.~\cite{PatilPutzTrebstZhuLudwig})
are given by the following replica theory expression in the replica limit $R\rightarrow1$
\begin{align}
    \overline{
\langle { {\cal O}}_1\rangle_{\vec m}
\cdots
\langle {{\cal O}}_N\rangle_{\vec m}
}\propto   \lim_{R\rightarrow 1}
\sum_{\{\omega_i^{(a)}\}_{a=1}^R}\mathcal{O}_1^{(1)}\cdots \mathcal{O}_N^{(N)}e^{-\mathcal{H}^R[\{\omega_i^{(a)}\}_{a=1}^R]} \,,
\label{EqLatticeReplicaTheoryBetaV}
\end{align}
where $\mathcal{O}_k^{(a)}$ denotes observable $\mathcal{O}_k$ in the $a^{\text{th}}$ replica copy and the replica Hamiltonian is given by
\begin{align}
     &-\mathcal{H}^{R}[\{\omega_i^{(a)}\}_{a=1}^R]=\nonumber\\=&\sum_{\langle ij\rangle}\Big\{\frac{\beta}{2}\sum_{r=1}^{q-1}\sum_{a=1}^{R}[(\omega^{(a)}_i(\omega^{(a)}_j)^{-1})^{r}+((\omega^{(a)}_i)^{-1}\omega^{(a)}_j)^{r}]\nonumber\\&+\frac{\tilde{\gamma}^2}{2} \sum_{r=1}^{q-1}|\sum_{a=1}^R(\omega^{(a)}_i(\omega^{(a)}_j)^{-1})^{r}|^2\Big\}+ \mathcal{O}(\tilde{\gamma}^3) \,.
     \label{EqReplicatedPottsHamiltonianProtocolI}
\end{align}
We note that when $\beta=0$, the above replica Hamiltonian is invariant under a {\sl local} (`gauge')
transformation of changing the Potts spin at any given site
    \begin{align}
    &\omega_{i}^{(a)}\rightarrow\omega_{i}^{(a)}\Omega_i, \;\;\;\;\forall a=1, \ldots , R\label{EqGaugeTransformationRReplica} \\\big[\Omega_i= &1,\,e^{2\pi \dot{\iota}/q},\,e^{4\pi \dot{\iota}/q},\,\cdots,e^{2\pi (q-1)\dot{\iota}/q}\big],\nonumber
\end{align}
and this local symmetry exists also in the $R\rightarrow1$ replica limit.
The replica Hamiltonian on this $\beta=0$ line is nothing but a gauge-invariant formulation of the {\sl ordinary} Nishimori line from the random-bond Potts model~\cite{NishimoriStephen,JacobsenPicco}. The replica limit $R\rightarrow0$ replica theory for the random-bond Potts model on the ordinary Nishimori line can be obtained by `gauge-fixing' the above $R\rightarrow1$ replica Hamiltonian on the $\beta=0$ line. [See e.g. App $D$ of Ref.~\cite{PatilPutzTrebstZhuLudwig}, and also Refs.~\cite{ZhuTantivasadakarnVishwanthTrebstVerresen, NahumJacobsen,LeDoussalHarrisII}, for the discussion of this in the $q=2$ (Ising) case, which straightforwardly generalizes to arbitrary $q$.]

Let us now move to $\beta\neq 0$, and study the replica Hamiltonian in Eq.~\eqref{EqReplicatedPottsHamiltonianProtocolI} by first ignoring  the $\mathcal{O}(\tilde{\gamma}^3)$ terms and considering
the following truncated replica Hamiltonian at second-order,
\begin{align}
     &-{\mathcal{\tilde{H}}}^{R}[\{\omega_i^{(a)}\}_{a=1}^R]=
     \nonumber \\
   =&\sum_{\langle ij\rangle}\Big\{\frac{\beta}{2}\sum_{r=1}^{q-1}\sum_{a=1}^{R}[(\omega^{(a)}_i(\omega^{(a)}_j)^{-1})^{r}+((\omega^{(a)}_i)^{-1}\omega^{(a)}_j)^{r}]
     \nonumber \\
     &+\frac{\Delta}{2} \sum_{r=1}^{q-1}|\sum_{a=1}^R(\omega^{(a)}_i(\omega^{(a)}_j)^{-1})^{r}|^2\Big\} \,,
     \label{EqReplicatedPottsHamiltonianUngauged}
\end{align}
where we have defined $\Delta:=\tilde{\gamma}^2$.~\footnote{As discussed below, the replica theory in Eq.~\eqref{EqReplicatedPottsHamiltonianUngauged} is an exact description of the learning phase diagram for a specialized Gaussian measurement protocol. The latter protocol is discussed in detail in Sec.~\ref{SecSpecializedMeasurementProtocol}.}
Following the discussion in Ref.~\cite{PatilPutzTrebstZhuLudwig} for the replica theory of the Ising ($=2$-state Potts model) learning phase diagram, we consider 
the above replica theory on the line,
\begin{equation}
    \beta=\Delta : \text{``\higher  Nishimori line''} \,,
    \label{EqDefHigherNishimoriLine}
\end{equation}
where the above 
truncated
replica Hamiltonian can be written as
\begin{equation}
     -\tilde{\mathcal{H}}^{R}[\{\omega_i^{(a)}\}_{a=1}^R]=\frac{\beta}{2}\sum_{\langle ij\rangle}\sum_{r=1}^{q-1}\Big|\sum_{a=1}^{R}(\omega^{(a)}_i(\omega^{(a)}_j)^{-1})^{r}+1\Big|^2\,\label{EqReplicatedPottsHamiltonianUngaugedII}.
\end{equation}
Then by generalizing the discussion in Ref.~\cite{PatilPutzTrebstZhuLudwig,WieseDasNahum} for the Ising (=\,2-state) case, we can consider introducing an additional, $(R+1)^{\text{th}}$ replica copy $\omega_i^{(R+1)}=1,\,e^{2\pi \dot{\iota}/q},\,e^{4\pi \dot{\iota}/q},\,\cdots,e^{2\pi \dot{\iota}(q-1)/q}$ at each site via the following transformation 
\begin{equation}
    \omega_i^{(a)}\rightarrow\omega_i^{(a)}\left(\omega_{i}^{(R+1)}\right)^{-1} \;\;\; a=1,\ldots,  R \,.
    \label{EqGauging}
\end{equation}
Then 
the 
$R$-replica Hamiltonian on the \higher  Nishimori line $\beta=\Delta$ in Eq.~\eqref{EqReplicatedPottsHamiltonianUngaugedII} is 
equivalently written as the following $(R+1)$ replica Hamiltonian
\begin{equation}
     -\tilde{\mathcal{H}}^{R+1}[\{\omega_i^{(a)}\}_{a=1}^{R+1}]=\frac{\beta}{2} \sum_{\langle ij\rangle}\sum_{r=1}^{q-1}\Big|\sum_{a=1}^{R+1}(\omega^{(a)}_i(\omega^{(a)}_j)^{-1})^{r}\Big|^2 \,.
     \label{EqGaugedReplicatedPottsHamiltonian}
\end{equation}
The above $(R+1)$-replica Hamiltonian is invariant under the following gauge transformation~(contrast this with the local transformation of Eq.~\eqref{EqGaugeTransformationRReplica} that only has $R$ replicas)
\begin{align}
    &\omega_{i}^{(a)}\rightarrow\omega_{i}^{(a)}\Omega_i, \;\;\;\;\forall a=1,\dots , R+1
    \label{EqGaugeTransformation} \\\
    \big[&\Omega_i= 1,\,e^{2\pi \dot{\iota}/q},\,e^{4\pi \dot{\iota}/q},\,\cdots,e^{2\pi (q-1)\dot{\iota}/q}\big]\nonumber
\end{align}
and has an {\sl enlarged} replica symmetry under permutations of all
$R+1$ replicas. The gauge-invariant $(R+1)$-replica Hamiltonian in
Eq.~\eqref{EqGaugedReplicatedPottsHamiltonian} describes distinct
problems in different replica limits. 
For $(R+1)\to 0$, it 
governs the 
spin-glass problem for the Potts model;
for $(R+1)\to 1$, it yields the
gauge-invariant formulation of the
ordinary Nishimori line 
in the random-bond Potts model~\cite{NishimoriStephen}.
The latter gauge-invariant formulation of
the ordinary Nishimori line is the
infinite-temperature line $\beta=0$
in the learning phase diagram [Fig.~\ref{FigPottsLearningPhaseDiagram}(a)],
where the replica theory has a
local symmetry [cf. Eq.~\eqref{EqReplicatedPottsHamiltonianUngauged} with $\beta=0$].
For $(R+1)\rightarrow2$, the $(R+1)$-replica Hamiltonian in Eq.~\eqref{EqGaugedReplicatedPottsHamiltonian} provides the gauge-invariant formulation of the \higher  Nishimori line $\beta=\Delta$ [Eq.~\eqref{EqDefHigherNishimoriLine}] for the replica theory of Eq.~\eqref{EqReplicatedPottsHamiltonianUngauged} in the physical `learning' replica limit $R\rightarrow1$.

The $(R+1)$-replica Hamiltonian formulation in Eq.~\eqref{EqGaugedReplicatedPottsHamiltonian} 
of the 
truncated replica Hamiltonian
in Eq.~\eqref{EqReplicatedPottsHamiltonianUngauged} on the higher Nishimori line $\beta=\Delta$ allows us to obtain a variety of non-trivial exact
results.
For example, by realizing this higher replica symmetric gauge-invariant formulation, one can derive
the following equality between the correlation functions on the 
\higher  Nishimori line
\begin{align}
&\langle (\omega_{i_1}^{(a_1)}\cdots\omega_{i_k}^{(a_1)})\cdots(\omega_{i_1}^{(a_n)}\cdots\omega_{i_k}^{(a_n)})\rangle_{R}=\nonumber\\
&
\langle (\omega_{i_1}^{(a_1)}\cdot\cdot\omega_{i_k}^{(a_1)})\cdots(\omega_{i_1}^{(a_n)}\cdot\cdot\omega_{i_k}^{(a_n)})[(\omega_{i_1}^{(a_{n+1})}\cdots\omega_{i_k}^{(a_{n+1})})^*]^n\rangle_{R},   \label{EqReplicatedNishimoriIdentityBulk} 
\end{align}
where replica indices $1\leq a_1,\,a_2,\,\ldots, a_{n},a_{n+1}\leq R$ are pairwise unequal and 
the 
correlation functions
are evaluated 
using the truncated
$R$-replica theory 
of Eq.~\eqref{EqReplicatedPottsHamiltonianUngauged} on the $\beta=\Delta$ line.

Modulo the ignored $\mathcal{O}(\tilde{\gamma}^3)$ terms from Eq.~\eqref{EqReplicatedPottsHamiltonianProtocolI}, the truncated $R$-replica Hamiltonian in Eq.~\eqref{EqReplicatedPottsHamiltonianUngauged} should allow us to calculate measurement-averaged moments.
In particular, the equality in Eq.~\eqref{EqReplicatedNishimoriIdentityBulk} in the $R\rightarrow 1$ replica limit then allows us to obtain non-trivial identities for measurement-averaged moments in the learning phase diagram. However, it
is indeed important
to note that we have ignored the terms of order $\tilde{\gamma}^3$ in Eq.~\eqref{EqReplicatedPottsHamiltonianProtocolI} while reaching the higher replica symmetric gauge-invariant formulation in Eq.~\eqref{EqGaugedReplicatedPottsHamiltonian}. 
The ignored $\mathcal{O}(\tilde{\gamma}^3)$ and higher terms explicitly break the $(R+1)$-replica permutation symmetry of Eq.~\eqref{EqGaugedReplicatedPottsHamiltonian}.
The physical question is therefore whether the higher-replica-symmetric formulation {\sl emerges} at long distances in the infrared or not.
An analogous 
question was investigated
in the case of the Ising learning phase diagram in Refs.~\cite{PatilPutzTrebstZhuLudwig,WieseDasNahum}, where 
the analogous
higher-replica symmetric gauge-invariant formulation at a lattice level 
is
absent for the binary measurement protocol due to the presence of 
similar
higher-order terms in measurement strength in the corresponding replica Hamiltonian.
However, via   extensive tensor network numerics with a large $512\times512$ system size, it was demonstrated in Ref.~\cite{PatilPutzTrebstZhuLudwig} that the long-distance behavior of the Edwards-Anderson correlator at the Ising learning tricritical point~\cite{PutzGarrattNishimoriTrebstZhu} is characterized by the
power-law exponent of the 
spin-spin correlation function at the unmeasured Ising critical point, which is a smoking gun signature of 
emergent 
higher Nishimori criticality.
We will now investigate numerically if the
higher replica symmetric formulation
is {\sl emergent} at
the tricritical point in the learning phase diagram for the $3$-state Potts model with the measurement protocol in Eq.~\eqref{EqMeasurementProtocolA}, and if the tricritical point
is a \higher  Nishimori critical point.
We will do this by comparing the
behavior 
of the 
Edwards-Anderson correlator at the tricritical point with that of the spin-spin correlator at the unmeasured Potts critical point.

Moreover, in Sec.~\ref{SecSpecializedMeasurementProtocol},
we will discuss a Bayesian inference problem for the Potts model with a
Gaussian protocol for bond-energy measurements fine-tuned such that the 
replica Hamiltonian is given precisely by the truncated replica Hamiltonian of
Eq.~\eqref{EqReplicatedPottsHamiltonianUngauged},
i.e.\ {\sl without} any terms of $\mathcal{O}(\tilde{\gamma}^3)$ order,
such that,
along the \higher  Nishimori line [Eq.~\eqref{EqDefHigherNishimoriLine}], the higher replica symmetric gauge-invariant formulation in Eq.~\eqref{EqGaugedReplicatedPottsHamiltonian} follows exactly.

\section{Numerical Results}
\label{SecNumericalResults}

The replica analysis of Sec.~\ref{SecReplicaTheoryAndHigherNishimoriLine}
suggests that the learning tricritical point of the Potts model may realize an emergent \higher Nishimori universality class.  
Here we address this question numerically for the $3$-state Potts model using 
a hybrid Monte-Carlo / tensor-network approach~\cite{PutzGarrattNishimoriTrebstZhu}
that allows us to probe systems up to lattice dimensions $256\times256$.
Technically, the Monte Carlo part of our hybrid approach samples configurations 
from the underlying thermal Potts ensemble and then samples measurement outcomes 
conditioned on those configurations. For each measurement record, tensor-network
contractions evaluate the conditioned observables in a second step. 
We determine the learning phase diagram, summarized in Fig.~\ref{fig:numerical_phase_diagram_withGaussian} below,
from finite-size scaling of the coherent information on systems with linear size $L=8$ to $128$, and measure the
Edwards--Anderson (EA) correlator directly on $256\times256$ systems,
averaging over approximately $10^5$ measurement realizations.

\begin{figure}[t]
    \includegraphics[width=\linewidth]{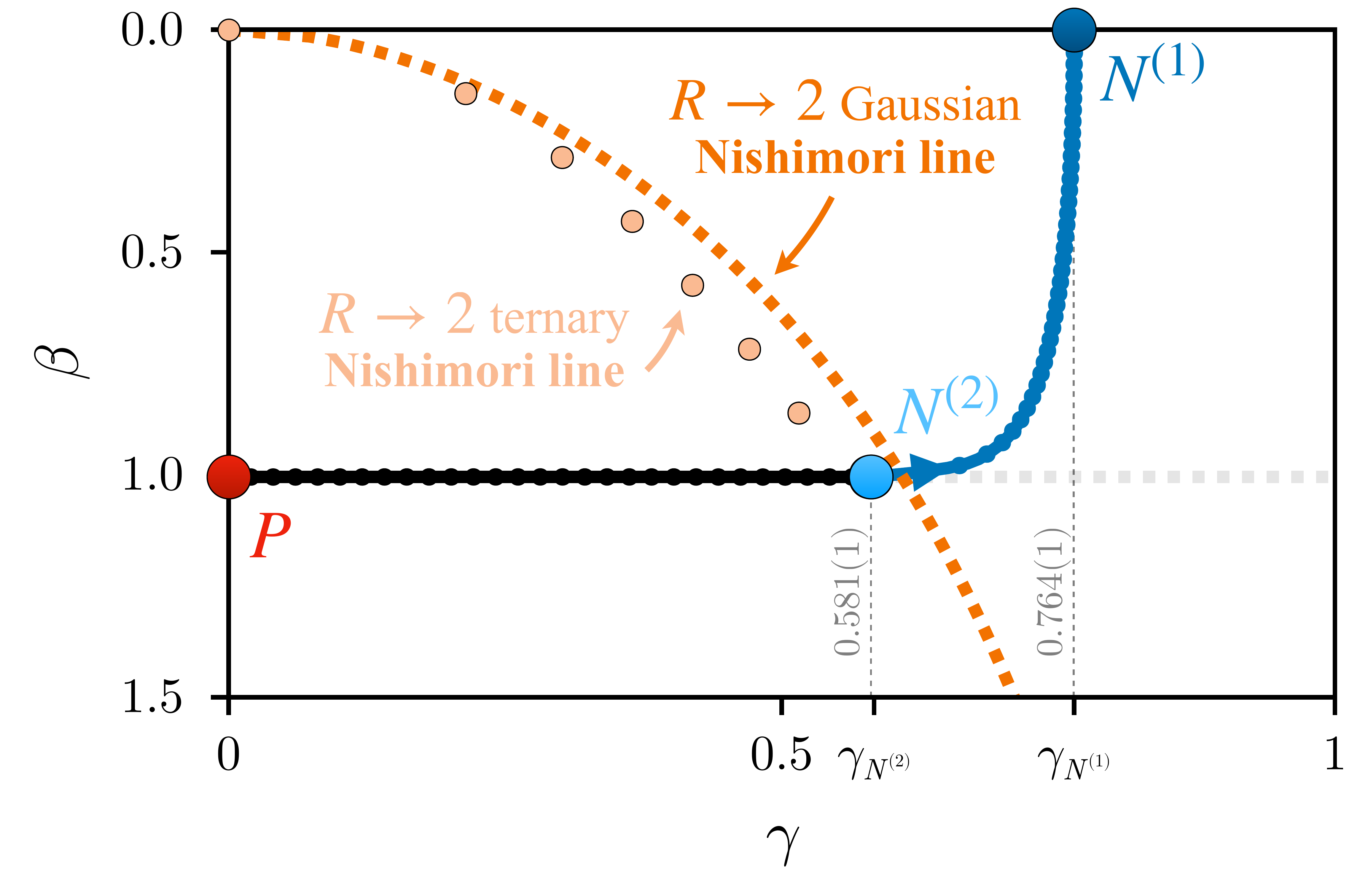}
    \caption{{\bf Numerical phase diagram.}
            Shown is the numerically obtained learning phase diagram of the $3$-state Potts model, employing the discrete $3$-state protocol [Eq.~\eqref{EqMeasurementProtocolA}].
            The phase boundaries are obtained from finite-size scaling of the coherent information [Fig.~\ref{fig:Ic_nishimori12}]. 
            The \higher Nishimori critical point $N^{(2)}$ is located at $\gamma_{N^{(2)}} = 0.581(1)$, 
            while the ordinary Nishimori critical point $N^{(1)}$  is located at $\gamma_{N^{(1)}} = 0.764(1)$.             
            The orange markers indicate the location of the  {\sl emergent} \higher Nishimori line, determined as shown in Fig.~\ref{fig:nishimori_condition},
            which appears in the vicinity of the \higher Nishimori line for the Gaussian protocol (dashed line).
            The slight shift of the {\sl emergent} \higher Nishimori line from the $N^{(2)}$ point
            is partially a numerical artifact as the (numerically determined) inverse correlation length 
            of the unmeasured model is slightly overestimated (see Fig.~\ref{fig:2X2_along_betac} for example), which in turn 
             leads to a slight underestimation of the measurement strength $\gamma$ at which the \higher  Nishimori condition is satisfied.
        }
    \label{fig:numerical_phase_diagram_withGaussian}
\end{figure}

\begin{figure}[h!]
    \includegraphics[width=\linewidth]{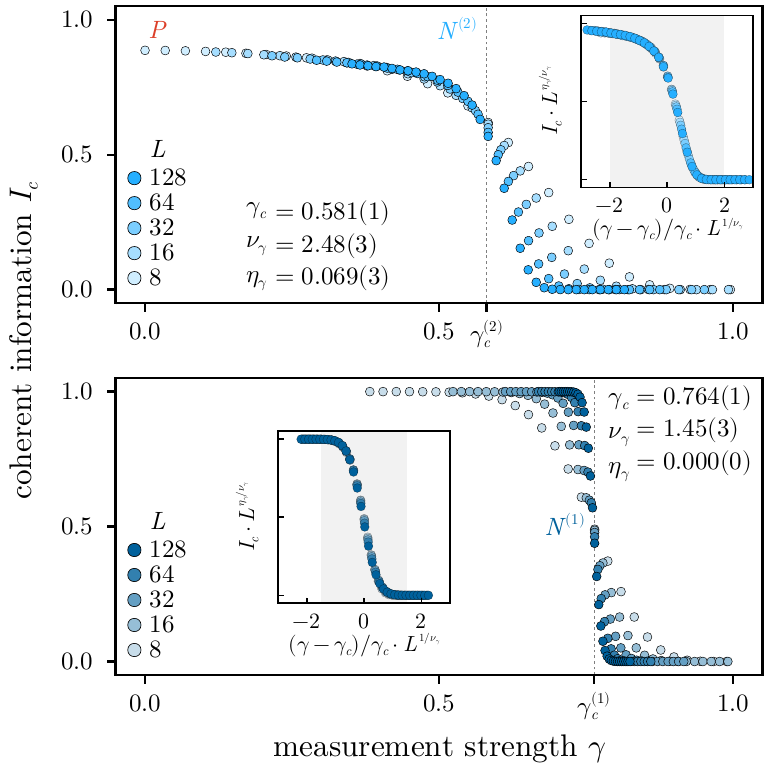}
    \caption{{\bf Coherent information along horizontal cuts through \higher  and {\sl ordinary} Nishimori points.}
            Shown is the coherent information $I_c$ as a function of $\gamma$ for learning the $q=3$ Potts model 
            at $\beta = \beta_c$ (top panel) and $\beta = 0$ (bottom panel). 
            The locations $\gamma_c$ of the critical points $N^{(2)}$ (top) and $N^{(1)}$ (bottom) are obtained from finite-size scaling 
            of the coherent information, see the scaling collapse shown in the insets. 
            Data points are obtained by averaging over $\sim$$100,000$ disorder realizations, making the error bars smaller than the marker size.
        }
    \label{fig:Ic_nishimori12}
\end{figure}

\subsubsection{Learning phase diagram}
To map out the learning phase diagram, we use the coherent information $I_c$ as a principal diagnostic 
of the different  information phases. 
In the thermodynamic limit $(L\to\infty)$, it approaches quantized limiting
values, $I_c=1$ in the paramagnetic
phases or $0$ in the ferromagnetic 
and `spin-glass' phases.
At a phase transition, the finite-size crossovers between these limiting values sharpen,
see Fig.~\ref{fig:Ic_nishimori12} for two horizontal cuts in the phase diagram where the transition points 
correspond to the \higher and ordinary Nishimori points $N^{(2)}$ and $N^{(1)}$, respectively.

The phase diagram obtained in this way is
shown in Fig.~\ref{fig:numerical_phase_diagram_withGaussian}.  Along the
infinite-temperature line $\beta=0$, we locate the ordinary Nishimori
critical point at
\begin{equation}
    \gamma_{N^{(1)}}=0.764(1),
\end{equation}
separating the paramagnetic phase from the phase with long-range order
in the EA correlator and short-range order in the first moment of the spin-spin correlator, 
which we refer to as the `spin-glass' phase.  
This value is consistent with previous numerical work on the $3$-state
random-bond Potts model on the Nishimori line~\cite{JacobsenPicco,HoneckerJacobsenPiccoPujol,PhysRevA.91.042331} 
as well as a recent analytical (minimal replica projection) estimate~\cite{mukherjee2026nishimorithresholdestimationbayesian}.
Along the critical-temperature line $\beta=\beta_c^{(3)}$, where
$\beta_c^{(3)}$ is the critical (inverse) temperature of the unmeasured
Potts model~\cite{Potts_1952},
\begin{equation}
    \beta_{c}^{(3)}=\frac{1}{3}\ln{(1+\sqrt{3})} = 0.3350\ldots \, ,
\end{equation}
we find a second transition at
\begin{equation}
    \gamma_{N^{(2)}}=0.581(1) .
\end{equation}
It is the above learning tricritical point
at which the paramagnetic, ferromagnetic, and `spin-glass' phases meet.
As we will show below, this is indeed the \higher Nishimori  critical point of the $3$-state Potts phase diagram,
which we demonstrate (in the next subsection) by showing that its EA correlations fulfill the {\sl \higher Nishimori condition}
\begin{equation}
  \overline{\left|
  \langle\omega_i\omega_j^{-1}\rangle_{\vec m}\right|^2}
  \sim \frac{1}{|i-j|^{4/15}} \,.
\end{equation}
Concerning its precise location, we note that the numerical estimate from the coherent information data collapse 
[Fig.~\ref{fig:Ic_nishimori12} upper panel] is slightly below 
the estimate obtained from the truncated replica theory ($\gamma=0.60922$); see Appendix~\ref{Appendix_Numerical}. 

The numerical validation that the \higher Nishimori point $N^{(2)}$ is part of an emergent \higher Nishimori line
comes from a careful analysis of the decay of the EA correlator within the paramagnetic phase (as detailed in 
the next subsection) that allows us to track the line via the orange dots in the phase diagram of Fig.~\ref{fig:numerical_phase_diagram_withGaussian}.
As can be seen, these points are in close vicinity of the exact \higher Nishimori line for Gaussian 
measurement outcomes (dashed orange line).

The \higher and ordinary Nishimori transitions are connected
by the phase boundary between the paramagnetic and `spin glass'
phase, see Fig.~\ref{fig:numerical_phase_diagram_withGaussian}. 
Along this line, the RG flow is from the 
\higher to the ordinary Nishimori universality class, as we show
below (in the next-next subsection) by tracking the change of critical 
exponents along this line of phase transitions [Fig.~\ref{fig:nu_vs_β}].

\begin{figure}[t]
    \includegraphics[width=\linewidth]{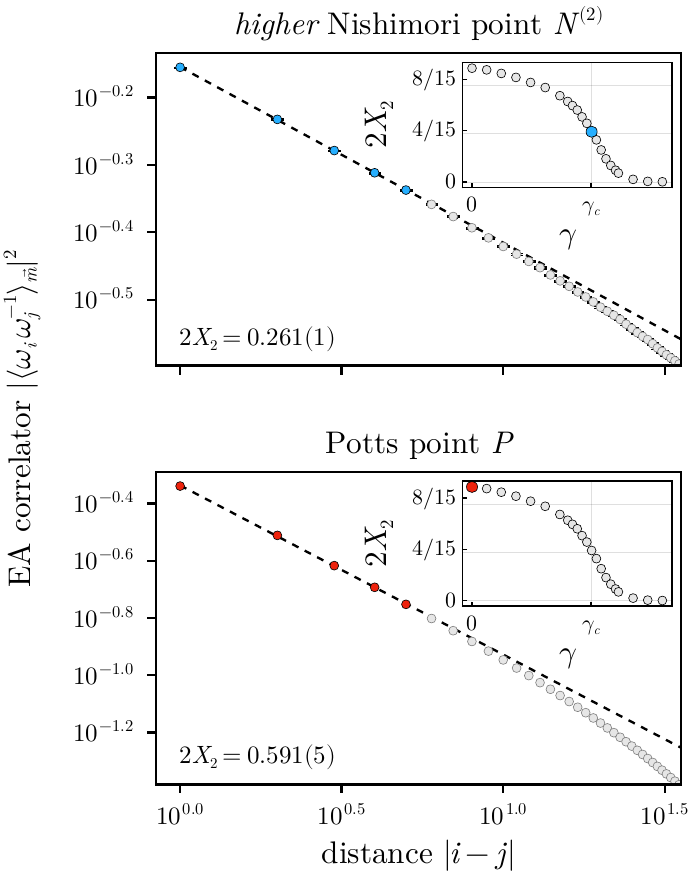}
    \caption{{\bf Edwards-Anderson correlator at critical Points.}
            This plot shows the power-law fit of the Edwards-Anderson correlator at the \higher  Nishimori point (top panel) and the Potts critical point (bottom panel) for a $256\times 256$ system.
    }
    \label{fig:oioj_higherNishimori_Potts}
\end{figure}

\subsubsection{Emergent higher Nishimori criticality}

The decisive test of emergent higher Nishimori criticality is provided by the long-distance decay of the EA correlator,
that is expected to fulfill the \higher Nishimori condition
\begin{equation}
  \overline{\left|
  \langle\omega_i\omega_j^{-1}\rangle_{\vec m}\right|^2}
  \sim \frac{1}{|i-j|^{2X_2}} \,.
  \label{eq:NishimoriCondition}
\end{equation}
As shown below, the higher Nishimori identity in Eq.~\eqref{EqPottsEquality} implies that this correlator decays at the learning tricritical point with the exponent of the spin correlator of the unmeasured
critical Potts model.  For $q=3$, the spin scaling dimension is fixed by Eq.~\eqref{EqSpinScalingDimensionPottsModel}, giving
\begin{equation}
    2X_2=2X_\sigma=\frac{4}{15}\simeq0.267 \,.
\end{equation}
To assess finite-size effects, we first extract the corresponding effective exponent at the unmeasured Potts critical point $P$.  There,
the EA correlator is the square of the spin correlator and hence must asymptotically decay with exponent
$4X_\sigma=8/15\simeq0.533$.  On our $256\times256$ systems, however, we obtain $2X_2=0.591(5)$; see
the lower panel of Fig.~\ref{fig:oioj_higherNishimori_Potts}.  This upward shift of about $0.06$ sets the scale of finite-size corrections in our simulations and is comparable to that observed in the Ising learning problem~\cite{PatilPutzTrebstZhuLudwig}. At the learning tricritical point, in contrast, we find
\begin{equation}
    2X_2=0.261(1),
\end{equation}
as shown in the upper panel of Fig.~\ref{fig:oioj_higherNishimori_Potts}.  The measured value agrees with the higher Nishimori prediction $4/15$ within the
systematic scale set by the calibration at $P$.  This agreement is the central numerical evidence for emergent higher Nishimori criticality 
at the tricritical point in the learning Potts model at hand: 
The microscopic discrete protocol explicitly breaks the enlarged $S_{R+1}$ replica permutation symmetry in the gauge-invariant formulation, 
yet its EA correlator is governed at long distances by the exponent dictated by that symmetry.  
We therefore identify the learning tricritical point of the $3$-state Potts model with the higher Nishimori universality class,
in direct analogy with the Ising case~\cite{PatilPutzTrebstZhuLudwig,WieseDasNahum}. 

\begin{figure}[t]
    \includegraphics[width=\linewidth]{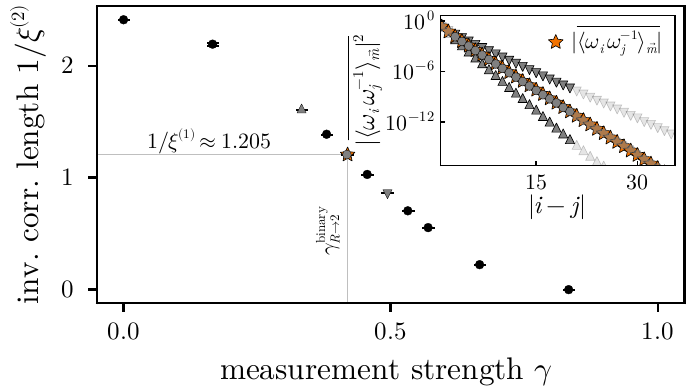}
    \caption{{\bf \textbf{\textit{Higher}}  Nishimori condition} along a high-temperature cut $\beta = (4 /7 )\beta_c$ in the paramagnetic phase above criticality. 
    Shown is the inverse correlation length $1/\xi^{(2)}$ extracted from the EA correlator
    as a function of the measurement strength $\gamma$. 
    The (inverse) EA correlation length is extracted from a fit, Eq.~\eqref{eq:NishimoriLine},
     and compared to the (inverse) spin-spin correlation length $1/\xi^{(1)}$ of the unmeasured Potts model at $\beta = (4 /7 )\beta_c$.
    }
    \label{fig:nishimori_condition}
\end{figure}

Off criticality, inside the paramagnetic phase, a similar approach of matching the correlation length $\xi^{(2)}$ capturing
the exponential decay of the EA correlator in this phase 
\begin{equation}
	\overline{|\langle \omega_{i}\omega_{j}^{-1}\rangle_{\vec{m}}|^{2}}\propto e^{-|i-j|/\xi^{(2)}}
	\label{eq:NishimoriLine}
\end{equation}
to the 
correlation length $\xi^{(1)}$ of the unmeasured model,
allows us to track the \higher Nishimori  line {\sl within} the paramagnetic phase. 
In Fig.~\ref{fig:nishimori_condition}, we show data for the inverse correlation length $1/\xi^{(2)}$
for a horizontal cut through the phase diagram at $\beta=4/7 \;\beta_c$. 
For the unmeasured Potts model (at this temperature) we extract $1/\xi^{(1)}\simeq1.205$ 
and by locating the measurement strength, for which $\xi^{(2)} = \xi^{(1)}$, we arrive at 
a numerical estimate for a point on the \higher Nishimori line $(\beta_N, \gamma_N)$. 
The so-determined points are given by orange circles in the phase diagram of Fig.~\ref{fig:numerical_phase_diagram_withGaussian}
for various temperatures.

\subsubsection{Intermediate \texorpdfstring{$L^{(1)}$}{Lg} fixed point}

\begin{figure}[t]
    \includegraphics[width=\linewidth]{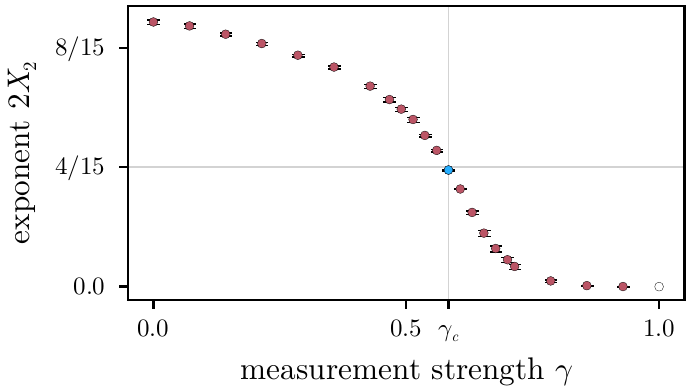}
    \caption{{\bf $2X_2$ along the $\beta = \beta_c$ line.}
    This plot shows the exponent $2X_2$ of the Edwards-Anderson correlator 
    along the $\beta = \beta_c$ line in the learning phase diagram of the $3$-state Potts model. 
    Data are obtained by fitting $\overline{|\langle \omega_{i}\omega_{j}^{-1}\rangle_{\vec{m}}|^{2}} \sim  \frac{1}{|i-j|^{2X_2}}$
    in the middle of a $256\times 256$ system.
    Note that the deviation at the clean Potts critical point from the expected value of $8/15$ is of similar magnitude 
    as it was in the Ising case~\cite{PatilPutzTrebstZhuLudwig} for $256\times256$ systems.
    The vertical line of $\gamma_c=\gamma_c^{(2)}$ is adopted from the finite-size scaling data collapse of the coherent information 
    shown in the top panel of Fig.~\ref{fig:Ic_nishimori12}, at which $2X_2$ agrees well with the analytic expected value. 
    }
     \label{fig:2X2_along_betac}
\end{figure}

Returning to the  critical-temperature $\beta=\beta_c$ line, Fig.~\ref{fig:2X2_along_betac} 
shows the EA exponent as a function of measurement strength.  
It evolves from the clean Potts value $8/15$ at $\gamma=0$ towards smaller values beyond the tricritical point.  The
3-loop
$\epsilon$-expansion prediction $2X_2\simeq0.315$ for the attractive
fixed point $L^{(1)}$, discussed in Ref.~\cite{NahumJacobsen}, is not resolved as a plateau at the available system sizes.  Instead, the continuous drift of the 
EA exponent is consistent with a slow crossover towards this weakly attractive fixed point.
The absence of a resolvable plateau in our simulations is not completely unexpected given the long crossover
scale. The measurement-induced disorder coupling $\Delta=
\tilde{\gamma}^2$ has an RG eigenvalue $y=\alpha/\nu=2/5$ at the clean Potts point, 
which yields an estimate of the crossover length scale $\ell_{\rm cross}\sim 
\Delta^{-5/2}$. For example, at the representative measurement strength
$\gamma=0.30$, the relation
$\widetilde{\gamma}=\frac{1}{3}
\ln\!\left(\frac{1+2\gamma}{1-\gamma}\right)$
gives $\widetilde{\gamma}=0.2756$ and
$\Delta=0.0759$, and hence
$\ell_{\mathrm{cross}}\sim6.3\times10^{2}$, 
which is of the same order of magnitude as
our maximum linear system size $L=256$. Equivalently, in our numerics, the crossover
variable $\Delta \ell^{2/5}$, where $\ell<L$ is the distance between two points in a correlator, seems to have reached only an $\mathcal{O}(1)$ value.
At stronger measurement the crossover away from the clean Potts point becomes shorter,
but the flow is then increasingly influenced by the nearby tricritical endpoint $N^{(2)}$, leaving no broad intermediate
window governed solely by the attractive $L^{(1)}$ fixed point. For accessible system sizes, the flow therefore remains
preasymptotic near the tricritical region. 

\subsubsection{RG crossover along the critical boundary}

\begin{figure}[t]
    \includegraphics[width=\linewidth]{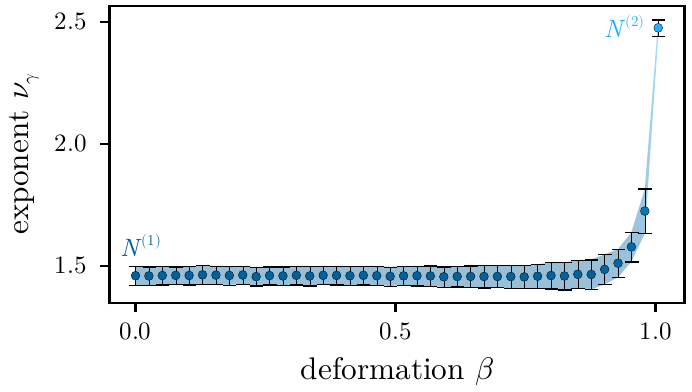}
    \caption{{\bf RG-Flow from \higher  to {\sl ordinary} Nishimori point.} Shown is the numerically obtained correlation length exponent $\nu$ as a function of $\beta$ for learning the $q=3$ Potts model. The exponent $\nu$ is obtained from finite-size scaling of the coherent information for system sizes $L = 8$ to $64$ except for the \higher  Nishimori point where we use $L = 16$ to $128$.}
    \label{fig:nu_vs_β}
\end{figure}

The phase boundary (between the paramagnetic and `spin glass' phases) emanating from the \higher Nishimori point $N^{(2)}$ 
bends towards the ordinary Nishimori point $N^{(1)}$ on the $\beta=0$ axis, as anticipated from the
replica analysis; see Fig.~\ref{fig:numerical_phase_diagram_withGaussian}.  Figure \ref{fig:nu_vs_β} shows the effective correlation-length exponent extracted from coherent-information scaling along this boundary.  Away from the tricritical region, it is close to the ordinary Nishimori value
$\nu_\gamma\simeq1.45$, while it rises towards
$\nu_\gamma\simeq2.5$ near $N^{(2)}$.  Within the accessible system sizes, this evolution is consistent with a crossover away from an unstable higher Nishimori multicritical point and towards the stable ordinary Nishimori critical point governing the generic transition.

\section{Exact Higher Nishimori Physics in a Specialized Measurement Protocol}
\label{SecSpecializedMeasurementProtocol}
In this section, we will consider a  Gaussian measurement protocol which yields the replica theory in Eq.~\eqref{EqReplicatedPottsHamiltonianUngauged}, 
in which the  higher-replica symmetric formulation [Eq.~\eqref{EqGaugedReplicatedPottsHamiltonian}]
on the \higher  Nishimori line $\beta=\Delta$ holds {\sl exactly} at the microscopic level. 
In this measurement protocol, we will replace the discrete $q$-state measurement protocol
of Eq.~\eqref{EqMeasurementProtocolA} by a 
complex continuous Gaussian measurement operation~\footnote{
Modulo the specific form of the measurement operation in Eq.~\eqref{eq:PmsigmaGaussianPotts}, in general, one might expect a Gaussian distribution with `{\sl blurred}' measurement outcomes 
to more closely model an experiment than the sharp Kronecker-delta measurement outcomes in Eq.~\eqref{EqMeasurementProtocolA}.
For a quantum measurement setup, 
a protocol to perform exact Gaussian quantum weak measurements was discussed in Ref.~\cite{GarrattWeinsteinAltman2022} by weakly coupling the measurement operator to a quantum harmonic oscillator.
} 
given by
\begin{equation}
	P((m_r)_{ij}|\{\omega_{i}\}) \propto  e^{-
        \frac{\Delta}{2} 
       \sum_{r=1}^{q-1}|{(m_r)}_{ij}-(\omega_i\omega^{-1}_j)^r|^2} \, ,
       \label{eq:PmsigmaGaussianPotts} 
\end{equation}
where 
${(m_r)}_{ij}\in \mathbb{C}$ \,
($r=1,...,q-1$) are the measurement outcomes  and since 
\begin{equation}
    (\omega_i^r\omega^{-r}_j)^{*}=\omega_i^{-r}\omega_{j}^r=\omega_i^{q-r}\omega_{j}^{-(q-r)},\;\dots\;[\omega_i^q=1]
\end{equation}
the measurement outcomes are also taken to satisfy 
\begin{equation}
    (m_r)_{ij}^{*}=(m_{q-r})_{ij}\,\;\;\;\; r=1, \cdots,q-1.\label{EqMeasurementOutcomesRelatedByComplexConjugation}
\end{equation}
Of course, the measurement protocol can be 
translated into that of measuring the corresponding real and imaginary parts of $(\omega_i\omega^{-1}_j)^r$ ($r=1, \cdots,q-1$), where Eq.~\eqref{EqMeasurementOutcomesRelatedByComplexConjugation} is naturally satisfied.
This measurement protocol is inspired 
by Ref.~\cite{NishimoriStephen} which identified the ordinary Nishimori line (gauge-invariant limit $R\rightarrow1$) 
in the random-bond Potts model (replica limit $R\rightarrow0$), which makes an appearance on the infinite temperature $\beta=0$ line in our learning phase diagram.
This Gaussian measurement protocol allows us to unambiguously identify the \higher  Nishimori line (gauge-invariant limit $R\rightarrow2$) in the learning phase diagram of the Potts model (replica limit $R\rightarrow1$). 

Unlike the $q$-state measurement protocol 
[Eq.~\eqref{EqMeasurementProtocolA}] in 
the 
previous sections, 
it can be easily verified (see, e.g., App.~C of Ref.~\cite{PatilPutzTrebstZhuLudwig})
that the replica theory for the above measurement protocol is given by Eq.~\eqref{EqReplicatedPottsHamiltonianUngauged} {\sl without} the higher-order $\mathcal{O}(\tilde{\gamma}^3)$ terms in Eq.~\eqref{EqReplicatedPottsHamiltonianProtocolI}.
In particular, along the 
\higher  Nishimori line [Eq.~\eqref{EqDefHigherNishimoriLine}] given by 
\begin{equation}
    \beta=\Delta \,,
\end{equation}
the replica Hamiltonian for the learning phase diagram with the above measurement protocol is precisely given by the Hamiltonian in Eq.~\eqref{EqReplicatedPottsHamiltonianUngaugedII}, which admits the higher replica symmetric gauge-invariant formulation in Eq.~\eqref{EqGaugedReplicatedPottsHamiltonian}.
Then using Eq.~\eqref{EqReplicatedNishimoriIdentityBulk} [for 
a derivation, see App.~\ref{AppDerivationOfIdentityForMoments}] in the $R\rightarrow1$ replica limit and for $n=1$, the Potts correlation functions satisfy the following equality along the \higher Nishimori line $\beta=\Delta$
\begin{equation}
    \overline{|\langle \omega_{i_1}^{l_1}\omega_{i_2}^{l_2}\cdots\omega^{l_k}_{i_k}\rangle_{\vec{m}}|^{2}}=\overline{\langle \omega_{i_1}^{l_1}\omega_{i_2}^{l_2}\cdots\omega^{l_k}_{i_k}\rangle_{\vec{m}}},\;\;\;\;
    [l_1, \cdots,l_k\in \mathbb{Z}].\label{EqPottsK-pointEquality}
\end{equation}
In particular, since measurement-averaged first moments are unaffected by conditioning on measurement outcomes and
the first moment average is equal to the unmeasured correlation function~\footnote{As discussed in Sec.~\ref{SecQuantumProtocol}, 
the Bayesian inference problem can be formulated as a quantum Born-rule measurement problem on the Rokhsar-Kivelson wavefunction for the stat-mech model~\cite{PutzGarrattNishimoriTrebstZhu,PatilLudwig20251}. 
In the quantum formulation, the second equality in  Eq.~\eqref{EqPottsEqualityLongDistance} can be understood as a simple consequence of the POVM condition, 
$\sum_{\vec{m}}\hat{K}_{\vec{m}}^{\dagger}\hat{K}_{\vec{m}}=1$.
Note that
\begin{align*}
&\overline{\langle \hat{\mathcal{O}}_1\rangle_{\vec{m}}}=\sum_{\vec{m}}P(\vec{m})\bra{\Psi_{\vec{m}}}\hat{\mathcal{O}}_1\ket{\Psi_{\vec{m}}}
\\&=\sum_{\vec{m}}\bra{RK}\hat{K}^{\dagger}_{\vec{m}}\hat{K}_{\vec{m}}\ket{RK}\frac{\bra{RK}\hat{K}^{\dagger}_{\vec{m}}\hat{\mathcal{O}}_1\hat{K}_{\vec{m}}\ket{RK}}{\bra{RK}\hat{K}^{\dagger}_{\vec{m}}\hat{K}_{\vec{m}}\ket{RK}}\\
&=\sum_{\vec{m}}{\bra{RK}\hat{K}^{\dagger}_{\vec{m}}\hat{\mathcal{O}}_1\hat{K}_{\vec{m}}\ket{RK}}\\&={\bra{RK}\hat{\mathcal{O}}_1\sum_{\vec{m}}\hat{K}^{\dagger}_{\vec{m}}\hat{K}_{\vec{m}}\ket{RK}}={\bra{RK}\hat{\mathcal{O}}_1\ket{RK}}=\langle \hat{\mathcal{O}}_1\rangle \,,
\end{align*}
where we have commuted $\hat{\mathcal{O}}_1$ through the Kraus operator $\hat{K}_{\vec{m}}$, since both are diagonal in the basis defined by stat-mech configurations, and {then} used the POVM condition.}, 
we obtain that along the \higher  Nishimori line $\beta=\Delta$
\begin{equation}
    \overline{|\langle \omega_{i}\omega_{j}^{-1}\rangle_{\vec{m}}|^{2}}=\overline{\langle \omega_{i}\omega_{j}^{-1}\rangle_{\vec{m}}}=\langle \omega_{i}\omega_{j}^{-1}\rangle\,, 
    \label{EqPottsEquality}
\end{equation}
where the correlation function $\langle \cdots\rangle$ is evaluated in the unmeasured $q$-state Potts model [Eq.~\eqref{EqPottsHamiltonian}] at (inverse) temperature $\beta$.
The critical point of the $q$-state Potts model in Eq.~\eqref{EqPottsHamiltonian} for $2<q\leq 4$ is known to occur~\cite{Potts_1952} at
\begin{equation}
    \label{EqLocationOfPottsCriticalPt}
    \beta_{c}^{(q)}=\frac{1}{q}\ln{(1+\sqrt{q})} \, .
\end{equation}
Then, using Eq.~\eqref{EqPottsEquality},
which implies an identical critical scaling behavior for
the first and the second measurement-averaged moment of the Potts spin correlator, 
we see that the $q$-state Potts learning phase diagram ($2<q\leq 4$) with the measurement protocol in Eq.~\eqref{eq:PmsigmaGaussianPotts} should have a distinct critical point lying at
\begin{equation}
    \label{EqPottsHigherNishimori}\beta=\Delta=\beta_{c}^{(q)}=\frac{1}{q}\ln{(1+\sqrt{q})}.
\end{equation}
Since the critical point lies on the \higher  Nishimori line $\beta=\Delta$, we will refer to it as the \higher  Nishimori critical point. 
In fact, following exactly the same argument as that given for the ordinary Nishimori critical point in the phase diagram of the RBIM~\cite{LeDoussalHarrisI}, due to the equality 
in Eq.~\eqref{EqPottsEquality} that is satisfied on the entire \higher  Nishimori line, the higher Nishimori critical point in the learning phase diagram must lie 
at the intersection of the ferromagnet-to-paramagnet phase boundary and
the paramagnet-to-`spin-glass'~\footnote{The `spin-glass' phase
is characterized by 
long-range order in the measurement-averaged second moment of the spin-spin correlator while its measurement-averaged first moment remains short-ranged.} 
phase boundary, i.e. it is a tricritical point where the three phases meet.

\section{Exact Universal Properties and the Phase Diagram}
In this section, we will discuss a number of exact universal properties of the critical points in the Potts learning phase diagram. This includes a discussion of exact results at the \higher  Nishimori critical point, and a general argument for stability of the ordinary Nishimori critical universality class in monitored problems. 
We will also discuss the Casimir effective central charges of the critical points and 
their monotonic {\sl decrease} under the 
measurement-induced RG flows that follows  from the $c$-effective theorem~\cite{PatilLudwig20251}.
We will  contrast this with
the Casimir effective central charges of the corresponding critical points in the random-bond Potts model phase diagram, 
and their corresponding {\sl increase} along RG flows.

\subsection{Learning Phase Diagram}

\subsubsection{The \higher  Nishimori universality class \texorpdfstring{$N^{(2)}$}{Lg}}

A sketch of different critical points and RG flows for the learning problem of the $q$-state ($2<q\leq 4$) Potts model with the measurement protocol in Eq.~\eqref{eq:PmsigmaGaussianPotts} is presented in Fig.~\ref{FigPottsLearningPhaseDiagram}(a). 
Given global symmetries, one typically expects the long-distance behavior of the critical points in the learning phase diagram to be largely independent of microscopic differences in energy-density measurement protocols.  
For example, we discussed the $q$-state
measurement protocol in Eq.~\eqref{EqMeasurementProtocolA} in Sec.~\ref{SecReplicaTheoryAndHigherNishimoriLine}, and 
at a technical level
the terms of $\mathcal{O}(\tilde{\gamma}^3)$ in the
replica theory [Eq.~\eqref{EqReplicatedPottsHamiltonianProtocolI}] 
quantify the difference between the $q$-state measurement protocol 
and the Gaussian protocol in Eq.~\eqref{eq:PmsigmaGaussianPotts}. 
These higher-order $\mathcal{O}(\tilde{\gamma}^3)$ terms precisely prohibit the existence of an exact \higher  Nishimori line, which has a higher-replica-symmetric gauge-invariant formulation, in the learning phase diagram with the measurement protocol in Eq.~\eqref{EqMeasurementProtocolA}.
However, if the above discussed  $\mathcal{O}(\tilde{\gamma}^3)$ terms are irrelevant at the tricritical point in the RG sense,
the higher replica symmetric structure should emerge in the infrared. The equality in Eq.~\eqref{EqPottsEquality} would then be replaced by an equivalence at large distances, i.e.
\begin{equation}
    \overline{|\langle \omega_{i}\omega_{j}^{-1}\rangle_{\vec{m}}|^{2}}\sim\overline{\langle \omega_{i}\omega_{j}^{-1}\rangle_{\vec{m}}}=\langle \omega_{i}\omega_{j}^{-1}\rangle,\label{EqPottsEqualityLongDistance}
\end{equation}
where $\sim$ denotes equivalence at long distances~\footnote{Note that the second equality in the above equation remains exact because measurement-averaged first moments are unaffected by conditioning on measurement outcomes~\cite{Note7}.}.
As discussed in Sec.~\ref{SecNumericalResults}, by comparing correlators in Eq.~\eqref{EqPottsEqualityLongDistance}, we have found numerical evidence for the {\sl emergence} of \higher  Nishimori criticality at the tricritical point in the learning phase diagram with the $q$-state
measurement protocol [Eq.~\eqref{EqMeasurementProtocolA}].

\subsubsection{Stability of ordinary Nishimori critical universality class \texorpdfstring{$N^{(1)}$}{Lg}}

The ordinary Nishimori line and the ordinary Nishimori critical point~\cite{NishimoriStephen,JacobsenPicco,HoneckerJacobsenPiccoPujol} from the random-bond Potts model (replica limit $R\rightarrow0$) appear  in their `gauge-invariant' formulation as a replica theory in the $R\rightarrow1$ replica limit with local symmetry, respectively, as the $\beta=0$ line and the critical point on the $\beta=0$ line in the learning phase diagram.
[Note that when $\beta=0$, the $R$-replica Hamiltonian in Eq.~\eqref{EqReplicatedPottsHamiltonianProtocolI} has local symmetry in Eq.~\eqref{EqGaugeTransformationRReplica},
which exists also in the replica $R\rightarrow1$ limit.]
Analogous to the Ising learning phase diagram~\cite{PutzGarrattNishimoriTrebstZhu}, we expect an RG flow from the \higher  Nishimori critical point identified in the previous section to the ordinary Nishimori critical point sitting at $\beta=0$.
This is because the ordinary Nishimori critical point in the learning phase diagram at $\beta=0$ describes the paramagnet to `spin-glass' transition, where the `spin-glass' phase is characterized by ordering of the measurement-averaged second moment of the spin-spin correlation function, i.e.\ the EA correlator.
Since at inverse temperatures $\beta<\beta_c$, depending on the measurement strength, we can again have either a paramagnet or a `spin-glass' phase, the phase transition between them is 
characterized by ordering of the EA parameter, while the spin-spin correlation remains short-ranged.
Hence, 
the learning transition for $\beta<\beta_c$ is
expected to be in the same universality class as the ordinary Nishimori critical point on the $\beta=0$ line.
This suggests an RG flow from the \higher  Nishimori critical point to the ordinary Nishimori critical point on the $\beta=0$ line.

We have observed this RG flow numerically in Sec.~\ref{SecNumericalResults} with the $q$-state protocol 
[Eq.~\eqref{EqMeasurementProtocolA}].

In fact, it can be argued that reducing temperature at the ordinary Nishimori critical point on the infinite-temperature ($\beta=0$) line must be trivially an RG irrelevant perturbation.
This is because at infinite temperature ($\beta=0$) the $R$-replica theory in Eq.~\eqref{EqReplicatedPottsHamiltonianProtocolI} has a local symmetry under simultaneous flip of spins in all replica copies at any given site [Eq.~\eqref{EqGaugeTransformationRReplica}].
All the non-trivial (scalar) scaling fields of the ordinary Nishimori critical point should be
invariant under this local symmetry. 
Reducing the temperature introduces the finite-$\beta$ term in Eq.~\eqref{EqReplicatedPottsHamiltonianProtocolI}, which preserves the global $\mathbb{Z}_q$ symmetry but 
does not respect the above local symmetry. 
A simple calculation based on Elitzur's theorem then shows that the transition at small $\beta\neq 0$ is in the same universality class as at $\beta=0$~\footnote{Most of the terms in the Taylor expansion about the unperturbed ordinary Nishimori transition at $\beta=0$ [cf. Eq.~\eqref{EqReplicatedPottsHamiltonianProtocolI}], in terms of the correlation functions of the perturbation, vanish identically due to the local symmetry, by Elitzur's theorem~\cite{Elitzur1975}. This is except for `contact-terms' where terms in a multi-bond correlator (calculated with the unperturbed theory) appear in combinations invariant under
the above local symmetry at any given bond $\langle ij\rangle$. The latter terms, which are invariant under the above local symmetry as well as (by construction) under
the global $\mathbb{Z}_q$ symmetry, are the same as the (local and global) symmetry allowed terms that are already present on the ordinary Nishimori line $\beta=0$ [Eq.~\eqref{EqReplicatedPottsHamiltonianProtocolI}]. These terms, therefore, only shift the (non-universal) location of the transition
along the ordinary Nishimori line, and do not introduce additional relevant scaling fields at the critical point.}.
That is, the thermal  perturbation is trivially irrelevant under RG. 
The argument 
is a variation of the seminal argument
for stability of pure-gauge transitions in Refs.~\cite{Wegner1971,FradkinShenker1979}.
Here, we have 
applied 
the
argument
based on Elitzur's theorem
to the replica $R\rightarrow1$ theory of the ordinary Nishimori critical point, which, in contrast to its random-bond Potts (replica $R\rightarrow0$) formulation, 
exhibits local (`gauge') symmetry.

This is a rather general argument that explains the stability of the (ordinary) Nishimori critical
universality class in other monitored/learning
problems that are governed by the replica $R\rightarrow1$ limit. 
For example, the same argument applies to the ordinary Nishimori critical point in the classical Bayesian inference phase diagram for the Ising model considered in Ref.~\cite{PutzGarrattNishimoriTrebstZhu}, equivalent to the learning phase diagram of the wavefunction-deformed toric code~\cite{CastelnovoChamon,ArdonneFedleyFradkin,PapanikolaouRamanFradkin,IsakovFendleyLudwigTrebstTroyer,Zhu19deform,Verresen25deform}.
In the toric code learning problem, in contrast to its replica $R\rightarrow0$ formulation in the random-bond Ising model, the ordinary Nishimori critical point appears as a replica theory in the $R\rightarrow1$ replica limit with
a local symmetry, and hence following the above argument,
the wavefunction-deformation 
should be trivially RG irrelevant. 
It also suggests the stability of the ordinary Nishimori transition in the
learning
phase diagram of the transverse-field Hamiltonian-deformed toric code~\cite{FradkinShenker1979,TrebstWernerTroyerShtengelNayak,TupitsynKitaevProkofevStamp}.
(A related but different phase diagram for the optimal decoding problem of the Hamiltonian-deformed toric code was considered in Ref.~\cite{WeinsteinGarratt2026}.)

\subsubsection{An attractive fixed point \texorpdfstring{$L^{(1)}$}{Lg}\label{SubSecAttractiveFixedPoint}}

Besides the higher Nishimori critical point $N^{(2)}$ and the ordinary Nishimori critical point $N^{(1)}$, for $2<q\leq 4$,
there exists another (finite-measurement) fixed point in the learning phase diagram of the $q$-state Potts model, which is an {\sl attractive} fixed point that is obtained by performing 
weak energy-density measurements on the unmeasured critical point. 
The existence of this fixed point was demonstrated 
in Ref.~\cite{NahumJacobsen} by using the result of
a controlled epsilon expansion in $\epsilon=q-2$
developed to study the fixed point $L^{(0)}$ [Fig.~\ref{FigPottsLearningPhaseDiagram} (b)] in the
random-bond Potts models~\cite{Ludwig-Potts-OPE-RG-NPB285-1987-97,LUDWIGCardy} (generalized to higher
loop order in~\cite{DotsenkoPiccoPujol,DotsenkoJacobsenLewisPicco}) and by applying it to the $R\rightarrow1$
replica limit.
Since both the Gaussian [Eq.~\eqref{eq:PmsigmaGaussianPotts}]
and the $q$-state measurement protocol [Eq.~\eqref{EqMeasurementProtocolA}]
correspond to measuring bond energy densities [recall $\delta_{\omega_i,\omega_j}=\frac{1}{q}\sum_{r=0}^{q-1}(\omega_i\omega_j^{-1})^r$], for weak measurements at the unmeasured 
Potts critical point, we expect an RG flow to the above-discussed attractive fixed point. This fixed point is denoted by ${L}^{(1)}$ in Fig.~\ref{FigPottsLearningPhaseDiagram}(a).

\subsubsection{Ferromagnet-to-spin-glass transitions}

Finally, let us turn to the zero-temperature spin-glass critical point ($S^{(0)}$) in the RBPM phase diagram and the projective-measurement critical point ($S^{(1)}$) in the learning phase diagram 
[Fig.~\ref{FigPottsLearningPhaseDiagram}]. 
Both of these critical points capture the ordering transition of the first moment of the spin-spin correlator while the second moment remains ordered. Of course, the moments are evaluated by averaging over statistically uncorrelated disorder in the RBPM and by averaging over measurement outcomes in the learning phase diagram, which is precisely captured by taking the different replica limits $R\rightarrow0$ and  $R\rightarrow1$  for the respective critical points.
We note that as the replica limit is moved from $R\rightarrow1$ to $R\rightarrow0$, the `spin-glass' phase in the learning phase diagram seems to collapse to a part of the $T=0$ line in the RBPM phase diagram.
In a sense, this change in the replica limit from $R\rightarrow0$ to $R\rightarrow1$ is quite similar to, but more drastic than, moving from  spatial dimension $d=2$ to $d=3$~\cite{BinderYoungSpinGlass}, where the shared feature in both cases is that
we go from a spin-glass line to an extended `spin-glass' phase in the phase diagram.

We note that while the critical properties of the projective-measurement fixed point $S^{(1)}$ seem to be completely fixed by the unmeasured critical theory~\footnote{In the projective limit, the configuration of the underlying Potts spins is completely pinned (up to a $\mathbb{Z}_q$ flip) by the measurement outcomes. Then the measurement-averaged moments of the correlation functions are straightforwardly determined using the known correlation functions at the unmeasured critical point. This is in contrast with the $L^{(1)}$ and $N^{(2)}$ critical points [Fig.~\ref{FigPottsLearningPhaseDiagram}], where moments of correlation functions (except for one or two lowest moments) are expected to display completely distinct critical behavior from the unmeasured critical point.}
without frustration, the zero-temperature fixed point $S^{(0)}$, on the other hand,
is a minimal-weight perfect matching problem where frustrated vortices need to be paired 
(see, e.g., Ref.~\cite{PandeyMahadevanMiddletonFisher} for a recent development on $S^{(0)}$ in the Ising case~\footnote{We note that the strong-disorder RG for the Ising case in~\cite{PandeyMahadevanMiddletonFisher} proceeds via a mapping to a $2D$ quantum free-fermion Hamiltonian (see also Ref.~\cite{MotrunichDamleHuse}). For the $q>2$ state Potts model, there is no obvious free-fermion mapping, and it might be difficult to find a controlled strong-disorder RG analysis for the $S^{(0)}$ critical point in the RBPM [Fig.~\ref{FigPottsLearningPhaseDiagram}(b)].}).
Thus, in addition to destruction of the extended spin-glass phase, changing the replica limit
from $R\rightarrow1$
to $R\rightarrow0$ also seems to drastically change the physics of the ferromagnet-to-spin-glass critical point. This is quite different from the pair
 $L^{(0)}$ and $L^{(1)}$, or the pair  $N^{(1)}$ and $N^{(2)}$, in the RBPM and learning phase diagrams, where the universal critical properties are expected to smoothly change upon changing the replica limit from $R\rightarrow0$ to $R\rightarrow1$. Subtleties in using the replica trick to study the zero-temperature spin-glass critical point $S^{(0)}$ for the Ising case have been previously noted, e.g.\   Refs.~\cite{Aharony_1978,Domany_1979}.

\subsection{The Edwards-Anderson correlator and higher moments}
For $2<q\leq 4$, in the unmeasured critical $q$-state Potts model the correlation function $\langle \omega_{i}\omega_{j}^{-1}\rangle$ exhibits the following power-law decay
\begin{equation}
    \langle \omega_{i}\omega_{j}^{-1}\rangle\sim \frac{1}{|i-j|^{2X_\sigma}},\;\;\;|i-j|\gg1
\end{equation}
where $X_\sigma$ is the scaling dimension of the spin field in the $q$-state Potts CFT~\cite{DotsenkoFateev}
\begin{align}
    X_\sigma=&\frac{y^2-1}{4y(2y-1)},\;\;\text{where}\;\;
    y=\frac{\pi}{2\arccos(\sqrt{q}/2)}.\label{EqSpinScalingDimensionPottsModel}
\end{align}
Consequently, at the \higher  Nishimori point of the $q$-state Potts learning phase diagram 
with the Gaussian protocol [Eq.~\eqref{eq:PmsigmaGaussianPotts}]
we obtain, from Eq.~\eqref{EqPottsEqualityLongDistance}, that
\begin{equation}
    \overline{|\langle \omega_{i}\omega_{j}^{-1}\rangle_{\vec{m}}|^{2}}=
    \overline{\langle \omega_{i}\omega_{j}^{-1}\rangle_{\vec{m}}}\sim  \frac{1}{|i-j|^{2X_\sigma}},
\end{equation}
where $X_{\sigma}$ is given in Eq.~\eqref{EqSpinScalingDimensionPottsModel}. 
For the $3$-state Potts model 
\begin{equation}
    X_{\sigma}=2/15 \,,
\end{equation}
and the Edwards-Anderson correlator should decay with the same power-law exponent at the \higher  Nishimori critical point in the $3$-state Potts learning phase diagram. 
As discussed above, the exponent numerically obtained with the $q$-state protocol 
[Eq.~\eqref{EqMeasurementProtocolA}] is also consistent with the above value  (see Sec.~\ref{SecNumericalResults}).

Moreover, we note that it follows from Eq.~\eqref{EqPottsEquality}, and the identity between the first moment correlation functions and 
the unmeasured correlation functions~\cite{Note7}, that at the higher Nishimori critical point the measurement-average of the modulus square of any correlation function is equal to the
correlation function at the unmeasured critical point
\begin{align}
    \overline{|\langle \omega_{i_1}^{l_1}\omega_{i_2}^{l_2}\cdots\omega^{l_k}_{i_k}\rangle_{\vec{m}}|^{2}}={\langle \omega_{i_1}^{l_1}\omega_{i_2}^{l_2}\cdots\omega^{l_k}_{i_k}\rangle},\;\;\;\;
    [l_1, \cdots,l_k\in \mathbb{Z}]
\end{align}
where the long-distance properties of the unmeasured correlation function ${\langle \omega_{i_1}^{l_1}\omega_{i_2}^{l_2}\cdots\omega^{l_k}_{i_k}\rangle}$ at the $2D$ Potts critical point are known exactly from 
$2D$ rational CFT literature~\cite{BelavinPolyakovZamolodchikov,DOTSENKO1984Potts,DotsenkoFateev}.
A more general correlation function equality satisfied at the higher Nishimori critical point is obtained by taking $R\rightarrow1$ replica limit of Eq.~\eqref{EqReplicatedNishimoriIdentityBulk} [for derivation, see App.~\ref{AppDerivationOfIdentityForMoments}],
\begin{align}
    &\overline{\langle \omega_{i_1}^{l_1}\omega_{i_2}^{l_2}\cdots\omega^{l_k}_{i_k}\rangle_{\vec{m}}^{n}}=\nonumber\\
    &=\overline{\langle \omega_{i_1}^{l_1}\omega_{i_2}^{l_2}\cdots\omega^{l_k}_{i_k}\rangle_{\vec{m}}^{n}{\langle \omega_{i_1}^{-nl_1}\omega_{i_2}^{-nl_2}\cdots\omega^{-nl_k}_{i_k}\rangle}_{\vec{m}}},\label{EqGeneralHigherNishimoriEquality}\\
    &\;\;\;\;\;\;\;\;\;\;[l_1, \cdots,l_k\in \{0,1,\cdots, q-1\}].\nonumber
\end{align}

It is rather surprising that our identification of the higher Nishimori structure of the learning tricritical point allows us to obtain
a number of non-trivial exact results, e.g. the Edwards-Anderson correlator,
at this strongly-disordered and frustrated critical point.  
Finally, analogous to the discussion in Ref.~\cite{PatilPutzTrebstZhuLudwig}, using the 
known exponents for the first and second moments, we can obtain various rigorous bounds on the power-law exponents for higher ($>2$) measurement-averaged moments. For example, it is easily demonstrated that the measurement-averaged $(2n)^{\text{th}}$ moment of the $|\langle \omega_{i}\omega_{j}^{-1}\rangle_{\vec{m}}|$ correlation function decays as a power law,
\begin{equation}
    \overline{|\langle \omega_{i}\omega_{j}^{-1}\rangle_{\vec{m}}|^{2n}}\sim  \frac{1}{|i-j|^{2X_{2n}}}
\end{equation}
and the power-law exponent 
satisfies
\begin{equation}
    X_{\sigma}\leq X_{2n}\leq n X_\sigma
\end{equation}
with $X_\sigma$ given in Eq.~\eqref{EqSpinScalingDimensionPottsModel}.
\par
As discussed in Sec.~\ref{SubSecAttractiveFixedPoint}, the universal properties of the attractive fixed point $L^{(1)}$ can be studied in a perturbative epsilon expansion in $q-2$~\cite{NahumJacobsen,Ludwig-Potts-OPE-RG-NPB285-1987-97,LUDWIGCardy,LUDWIG1990infinitehierarchy,DotsenkoPiccoPujol,DotsenkoJacobsenLewisPicco}. For example, as obtained in Ref.~\cite{NahumJacobsen} (also see~\cite{Ludwig-Potts-OPE-RG-NPB285-1987-97,LUDWIG1990infinitehierarchy,DotsenkoJacobsenLewisPicco}) at the attractive fixed point $L^{(1)}$ in the $3$-state Potts Bayesian inference/learning phase diagram, the Edwards-Anderson correlator decays with the power-law exponent
\begin{equation}
    X_2\approx0.1577, \quad \text{(3-loop \ order \ result)}\,.
\end{equation} 
Since all measurement strengths in Fig.~\ref{FigPottsLearningPhaseDiagram}(a)
between $0<\Delta<\beta_c$ at the 
unmeasured Potts critical point $P$ 
flow under RG
to the attractive fixed point $L^{(1)}$,
the asymptotic long-distance behavior of the Edwards-Anderson correlator in the entire $0<\Delta<\beta_c$ interval should be characterized by the above exponent.

\subsection{Casimir effective central charge and the \texorpdfstring{$\bf c$}{Lg}-effective theorem}

Since weak measurements of energy-densities take us under RG flow to the critical point $L^{(1)}$ in Fig.~\ref{FigPottsLearningPhaseDiagram}(a), using the $c$-effective theorem in Ref.~\cite{PatilLudwig20251},
the Casimir effective central charge at $L^{(1)}$
must be {\sl less} than the central charge of the unmeasured Potts critical point
\begin{equation}
    c_{\rm Potts}>c_{\text{eff}}^{L^{(1)}} \, .
    \label{EqCeffL(1)andPotts}
\end{equation}
As mentioned
above, the universal properties of the $L^{(1)}$ fixed point can be studied perturbatively~\cite{NahumJacobsen} by using the result of a controlled epsilon expansion in
$\epsilon=q-2$ developed for the random-bond Potts models~\cite{Ludwig-Potts-OPE-RG-NPB285-1987-97,LUDWIGCardy}
(generalized to higher loop order in~\cite{DotsenkoPiccoPujol,DotsenkoJacobsenLewisPicco}) and by applying it
to the $R\rightarrow1$ replica limit. The Casimir effective central charge of the $L^{(1)}$ fixed point in this $\epsilon=q-2$ expansion is given by~\cite{NahumJacobsen}
\begin{equation}
     c_{\text{eff}}^{L^{(1)}}-c_{\rm Potts}=-\frac{\alpha^3}{8}-\frac{3\alpha^4}{16}+\mathcal{O}(\alpha^5),
\label{EqCeffEpsilonExpansionResult}
\end{equation}
where $\alpha=\frac{4\epsilon}{3\pi}-\frac{4\epsilon^2}{9\pi^2}+\mathcal{O}(\epsilon^3)>0$ and $c=1-\frac{3}{y(2y-1)}$, with $y$ given in Eq.~\eqref{EqSpinScalingDimensionPottsModel}, is the central charge 
of the unmeasured $q$-state Potts critical point~\cite{DotsenkoFateev}.
For example, for the unmeasured $3$-state Potts critical point
\[
	c_{\rm Potts} = 4/5 \,.
\]

As noted in Ref.~\cite{PatilLudwig20251}, from the above epsilon expansion result, it is clear that the effective central charge for the $L^{(1)}$ attractive fixed point,
at least for small values of $\epsilon=q-2$, is
less than the central charge $c_{\rm Potts}$ of the unmeasured critical
$q$-state Potts model, which is consistent with the general $c$-effective theorem. 
In particular, for the fixed point $L^{(1)}$ in the $3$-state Potts model learning phase diagram [Fig.~\ref{FigPottsLearningPhaseDiagram}(a)], the epsilon expansion result in Eq.~\eqref{EqCeffEpsilonExpansionResult} gives 
\[
	c_{\text{eff}}^{L^{(1)}}\approx0.79 \,, 
\]
which is very close to, {\sl but below}, the central charge of the unmeasured critical point.

\begin{figure*}[t]
	\includegraphics[width=\linewidth]{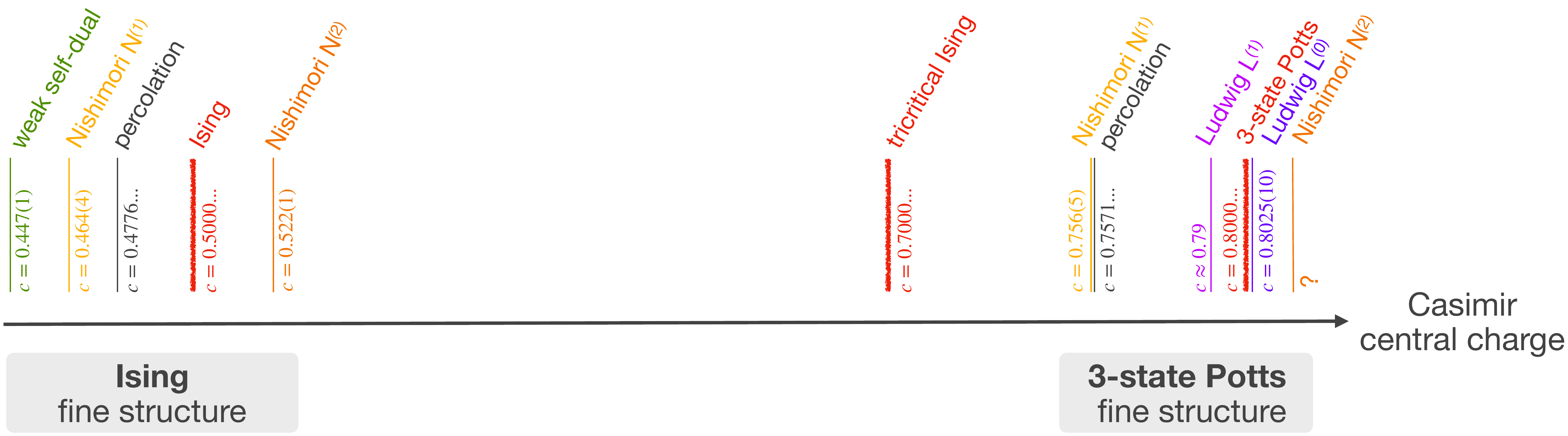}
	\caption{{\bf Central charge fine structure.} 
		The plot shows, to scale, the ``fine structure splitting" of the Ising and $3$-state Potts universality classes and their non-unitary daughter theories, 
		which have Casimir effective central charges near their corresponding unitary parent theory.
		Most numerical estimates for the Potts case come from Ref.~\cite{JacobsenPicco}.
        		Due to the $c$-effective theorem, the location of the higher Nishimori Potts point should be above the attractive fixed point $L^{(1)}$,
		 but the placement above the clean Potts (and $L^{(0)}$) is entirely speculative.
	\label{fig:CentralChargeLadder}}
\end{figure*}

Moreover, 
using 
the extension of the $c$-effective theorem proved in Refs.~\cite{PatilLudwig20262,PatilPutzTrebstZhuLudwig}, it also follows that the Casimir effective central charge decreases under the RG flow from the higher Nishimori critical point $N^{(2)}$ to the attractive fixed point $L^{(1)}$. 
Therefore, the Casimir effective central charge $c_{\text{eff}}^{N^{(2)}}$ of the higher Nishimori critical point in the Potts model learning phase diagram [Fig.~\ref{FigPottsLearningPhaseDiagram}(a)] must be greater than the 
effective
central charge of the
attractive fixed
point $L^{(1)}$, i.e.,
\begin{equation}
    c_{\text{eff}}^{N^{(2)}} > c_{\text{eff}}^{L^{(1)}}.\label{EqCeffN(2)andL(1)}
\end{equation}
As discussed in Ref.~\cite{PatilLudwig20251}, the decrease of the Casimir effective central charge demonstrated non-perturbatively by the $c$-effective theorem for RG flows in the $R\rightarrow1$ replica limit, can be used to argue for monotonic {\sl increase} of the Casimir effective central charge for RG flows in the $R\rightarrow0$ replica limit, which corresponds to systems with quenched impurity-type disorder. 
The latter argument follows straightforwardly from combining the result of the $c$-effective theorem with a natural, physically motivated assumption about 
the 
central charge 
of the replica theory (see Refs.~\cite{PatilLudwig20251,PatilPutzTrebstZhuLudwig}). 
In the present case, the argument implies that for the corresponding fixed points $P$, $L^{(0)}$, and $N^{(1)}$ in the random-bond Potts model [replica limit $R\rightarrow0$, Fig.~\ref{FigPottsLearningPhaseDiagram}(b)], the inequalities in Eq.~\eqref{EqCeffL(1)andPotts} and Eq.~\eqref{EqCeffN(2)andL(1)} should be reversed, i.e.\ 
\[
	c_{\rm Potts} < c_{\text{eff}}^{L^{(0)}}
\] 
and 
\[ 
	c_{\text{eff}}^{N^{(1)}} < c_{\text{eff}}^{L^{(0)}} \,.
\]
Indeed, in the phase diagram of the random-bond Potts model
it has long been known numerically for the $3$-state case~\cite{JacobsenPicco}
that the Casimir effective central charge increases under the RG flow from the {ordinary} Nishimori critical point $N^{(1)}$ to the attractive ferromagnetic fixed point $L^{(0)}$, and that it also increases under the RG flow from the clean Potts critical point to the attractive ferromagnetic fixed point $L^{(0)}$.
The latter result was independently known, even before the numerical result, from 
the above-discussed epsilon expansion~\cite{Ludwig-Potts-OPE-RG-NPB285-1987-97,LUDWIGCardy} 
for the $L^{(0)}$ critical point in the random-bond Potts model.
\\

To summarize all these results, Fig.~\ref{fig:CentralChargeLadder} shows the Casimir effective central charges for various critical points in the phase diagrams 
of both the 
 $2D$ $3$-state Potts learning phase diagram, Fig.~\ref{FigPottsLearningPhaseDiagram}(a),
 and  
 the $2D$ random-bond $3$-state Potts model
 Fig.~\ref{FigPottsLearningPhaseDiagram}(b) 
 along with the corresponding information for the learning and random-bond Ising models.
 What this plot strikingly illustrates is a `fine structure' of non-unitary daughter theories above and below the original unitary Ising and Potts models
 with central charges $c=1/2$ and $c=4/5$, respectively. We will return to this observation in the discussion section below.

\section{Discussion and Outlook}

In this work, we studied the learning phase diagram of the $2D$ $q$-state Potts model ($2<q\leq 4$) subjected to bond-energy measurements,
conceptually summarized in Fig.~\ref{FigPottsLearningPhaseDiagram} and quantitatively mapped out via numerical tensor network simulations in Fig.~\ref{fig:numerical_phase_diagram_withGaussian}. 
A summary of our results can be found in Sec.~\ref{SecIntro}.

Probably most striking is the resemblance of our results to the previously determined learning phase diagram of the Ising model \cite{PutzGarrattNishimoriTrebstZhu,NahumJacobsen,PatilPutzTrebstZhuLudwig,WieseDasNahum}, which points to a more 
general framing of our results: For both the Potts and the Ising model, the key distinction between the learning and conventional 
random-bond phase diagrams is the nature of the underlying disorder; 
a correlated disorder arising from the conditional 
probability distributions in the inference (or measurement)
setting versus completely uncorrelated quenched disorder for the random-bond model.
While the former is best categorized via its description by an $R\to1$ replica limit, the latter is captured by the conventional $R\to0$ replica limit.
This difference (of one) in these replica limits then carries over as a distinction for the emergent critical theories in the respective 
phase diagrams -- they are described by the {\sl same} replica theories, but {\sl different} replica limits. 
This was first observed for the multicritical Nishimori points in the random-bond versus learning phase diagrams of the Ising model, 
but recurs in precisely the same manner for the Potts model: while the random-bond model exhibits an ordinary Nishimori line, and critical point $N^{(1)}$,
with an elevated replica symmetry, captured by 
a gauge-invariant replica formulation in the $R\to1$ replica limit, 
the learning phase diagram exhibits a \higher Nishimori line, and critical point $N^{(2)}$,
which is captured by 
a gauge-invariant formulation in the $R\to2$ replica limit. 
In addition, the learning phase diagram also exhibits the ordinary Nishimori line and critical point $N^{(1)}$ at infinite temperature $\beta=0$, precisely occurring in their above-mentioned gauge-invariant replica $R\rightarrow1$ formulation.
For the Potts model this replica limit distinction further carries over to the additional intermediate 
fixed point between the critical Potts and multicritical Nishimori point, that is present in both, the conventional random-bond and learning phase diagrams, 
but which is described by an $R\to0$ limit in the former and an $R\to1$ limit in the latter -- these are the $L^{(0)}$ and $L^{(1)}$ points, respectively.

But there is even more structure to this multitude of disorder-induced critical theories, which can arise due to quenched bond-disorder or measurement-induced randomness
in the context of the Potts and Ising models, respectively.
They are all described by non-unitary conformal field theories, which can be sorted by their Casimir effective
central charges as illustrated in Fig.~\ref{fig:CentralChargeLadder}.
What the plot nicely exhibits is an emergent `fine structure' of non-unitary daughter theories for a given unitary parent theory -- in our case the Ising model ($c=1/2$)
and the $3$-state  Potts model ($c=4/5$), which we have further augmented here by the Casimir effective central charge of 
the zero-temperature percolation limit of the  self-dual random-bond $q$-state Potts model~\cite{JacobsenCardy}.
Notably, the same hierarchy of theories is found when going from the left to the right: 
 ordinary Nishimori $N^{(1)}$, percolation, clean theory, \higher Nishimori $N^{(2)}$. 
Part of  the order in Fig.~\ref{fig:CentralChargeLadder} is required by the $c$-effective theorem \cite{PatilLudwig20251} (and its extensions~\cite{PatilPutzTrebstZhuLudwig,PatilLudwig20262}) for the RG flows in the phase 
diagram of the $R\to1$ learning model and, given an additional assumption on the replica central charge, 
for the RG flows in the phase diagram of its $R\to0$ random-bond cousin
-- Figs.~\ref{FigPottsLearningPhaseDiagram}(a) and (b), respectively.
The fine-structure splitting of the Potts model includes the two intermediate fixed points $L^{(0)}$ and $L^{(1)}$, which are known to have no analogue 
in the Ising case. 
But the weak self-dual critical point of the $\mathbb{Z}_{\textbf{2}}$ toric code~\cite{Eckstein2024,Eckstein2025,Wang2025} on the far left
should have a (far-left) analogue in the Potts fine structure, 
accessible by projectively measuring out the $\mathbb{Z}_{\textbf{3}}$ toric code along the self-dual 
basis
(an exercise, which we leave to future research).
These observations also guide one to conjecture a similar fine structure of daughter theories for the $4$-state Potts model (with $c=1$) extending this diagram to the right
and around the clean tricritical Ising point ($c=7/10$) that sits,
in the sequence of minimal models of unitary CFTs, between the Ising and $3$-state Potts theories. 
While for the $4$-state Potts model we expect a similar `splitting' as for the $3$-state Potts model shown here (including the two intermediate fixed points $L^{(0)}$ and $L^{(1)}$),
hitherto no non-unitary daughter CFTs have been identified for the tricritical Ising theory in any microscopic models.
But, of course, one could easily imagine settings similar to the random-bond and learning models discussed here.

We expect the learning phase diagram in Fig.~\ref{FigPottsLearningPhaseDiagram}(a) to hold for the Potts model with number of Potts states $q$ between two and four.
For $q>4$, the phase transition in the unmeasured model is a first-order transition, and since measurement-averaged first moments are equal to the unmeasured moments~\cite{Note7}, irrespective of the measurement strength, the first moment of the spin-spin correlator should be ordered and cannot be power-law decaying on the line $\beta=\beta_c$ in the learning phase diagram~\cite{NahumJacobsen}.
Moreover, using Jensen's inequality~\cite{fellerintroduction}, then any higher measurement-averaged moment of the absolute value
of the spin-spin correlator has to be ordered on the $\beta=\beta_c$ line in the learning phase diagram.
This first-order phase transition behavior in, at least, a subset of
correlators at the learning transition is in contrast with the corresponding behavior in the $2D$ RBPM with $q>4$, where due to the Aizenman-Wehr theorem~\cite{AizenmanWehr} the hard first-order phase transition in the clean model gets rounded off due to quenched bond-randomness, and becomes a second-order phase transition with power-law decay for all  moments of the spin-spin correlation function~\cite{JacobsenCardy,JacobsenPiccoLargeq,Palagyi2000,Chatelain_2001}.
This contrast with the Aizenman-Wehr theorem is a consequence of correlations in measurement outcomes governed by the Born/Bayes rule that are absent in systems with quenched disorder, where the Born-rule average is replaced by uncorrelated noise.
It would nevertheless be interesting to investigate the phase diagram of the learning Potts model for large $q>4$ and its applications,
e.g.\ in the context of learning and quantum-error correcting codes.

A related and particularly rich arena in which to explore this structure is the family of $\mathbb{Z}_q$ clock models, for
which the dual description again involves a $\mathbb{Z}_q$ toric code. 
For $q\geq 5$, the clean model has two BKT transitions bounding a critical phase~\cite{Jose1977}. 
We therefore expect the \higher 
Nishimori line to produce a critical segment bounded by {\sl two} multicritical $N^{(2)}$ points. 
Along this segment, the Edwards-Anderson correlator should inherit the continuously varying exponent of the
clean clock-model critical phase. This constitutes the higher-Nishimori analog of the two ordinary Nishimori transitions found in decohered
$\mathbb{Z}_q$ toric codes~\cite{vijay2025,mukherjee2026nishimorithresholdestimationbayesian}, and whose investigation we also leave
to future work.\\

{\it Data availability}.-- 
The numerical data shown in the figures is available on Zenodo~\cite{zenodo_potts_nishimori}.

\acknowledgments
%
We are grateful to the Kavli Institute for Theoretical Physics (KITP), which is supported by the National Science Foundation by grant NSF PHY-2309135, 
and the KITP programs ``Noise-robust Phases of Quantum Matter" and ``Learning the Fine Structure
of Quantum Dynamics in Programmable Quantum Matter" in summer 2025, where part of this work was initiated. 
We also acknowledge hospitality of Nordita, the Nordic Institute for Theoretical Physics, during the summer 2026 workshop 
``Entanglement dynamics: Open quantum systems, monitored circuits, and topological order" where this work was continued.
R.A.P. acknowledges support from the African and Asian Students in STEM Scholarship, administered by the Office of International Students and Scholars at the University of California, Santa Barbara, and thanks the Institute for Theoretical Physics, Cologne, for its hospitality during part of this work.
The Cologne research group is supported, in part, by the Deutsche Forschungsgemeinschaft (DFG, German Research Foundation) 
under Germany’s Excellence Strategy—Cluster of Excellence Matter
and Light for Quantum Computing (ML4Q) EXC 2004/1 -- 390534769 as well as within the CRC network TR 183 (Project Grant
No. 277101999) as part of subproject B01.
G.Y.Z. acknowledges the support of NSFC-Young Scientists Fund (grant no.~12504181) and Start-up Fund of HKUST(GZ) (grant no.~G0101000221) 
and Guangdong provincial project (grant no.~2024QN11X201) and the Guangdong Basic and Applied Basic Research Foundation (grant no.~2026A1515010965). R.M.\ acknowledges support from a postdoctoral fellowship from the Alexander von Humboldt Foundation.
Our numerical simulations were performed on the Otus cluster at PC2 in Paderborn and the RAMSES cluster at RRZK Cologne.
%

\appendix

\section{Noise Channel obtained by Tracing out the Measurement Record
\label{app:noise}}

Here we derive the unconditioned noise channel associated with the
measurement protocol in Eq.~\eqref{EqKraus}, following the analogous
construction in Ref.~\cite{Putz25nishimori}.  Let $\rho_0$ denote the
deformed toric-code state before the measurements and
$\hat K_{\vec m}=\prod_{\langle ij\rangle}\hat K_{m_{ij}}$.
Keeping a classical register for the measurement outcomes gives the
hybrid classical-quantum state~\cite{Wang2025}
\begin{equation}
 \rho_{\mathrm{cq}}
 =\sum_{\vec m}\ket{\vec m}\bra{\vec m}\otimes
 \hat K_{\vec m}\rho_0\hat K_{\vec m}^{\dagger} \,,
 \label{EqClassicalQuantumMeasurementState}
\end{equation}
where $\ket{\vec{m}}$ is the state of the classical register.
Erasing the measurement record, i.e.\ tracing over this register,
therefore produces
\begin{equation}
 \widetilde\rho
 =\operatorname{Tr}_{\vec m}\rho_{\mathrm{cq}}
 =\sum_{\vec m}\hat K_{\vec m}\rho_0\hat K_{\vec m}^{\dagger}
 =\left(\prod_{\langle ij\rangle}\mathcal N_{ij}\right)(\rho_0) \,,
 \label{EqMeasurementRecordTraced}
\end{equation}
where $\mathcal N_{ij}$ is the single-link channel
$\mathcal N(\rho)=\sum_{m\in\mathbb Z_q}\hat K_m\rho\hat K_m^\dagger$.

To evaluate this channel, write
$D=(q-1)+e^{q\tilde\gamma}$.  In the basis used in
Eq.~\eqref{EqKraus}, its action on a matrix unit is
\begin{align}
 \mathcal N(\ket g\bra h)
 &=\frac{1}{D}\sum_{m\in\mathbb Z_q}
 e^{\frac{q\tilde\gamma}{2}
 (\delta_{m,g}+\delta_{m,h})}\ket g\bra h \nonumber\\
 &=
 \begin{cases}
 \ket g\bra g\,, & g=h \,,\\
 \lambda\,\ket g\bra h\,, & g\ne h \,,
 \end{cases}
 \label{EqNoiseChannelMatrixElements}\\
 \lambda&=\frac{q-2+2e^{q\tilde\gamma/2}}
 {q-1+e^{q\tilde\gamma}} \, ,
 \label{EqNoiseCoherenceFactor}
\end{align}
where $\lambda$ monotonically decreases with increasing measurement strength $\tilde{\gamma}$: $\lambda=1$ for $\tilde{\gamma}=0$, while $\lambda\to 0$ for $\tilde{\gamma}\to\infty$.
Thus tracing out the outcomes leaves all diagonal matrix elements
unchanged and uniformly suppresses every off-diagonal matrix element.
Equivalently, defining the complete dephasing map in the $\hat Z$
eigenbasis by
\begin{equation}
 \mathcal D_Z(\rho)=\sum_{g\in\mathbb Z_q}\ket g\bra g\rho\ket g\bra g
 =\frac{1}{q}\sum_{n=0}^{q-1}\hat Z^n\rho\hat Z^{-n},
\end{equation}
the resulting noise channel is
\begin{align}
 \mathcal N(\rho)
 &=\lambda\rho+(1-\lambda)\mathcal D_Z(\rho)\nonumber\\
 &=(1-p)\rho+\frac{p}{q-1}\sum_{n=1}^{q-1}
 \hat Z^n\rho\hat Z^{-n}  \,,
 \label{EqDephasingChannel}\\
 p&=\frac{q-1}{q}(1-\lambda)
 =\frac{q-1}{q}\,
 \frac{\left(e^{q\tilde\gamma/2}-1\right)^2}
 {q-1+e^{q\tilde\gamma}}\\
 &=\frac{q-1}{q^2}
 \left(\sqrt{1+(q-1)\gamma}-\sqrt{1-\gamma}\right)^2 \,.
 \label{EqDephasingProbability}
\end{align}
Here $p$ is the total probability of a nontrivial generalized phase
error, with each $\hat Z^n$, $n=1,\ldots,q-1$, occurring with
probability $p/(q-1)$.  In particular, for the $\mathbb Z_3$ toric code,
\begin{equation}
 \mathcal N(\rho)=(1-p)\rho+\frac{p}{2}
 \left(\hat Z\rho\hat Z^\dagger+
 \hat Z^2\rho(\hat Z^\dagger)^2\right) \,,
\end{equation}
with
$p=\frac{2}{9}\left(\sqrt{1+2\gamma}-\sqrt{1-\gamma}\right)^2$.
The limits $\gamma=0$ and $\gamma=1$ give,
respectively, the identity channel and complete dephasing, as expected.

Finally, erasing the classical record is itself a local
completely-positive trace-preserving map,
$\rho_{\mathrm{cq}}\mapsto\operatorname{Tr}_{\vec m}\rho_{\mathrm{cq}}$.
The data-processing principle therefore implies that it cannot improve
the recoverability of the encoded quantum information: any recovery
map available after erasure could also be applied to
$\rho_{\mathrm{cq}}$ after first discarding its classical register,
whereas the converse need not hold~\cite{SchumacherNielsen1996,
WildeQuantumInformation,Putz25nishimori}.  Consequently, as the measurement strength is
increased, logical information can be lost under the dephasing channel
no later than in the trajectory ensemble that retains the outcomes.
In terms of the common parameter $\gamma$, the corresponding
critical strengths obey
\begin{equation}
\gamma_c^{\mathrm{noise}}
 \leq \gamma_c^{\mathrm{meas}} \,.
\end{equation}
For the undeformed $\mathbb Z_3$ toric code, the numerical dephasing (Nishimori) threshold is
$p_c^{\mathrm{noise}}=0.158(2)$, where $p$ is the total probability of
the two equally likely nontrivial errors $\hat Z$ and
$\hat Z^2$~\cite{JacobsenPicco,PhysRevA.91.042331,mukherjee2026nishimorithresholdestimationbayesian}. 
Equation~\eqref{EqDephasingProbability} maps this to
$\gamma_c^{\mathrm{noise}}=0.595(4)$.  By comparison, our
measurement Nishimori threshold $\gamma_c^{\mathrm{meas}}=0.764(1)$ corresponds to
$p_c^{\mathrm{meas}}=0.2709(8)$.
Thus both channels are symmetric between $\hat Z$ and $\hat Z^2$, falling into the ordinary Nishimori universality class,  but
$p_c^{\mathrm{meas}}>p_c^{\mathrm{noise}}$, explicitly
realizing the inequality above and showing that the measurement-induced
threshold upper-bounds, rather than necessarily equals, the
dephasing-noise threshold.
Note that the dephasing-noise threshold of the critically deformed ($\beta=\beta_c$) $\mathbb{Z}_3$ toric code is {\it unknown}. It is upper-bounded by our measurement threshold: $ \gamma_c^{\mathrm{noise}}
 \leq \gamma_c^{\mathrm{meas}} =0.581(1)$ and if translated to dephasing noise probability, $p_c^{\mathrm{noise}} \leq p_c^{\mathrm{meas}} =0.1505(5)$.

\section{Derivation of the Identity for Moments of Correlation Functions
\label{AppDerivationOfIdentityForMoments}}

In this section, we discuss the derivation of the equality in Eq.~\eqref{EqPottsK-pointEquality}, and more generally Eq.~\eqref{EqGeneralHigherNishimoriEquality}.
This is the analog, for the measurement-averaged moments on the \higher  Nishimori line in the learning phase diagram for the Potts model, of the  well-known equality for 
bond-randomness (quenched disorder) averaged moments on the ordinary Nishimori line in the random-bond Potts model~\cite{NishimoriStephen}. 
[As discussed in Sec.~\ref{SecReplicaTheoryAndHigherNishimoriLine}, 
the ordinary Nishimori line also appears in the learning phase diagram as the infinite temperature $\beta=0$ line in its `gauge-invariant' formulation (also see, e.g., the discussion in App.~D of Ref.~\cite{PatilPutzTrebstZhuLudwig}).]
The derivation we present is a replica-based way of deriving the required equality in the Potts learning phase diagram. 
This is obtained by generalizing the replica-based derivation presented in Refs.~\cite{LeDoussalHarrisII,GeorgesHanselLeDoussalMaillard,GRL2001} for the $R$-replica theory of the ordinary Nishimori line in the random-bond Ising model (replica $R\rightarrow0$ limit), which was subsequently used in the $R\rightarrow1$ replica limit for the higher Nishimori line in the Ising learning phase diagram in Ref.~\cite{PatilPutzTrebstZhuLudwig}. 
In this work, we generalize this replica-based derivation for the replica Potts Hamiltonian in Eq.~\eqref{EqReplicatedPottsHamiltonianUngauged} on the $\beta=\Delta$ line, which in the $R\rightarrow0$ replica limit corresponds to the ordinary Nishimori line in the random-bond Potts model~\cite{NishimoriStephen}, and in the $R\rightarrow1$ replica limit corresponds to the distinct \higher  Nishimori line in the learning phase diagram of the Potts model in Fig.~\ref{FigPottsLearningPhaseDiagram}(a).
Let us consider the following
correlation function for the {$R$-replica theory} in Eq.~\eqref{EqReplicatedPottsHamiltonianUngauged} on the 
$\beta=\Delta$ {line}, 
\begin{align}
    &\langle (\omega_{i_1}^{(a_1)}\cdots\omega_{i_k}^{(a_1)})\cdots(\omega_{i_1}^{(a_n)}\cdots\omega_{i_k}^{(a_n)})\rangle_{R}= \frac{1}{Z_R}\times\nonumber\\
   &\sum_{\{\omega_{i}^{(a)}\}_{a=1}^{R}}(\omega_{i_1}^{(a_1)}\cdots\omega_{i_k}^{(a_1)})\cdots(\omega_{i_1}^{(a_n)}\cdots\omega_{i_k}^{(a_n)})\times \nonumber\\
   &\quad\quad\quad\quad\quad\quad\quad\quad\quad\quad\times\exp{\{-\mathcal{\tilde{H}}^{R}[\{\omega_{i}^{(a)}\}_{a=1}^{R}]\}}
,\label{EqAppCorrelationFunctionFirstStep}
\end{align}
{where replica indices
$1 \leq a_1, a_2, ..., a_n \leq R$
are all pairwise unequal, and} $Z_{R}$ is the partition function of the $R$-replica Hamiltonian Eq.~\eqref{EqReplicatedPottsHamiltonianUngauged} on the 
$\beta=\Delta$ {line}. 
Note that until this point we have not introduced the extra replica $\omega_i^{(R+1)}$ into the above replica theory to write it as a theory {of $R+1$ replicas} in Eq.~\eqref{EqGaugedReplicatedPottsHamiltonian} with gauge {invariance}
and 
{enlarged}
{replica}
permutation symmetry. 
We will
do that now. 
Let us first change the sum over dummy spin variables $\omega_{i}^{(a)}$ to $\tilde{\omega}_{i}^{(a)}$, and analogous to Eq.~\eqref{EqGauging} define the following subsequent variable change 
\begin{equation}
   \tilde{\omega}_{i}^{(a)}= \omega_{i}^{(a)}(\omega^{(R+1)}_{i})^{-1}.
\end{equation}
Then dropping the tildes,
\begin{align}
    &\langle (\omega_{i_1}^{(a_1)}\cdots\omega_{i_k}^{(a_1)})\cdots(\omega_{i_1}^{(a_n)}\cdots\omega_{i_k}^{(a_n)})\rangle_{R}=\frac{1}{Z_R}\times\nonumber\\
    &\sum_{\{\omega_{i}^{(a)}\}_{a=1}^{R}}(\omega_{i_1}^{(a_1)}\cdots\omega_{i_k}^{(a_1)})\cdots(\omega_{i_1}^{(a_{n})}\cdots\omega_{i_k}^{(a_{n})})\times\nonumber\\
&\times\big(\omega_{i_1}^{(R+1)}\omega_{i_2}^{(R+1)}\dots \omega_{i_k}^{(R+1)}\big)^{-n}\exp{\big\{ -\mathcal{\tilde{H}}^{R+1}[\{\omega_{i}^{(a)}\}_{a=1}^{R+1}]
\big\}},
\end{align}
where the Hamiltonian $\mathcal{\tilde{H}}^{R+1}[\{\omega_{i}^{(a)}\}_{a=1}^{R+1}]$ is given in Eq.~\eqref{EqGaugedReplicatedPottsHamiltonian}. 
Summing over all possible configurations of {$\{\omega_i^{(R+1)}\}$,}
we obtain
\begin{align}
    &\langle (\omega_{i_1}^{(a_1)}\cdots\omega_{i_k}^{(a_1)})\cdots(\omega_{i_1}^{(a_n)}\cdots\omega_{i_k}^{(a_n)})\rangle_{R}=\frac{1/q^{N_{S}}}{Z_R}\times\nonumber\\
    &\sum_{\{\omega_{i}^{(a)}\}_{a=1}^{R+1}}(\omega_{i_1}^{(a_1)}\cdots\omega_{i_k}^{(a_1)})\cdots(\omega_{i_1}^{(a_n)}\cdots\omega_{i_k}^{(a_n)})\times\nonumber
    \\
&
\times
\big((\omega_{i_1}^{(R+1)})^*\cdots (\omega_{i_k}^{(R+1)})^*\big)^{n}\exp{\big\{ -\mathcal{\tilde{H}}^{R+1}[\{\omega_{i}^{(a)}\}_{a=1}^{R+1}]\big\}},
\end{align}
where we have used $(\omega^{(a)}_i)^{-1}=(\omega^{(a)}_i)^*$ and $N_S$ is the number of sites on the square lattice.
Then using the 
{enlarged}
permutation symmetry
{$S_{R+1}$}
of
{the $(R+1)$} replica copies for the gauge-invariant Hamiltonian in Eq.~\eqref{EqGaugedReplicatedPottsHamiltonian}, we can exchange
the 
$(R+1)^{\text{th}}$ copy with
{an $({n+1})^{\rm th}$ replica copy whose replica index $a_{n+1}$ 
(with $1 \leq a_{n+1} \leq R$) is unequal to any of the previous $(n)$ replica indices, i.e.\ {it} satisfies}
$a_{n+1}\neq a_1,a_2,\cdots,a_n$, {and} we obtain
\begin{align}
&\langle (\omega_{i_1}^{(a_1)}\cdots\omega_{i_k}^{(a_1)})\cdots(\omega_{i_1}^{(a_n)}\cdots\omega_{i_k}^{(a_n)})\rangle_{R}=\frac{1/q^{N_{S}}}{Z_R}\times \nonumber\\
&\sum_{\{\omega_{i}^{(a)}\}_{a=1}^{R+1}}(\omega_{i_1}^{(a_1)}\cdots\omega_{i_k}^{(a_1)})\cdots(\omega_{i_1}^{(a_n)}\cdots\omega_{i_k}^{(a_n)})\times\nonumber\\
&\times((\omega^{(a_{n+1})}_{i_1})^*\cdots (\omega^{(a_{n+1})}_{i_k})^*)^n\exp{\big\{ -\mathcal{\tilde{H}}^{R+1}[\{\omega_{i}^{(a)}\}_{a=1}^{R+1}]\big\}}.\label{EqOddToEvenMomentsPenultimateStep}
\end{align}
As noted already, the 
{Boltzmann}
weight in the 
{above}
partition function is gauge-invariant under local gauge transformation in Eq.~\eqref{EqGaugeTransformation}.
In particular, we can fix the gauge by setting the $(R+1)^{\text{th}}$ replica $\omega_i^{(R+1)}$ to $+1$ for all $i$. 
{The} replica Hamiltonian in the
Boltzmann weight {then} does not depend on $\omega_i^{(R+1)}$ for any $i$ and becomes the same as 
Eq.~\eqref{EqReplicatedPottsHamiltonianUngaugedII}. 
{This gauge fixing gets rid of the factor of $1/q^{N_S}$ appearing on the RHS of Eq.~\eqref{EqOddToEvenMomentsPenultimateStep}, and}
we obtain 
the result that on the 
$\beta=\Delta$  line,
\begin{align}
&\langle (\omega_{i_1}^{(a_1)}\cdots\omega_{i_k}^{(a_1)})\cdots(\omega_{i_1}^{(a_n)}\cdots\omega_{i_k}^{(a_n)})\rangle_{R}=\nonumber\\
&=\frac{1}{Z_R}\sum_{\{\omega_{i}^{(a)}\}_{a=1}^{R}}(\omega_{i_1}^{(a_1)}\cdots\omega_{i_k}^{(a_1)})\cdots(\omega_{i_1}^{(a_n)}\cdots\omega_{i_k}^{(a_n)})\times\nonumber\\
&\times[(\omega^{(a_{n+1})}_{i_1})^*\cdots (\omega^{(a_{n+1})}_{i_k})^*]^n\exp{\big\{ -\mathcal{\tilde{H}}^{R}[\{\omega_{i}^{(a)}\}_{a=1}^{R}]\big\}}\\
&=\langle (\omega_{i_1}^{(a_1)}\cdot\cdot\omega_{i_k}^{(a_1)})\cdots(\omega_{i_1}^{(a_n)}\cdot\cdot\omega_{i_k}^{(a_n)})[(\omega_{i_1}^{(a_{n+1})}\cdots\omega_{i_k}^{(a_{n+1})})^*]^n\rangle_{R}.   \label{EqReplicatedNishimoriIdentity} 
\end{align}
The $R\rightarrow1$ replica limit of the result in Eq.~\eqref{EqReplicatedNishimoriIdentity} gives the following relation between measurement-averaged moments on the \higher  Nishimori line 
$\beta=\Delta$
in the Potts learning phase diagram
\begin{equation}
    \overline{\langle\omega_{i_1}\omega_{i_2}\cdots\omega_{i_k}\rangle_{\vec{m}}^n}=\overline{\langle\omega_{i_1}\omega_{i_2}\cdots\omega_{i_k}\rangle_{\vec{m}}^n\langle(\omega^*_{i_1})^n(\omega^*_{i_2})^n\cdots(\omega^*_{i_k})^n\rangle_{\vec{m}}},\label{EqApp}
\end{equation}
where the $n=1$ version of this identity is given in Eq.~\eqref{EqPottsK-pointEquality}.
The above [Eq.~\eqref{EqApp}] $R\rightarrow1$ replica limit of the replica correlation function identity in Eq.~\eqref{EqReplicatedNishimoriIdentity}, which we use for the measurement-averaged moments on the \higher Nishimori line in the Potts learning phase diagram, should be distinguished from the $R\rightarrow0$ replica limit of the identity in Eq.~\eqref{EqReplicatedNishimoriIdentity}, which is relevant for the bond-randomness averaged moments on the ordinary Nishimori line in the random-bond Potts model~\cite{NishimoriStephen}.

\section{Details of the Numerical Simulations}
\label{Appendix_Numerical}

\subsubsection*{Setup}

All the numerical simulations were done for the standard $3$-state Potts Hamiltonian
\begin{equation}
    H[\{\omega_i\}]=-\sum_{\langle ij\rangle}\delta_{\omega_i, \omega_j} \,.
    \label{EqPottsHamiltoniannumerics}
\end{equation}
For our discrete $q$-state measurement protocol, the measurement model is given by
\begin{equation}
     P(\{m_{ij}\}|\{\omega_{i}\})=\frac{\exp{\{\tilde{\gamma} \delta_{m_{ij},\omega_i\omega_j^{-1}}\}}}{(q-1)+e^{\tilde{\gamma}}} \,, 
     \label{EqMeasurementProtocolAnumerics}
\end{equation}
with
\begin{equation}
   \tilde{\gamma}=\log \left(\frac{1+(q-1)\gamma}{1-\gamma}\right),
   \label{Eq:mappinggamma}
\end{equation}
where $\gamma$ is the measurement strength and takes values in $[0,1]$ (a similar parametrization was used in the numerical simulations of the Ising learning phase diagram~\cite{PutzGarrattNishimoriTrebstZhu}).

\subsubsection*{Hybrid sampling and tensor-network contraction}

Our numerical procedure separates the sampling of the measurement ensemble from the evaluation of observables in the conditioned model. At every point $(\beta,\gamma)$, we first use Monte Carlo to draw configurations from the thermal Potts distribution and subsequently draw a measurement record from Eq.~\eqref{EqMeasurementProtocolAnumerics}, conditioned on that configuration. For each such record, the relevant conditioned partition functions and observables are evaluated by tensor-network contraction.

Compared with the Ising calculation, the  tensor network contraction of the three-state Potts model is considerably more demanding. The bond dimension of the local tensor network increases from $D=2$ for Ising spins to $D=3$ for Potts spins. During the row-by-row contraction, this larger local bond dimension generates larger intermediate MPS tensors and a broader singular-value spectrum. To prevent exponential growth of the MPS bond dimension, we compress the MPS using a singular-value cutoff of $10^{-10}$, discarding singular values below this threshold.

\begin{figure}[h!]
    \includegraphics[width=\linewidth]{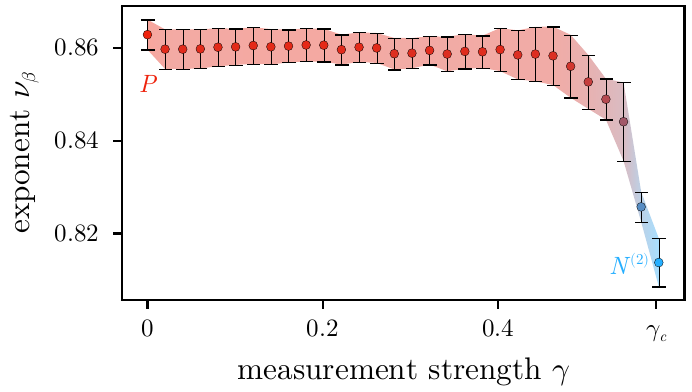}
    \caption{{\bf Critical exponent $\nu$ as a function of $\gamma$ along the critical line at $\beta = \beta_c$.}
            Shown is the numerically obtained correlation length exponent $\nu_\beta$ as a function of $\gamma$ for learning the $3$-state Potts model. 
            The critical exponent $\nu_\beta$ is obtained from finite-size scaling of the coherent information for linear system sizes $L = 16$ to $128$.
    }
    \label{fig:nu_vs_γ}
\end{figure}

For reference, the horizontal phase-diagram scan up to system size $L=64$, which yielded the data in Figs.~\ref{fig:nu_vs_β} and~\ref{fig:numerical_phase_diagram_withGaussian}, required approximately $100,000$ CPU core-hours. The corresponding vertical scan up to $L=128$, shown in Figs.~\ref{fig:nu_vs_γ} and~\ref{fig:numerical_phase_diagram_withGaussian}, required approximately $750,000$ CPU core-hours. Separate horizontal sweeps across the \textit{ordinary} and \textit{higher} Nishimori lines up to $L=128$ accounted for approximately $100,000$ CPU core-hours and produced the data in Fig.~\ref{fig:Ic_nishimori12}. Finally, the horizontal phase-diagram sweeps used to calculate the EA correlations up to $L=256$ required approximately $250,000$ CPU core-hours. The resulting data are shown in Figs.~\ref{fig:numerical_phase_diagram_withGaussian}, \ref{fig:oioj_higherNishimori_Potts}, \ref{fig:nishimori_condition}, and~\ref{fig:2X2_along_betac}.

\subsubsection*{Observables}

As discussed in Sec.~\ref{SecReplicaTheoryAndHigherNishimoriLine}, there is no {\sl exact} \higher  Nishimori line in the learning phase diagram with the above measurement protocol.
However, if we ignore the $\mathcal{O}(\tilde{\gamma}^3)$ terms in the corresponding replica theory [Eq.~\eqref{EqReplicatedPottsHamiltonianProtocolI}], the truncated replica theory has a
\higher Nishimori line given by
\begin{equation}
    \beta=\tilde{\gamma}^2/q \,.
\end{equation}
Inserting the value of $\beta_c$, the tricritical point for the truncated replica theory is located at
\begin{equation}
    \gamma_c=\gamma_{N^{(2)}}=0.60922 \,.
\end{equation}
Although this is not the exact transition location, which is a nonuniversal quantity, it is close to
the numerical value $\gamma_{N^{(2)}} = 0.581(1)$ that we find from Fig.~\ref{fig:Ic_nishimori12}.

\subsubsection*{Additional Numerical Data}
\label{App:Numdetails2}

For the sake of completeness, we present below numerical estimates for the correlation length exponent $\nu_\beta$ calculated from vertical cuts
through the critical temperature $\beta=\beta_c$ line as a function of $\gamma$. 
Similar to the estimation of  correlation length exponent $\nu_\gamma$ for horizontal cuts through our phase diagram for various levels of deformation $\beta$ in Fig.~\ref{fig:nu_vs_β},
this plot allows one to determine the direction of an RG flow. In this case, an RG flow out of the higher Nishimori point $N^{(2)}$
as one decreases the measurement strength.
What this plot does not show is any indication of an intermediate fixed point $L^{(1)}$, similar to our discussion of the results of Fig.~\ref{fig:2X2_along_betac}.
\\[5mm]

\newpage

\bibliography{PottsHigherNishimori}

\begin{thebibliography}{82}%
\makeatletter
\providecommand \@ifxundefined [1]{%
 \@ifx{#1\undefined}
}%
\providecommand \@ifnum [1]{%
 \ifnum #1\expandafter \@firstoftwo
 \else \expandafter \@secondoftwo
 \fi
}%
\providecommand \@ifx [1]{%
 \ifx #1\expandafter \@firstoftwo
 \else \expandafter \@secondoftwo
 \fi
}%
\providecommand \natexlab [1]{#1}%
\providecommand \enquote  [1]{``#1''}%
\providecommand \bibnamefont  [1]{#1}%
\providecommand \bibfnamefont [1]{#1}%
\providecommand \citenamefont [1]{#1}%
\providecommand \href@noop [0]{\@secondoftwo}%
\providecommand \href [0]{\begingroup \@sanitize@url \@href}%
\providecommand \@href[1]{\@@startlink{#1}\@@href}%
\providecommand \@@href[1]{\endgroup#1\@@endlink}%
\providecommand \@sanitize@url [0]{\catcode `\\12\catcode `\$12\catcode `\&12\catcode `\#12\catcode `\^12\catcode `\_12\catcode `\%12\relax}%
\providecommand \@@startlink[1]{}%
\providecommand \@@endlink[0]{}%
\providecommand \url  [0]{\begingroup\@sanitize@url \@url }%
\providecommand \@url [1]{\endgroup\@href {#1}{\urlprefix }}%
\providecommand \urlprefix  [0]{URL }%
\providecommand \Eprint [0]{\href }%
\providecommand \doibase [0]{https://doi.org/}%
\providecommand \selectlanguage [0]{\@gobble}%
\providecommand \bibinfo  [0]{\@secondoftwo}%
\providecommand \bibfield  [0]{\@secondoftwo}%
\providecommand \translation [1]{[#1]}%
\providecommand \BibitemOpen [0]{}%
\providecommand \bibitemStop [0]{}%
\providecommand \bibitemNoStop [0]{.\EOS\space}%
\providecommand \EOS [0]{\spacefactor3000\relax}%
\providecommand \BibitemShut  [1]{\csname bibitem#1\endcsname}%
\let\auto@bib@innerbib\@empty
\bibitem [{\citenamefont {Kitaev}(2003)}]{Kitaev2003}%
  \BibitemOpen
  \bibfield  {author} {\bibinfo {author} {\bibfnamefont {A.}~\bibnamefont {Kitaev}},\ }\bibfield  {title} {\bibinfo {title} {{Fault-tolerant quantum computation by anyons}},\ }\href {https://doi.org/https://doi.org/10.1016/S0003-4916(02)00018-0} {\bibfield  {journal} {\bibinfo  {journal} {Annals of Physics}\ }\textbf {\bibinfo {volume} {303}},\ \bibinfo {pages} {2} (\bibinfo {year} {2003})}\BibitemShut {NoStop}%
\bibitem [{\citenamefont {Tantivasadakarn}\ \emph {et~al.}(2024)\citenamefont {Tantivasadakarn}, \citenamefont {Thorngren}, \citenamefont {Vishwanath},\ and\ \citenamefont {Verresen}}]{Tantivasadakarn2024}%
  \BibitemOpen
  \bibfield  {author} {\bibinfo {author} {\bibfnamefont {N.}~\bibnamefont {Tantivasadakarn}}, \bibinfo {author} {\bibfnamefont {R.}~\bibnamefont {Thorngren}}, \bibinfo {author} {\bibfnamefont {A.}~\bibnamefont {Vishwanath}},\ and\ \bibinfo {author} {\bibfnamefont {R.}~\bibnamefont {Verresen}},\ }\bibfield  {title} {\bibinfo {title} {{Long-Range Entanglement from Measuring Symmetry-Protected Topological Phases}},\ }\href {https://doi.org/10.1103/PhysRevX.14.021040} {\bibfield  {journal} {\bibinfo  {journal} {Phys. Rev. X}\ }\textbf {\bibinfo {volume} {14}},\ \bibinfo {pages} {021040} (\bibinfo {year} {2024})}\BibitemShut {NoStop}%
\bibitem [{\citenamefont {Dennis}\ \emph {et~al.}(2002)\citenamefont {Dennis}, \citenamefont {Kitaev}, \citenamefont {Landahl},\ and\ \citenamefont {Preskill}}]{DennisKitaevLandahlPreskill}%
  \BibitemOpen
  \bibfield  {author} {\bibinfo {author} {\bibfnamefont {E.}~\bibnamefont {Dennis}}, \bibinfo {author} {\bibfnamefont {A.}~\bibnamefont {Kitaev}}, \bibinfo {author} {\bibfnamefont {A.}~\bibnamefont {Landahl}},\ and\ \bibinfo {author} {\bibfnamefont {J.}~\bibnamefont {Preskill}},\ }\bibfield  {title} {\bibinfo {title} {{Topological quantum memory}},\ }\href {https://doi.org/10.1063/1.1499754} {\bibfield  {journal} {\bibinfo  {journal} {Journal of Mathematical Physics}\ }\textbf {\bibinfo {volume} {43}},\ \bibinfo {pages} {4452} (\bibinfo {year} {2002})}\BibitemShut {NoStop}%
\bibitem [{\citenamefont {Eckstein}\ \emph {et~al.}(2024)\citenamefont {Eckstein}, \citenamefont {Han}, \citenamefont {Trebst},\ and\ \citenamefont {Zhu}}]{Eckstein2024}%
  \BibitemOpen
  \bibfield  {author} {\bibinfo {author} {\bibfnamefont {F.}~\bibnamefont {Eckstein}}, \bibinfo {author} {\bibfnamefont {B.}~\bibnamefont {Han}}, \bibinfo {author} {\bibfnamefont {S.}~\bibnamefont {Trebst}},\ and\ \bibinfo {author} {\bibfnamefont {G.-Y.}\ \bibnamefont {Zhu}},\ }\bibfield  {title} {\bibinfo {title} {{Robust Teleportation of a Surface Code and Cascade of Topological Quantum Phase Transitions}},\ }\href {https://doi.org/10.1103/PRXQuantum.5.040313} {\bibfield  {journal} {\bibinfo  {journal} {PRX Quantum}\ }\textbf {\bibinfo {volume} {5}},\ \bibinfo {pages} {040313} (\bibinfo {year} {2024})}\BibitemShut {NoStop}%
\bibitem [{\citenamefont {Eckstein}\ \emph {et~al.}(2025)\citenamefont {Eckstein}, \citenamefont {Han}, \citenamefont {Trebst},\ and\ \citenamefont {Zhu}}]{Eckstein2025}%
  \BibitemOpen
  \bibfield  {author} {\bibinfo {author} {\bibfnamefont {F.}~\bibnamefont {Eckstein}}, \bibinfo {author} {\bibfnamefont {B.}~\bibnamefont {Han}}, \bibinfo {author} {\bibfnamefont {S.}~\bibnamefont {Trebst}},\ and\ \bibinfo {author} {\bibfnamefont {G.-Y.}\ \bibnamefont {Zhu}},\ }\href {https://arxiv.org/abs/2512.19786} {\bibinfo {title} {{Learning transitions of topological surface codes}}} (\bibinfo {year} {2025}),\ \Eprint {https://arxiv.org/abs/2512.19786} {arXiv:2512.19786 [quant-ph]} \BibitemShut {NoStop}%
\bibitem [{\citenamefont {Nishimori}(1980)}]{Nishimori_1980}%
  \BibitemOpen
  \bibfield  {author} {\bibinfo {author} {\bibfnamefont {H.}~\bibnamefont {Nishimori}},\ }\bibfield  {title} {\bibinfo {title} {{Exact results and critical properties of the Ising model with competing interactions}},\ }\href {https://doi.org/10.1088/0022-3719/13/21/012} {\bibfield  {journal} {\bibinfo  {journal} {Journal of Physics C: Solid State Physics}\ }\textbf {\bibinfo {volume} {13}},\ \bibinfo {pages} {4071} (\bibinfo {year} {1980})}\BibitemShut {NoStop}%
\bibitem [{\citenamefont {Nishimori}(1981)}]{Nishimori1981}%
  \BibitemOpen
  \bibfield  {author} {\bibinfo {author} {\bibfnamefont {H.}~\bibnamefont {Nishimori}},\ }\bibfield  {title} {\bibinfo {title} {{Internal Energy, Specific Heat and Correlation Function of the Bond-Random Ising Model}},\ }\href {https://doi.org/10.1143/PTP.66.1169} {\bibfield  {journal} {\bibinfo  {journal} {Progress of Theoretical Physics}\ }\textbf {\bibinfo {volume} {66}},\ \bibinfo {pages} {1169} (\bibinfo {year} {1981})}\BibitemShut {NoStop}%
\bibitem [{\citenamefont {Iba}(1999)}]{Iba_1999}%
  \BibitemOpen
  \bibfield  {author} {\bibinfo {author} {\bibfnamefont {Y.}~\bibnamefont {Iba}},\ }\bibfield  {title} {\bibinfo {title} {{The Nishimori line and Bayesian statistics}},\ }\href {https://doi.org/10.1088/0305-4470/32/21/302} {\bibfield  {journal} {\bibinfo  {journal} {Journal of Physics A: Mathematical and General}\ }\textbf {\bibinfo {volume} {32}},\ \bibinfo {pages} {3875} (\bibinfo {year} {1999})}\BibitemShut {NoStop}%
\bibitem [{\citenamefont {Sourlas}(1994)}]{Sourlas_1994}%
  \BibitemOpen
  \bibfield  {author} {\bibinfo {author} {\bibfnamefont {N.}~\bibnamefont {Sourlas}},\ }\bibfield  {title} {\bibinfo {title} {{Spin Glasses, Error-Correcting Codes and Finite-Temperature Decoding}},\ }\href {https://doi.org/10.1209/0295-5075/25/3/001} {\bibfield  {journal} {\bibinfo  {journal} {Europhysics Letters}\ }\textbf {\bibinfo {volume} {25}},\ \bibinfo {pages} {159} (\bibinfo {year} {1994})}\BibitemShut {NoStop}%
\bibitem [{\citenamefont {Henley}(2004)}]{CLHenley_2004}%
  \BibitemOpen
  \bibfield  {author} {\bibinfo {author} {\bibfnamefont {C.~L.}\ \bibnamefont {Henley}},\ }\bibfield  {title} {\bibinfo {title} {{From classical to quantum dynamics at Rokhsar–Kivelson points}},\ }\href {https://doi.org/10.1088/0953-8984/16/11/045} {\bibfield  {journal} {\bibinfo  {journal} {Journal of Physics: Condensed Matter}\ }\textbf {\bibinfo {volume} {16}},\ \bibinfo {pages} {S891} (\bibinfo {year} {2004})}\BibitemShut {NoStop}%
\bibitem [{\citenamefont {Jian}\ \emph {et~al.}(2020)\citenamefont {Jian}, \citenamefont {You}, \citenamefont {Vasseur},\ and\ \citenamefont {Ludwig}}]{JianYouVasseurLudwig2019}%
  \BibitemOpen
  \bibfield  {author} {\bibinfo {author} {\bibfnamefont {C.-M.}\ \bibnamefont {Jian}}, \bibinfo {author} {\bibfnamefont {Y.-Z.}\ \bibnamefont {You}}, \bibinfo {author} {\bibfnamefont {R.}~\bibnamefont {Vasseur}},\ and\ \bibinfo {author} {\bibfnamefont {A.~W.~W.}\ \bibnamefont {Ludwig}},\ }\bibfield  {title} {\bibinfo {title} {{Measurement-induced criticality in random quantum circuits}},\ }\href {https://doi.org/10.1103/PhysRevB.101.104302} {\bibfield  {journal} {\bibinfo  {journal} {Phys. Rev. B}\ }\textbf {\bibinfo {volume} {101}},\ \bibinfo {pages} {104302} (\bibinfo {year} {2020})}\BibitemShut {NoStop}%
\bibitem [{\citenamefont {Bao}\ \emph {et~al.}(2020)\citenamefont {Bao}, \citenamefont {Choi},\ and\ \citenamefont {Altman}}]{BaoChoiAltman2019}%
  \BibitemOpen
  \bibfield  {author} {\bibinfo {author} {\bibfnamefont {Y.}~\bibnamefont {Bao}}, \bibinfo {author} {\bibfnamefont {S.}~\bibnamefont {Choi}},\ and\ \bibinfo {author} {\bibfnamefont {E.}~\bibnamefont {Altman}},\ }\bibfield  {title} {\bibinfo {title} {{Theory of the phase transition in random unitary circuits with measurements}},\ }\href {https://doi.org/10.1103/PhysRevB.101.104301} {\bibfield  {journal} {\bibinfo  {journal} {Phys. Rev. B}\ }\textbf {\bibinfo {volume} {101}},\ \bibinfo {pages} {104301} (\bibinfo {year} {2020})}\BibitemShut {NoStop}%
\bibitem [{\citenamefont {Patil}\ \emph {et~al.}(2026{\natexlab{a}})\citenamefont {Patil}, \citenamefont {Pütz}, \citenamefont {Trebst}, \citenamefont {Zhu},\ and\ \citenamefont {Ludwig}}]{PatilPutzTrebstZhuLudwig}%
  \BibitemOpen
  \bibfield  {author} {\bibinfo {author} {\bibfnamefont {R.~A.}\ \bibnamefont {Patil}}, \bibinfo {author} {\bibfnamefont {M.}~\bibnamefont {Pütz}}, \bibinfo {author} {\bibfnamefont {S.}~\bibnamefont {Trebst}}, \bibinfo {author} {\bibfnamefont {G.-Y.}\ \bibnamefont {Zhu}},\ and\ \bibinfo {author} {\bibfnamefont {A.~W.~W.}\ \bibnamefont {Ludwig}},\ }\href {https://arxiv.org/abs/2604.06324} {\bibinfo {title} {{Higher Nishimori Criticality and Exact Results at the Learning Transition of Deformed Toric Codes}}} (\bibinfo {year} {2026}{\natexlab{a}}),\ \Eprint {https://arxiv.org/abs/2604.06324} {arXiv:2604.06324 [cond-mat.stat-mech]} \BibitemShut {NoStop}%
\bibitem [{\citenamefont {P\"utz}\ \emph {et~al.}(2026)\citenamefont {P\"utz}, \citenamefont {Garratt}, \citenamefont {Nishimori}, \citenamefont {Trebst},\ and\ \citenamefont {Zhu}}]{PutzGarrattNishimoriTrebstZhu}%
  \BibitemOpen
  \bibfield  {author} {\bibinfo {author} {\bibfnamefont {M.}~\bibnamefont {P\"utz}}, \bibinfo {author} {\bibfnamefont {S.~J.}\ \bibnamefont {Garratt}}, \bibinfo {author} {\bibfnamefont {H.}~\bibnamefont {Nishimori}}, \bibinfo {author} {\bibfnamefont {S.}~\bibnamefont {Trebst}},\ and\ \bibinfo {author} {\bibfnamefont {G.-Y.}\ \bibnamefont {Zhu}},\ }\bibfield  {title} {\bibinfo {title} {{Learning Transitions in Classical Ising Models and Deformed Toric Codes}},\ }\href {https://doi.org/10.1103/4dwm-kn11} {\bibfield  {journal} {\bibinfo  {journal} {Phys. Rev. Lett.}\ }\textbf {\bibinfo {volume} {136}},\ \bibinfo {pages} {190402} (\bibinfo {year} {2026})}\BibitemShut {NoStop}%
\bibitem [{\citenamefont {Nahum}\ and\ \citenamefont {Jacobsen}(2025)}]{NahumJacobsen}%
  \BibitemOpen
  \bibfield  {author} {\bibinfo {author} {\bibfnamefont {A.}~\bibnamefont {Nahum}}\ and\ \bibinfo {author} {\bibfnamefont {J.~L.}\ \bibnamefont {Jacobsen}},\ }\bibfield  {title} {\bibinfo {title} {{Bayesian critical points in classical lattice models}},\ }\href {https://doi.org/10.1103/7dpt-d4s5} {\bibfield  {journal} {\bibinfo  {journal} {Phys. Rev. B}\ }\textbf {\bibinfo {volume} {112}},\ \bibinfo {pages} {235113} (\bibinfo {year} {2025})}\BibitemShut {NoStop}%
\bibitem [{\citenamefont {Castelnovo}\ and\ \citenamefont {Chamon}(2008)}]{CastelnovoChamon}%
  \BibitemOpen
  \bibfield  {author} {\bibinfo {author} {\bibfnamefont {C.}~\bibnamefont {Castelnovo}}\ and\ \bibinfo {author} {\bibfnamefont {C.}~\bibnamefont {Chamon}},\ }\bibfield  {title} {\bibinfo {title} {{Quantum topological phase transition at the microscopic level}},\ }\href {https://doi.org/10.1103/PhysRevB.77.054433} {\bibfield  {journal} {\bibinfo  {journal} {Phys. Rev. B}\ }\textbf {\bibinfo {volume} {77}},\ \bibinfo {pages} {054433} (\bibinfo {year} {2008})}\BibitemShut {NoStop}%
\bibitem [{\citenamefont {Ardonne}\ \emph {et~al.}(2004)\citenamefont {Ardonne}, \citenamefont {Fendley},\ and\ \citenamefont {Fradkin}}]{ArdonneFedleyFradkin}%
  \BibitemOpen
  \bibfield  {author} {\bibinfo {author} {\bibfnamefont {E.}~\bibnamefont {Ardonne}}, \bibinfo {author} {\bibfnamefont {P.}~\bibnamefont {Fendley}},\ and\ \bibinfo {author} {\bibfnamefont {E.}~\bibnamefont {Fradkin}},\ }\bibfield  {title} {\bibinfo {title} {{Topological order and conformal quantum critical points}},\ }\href {https://doi.org/https://doi.org/10.1016/j.aop.2004.01.004} {\bibfield  {journal} {\bibinfo  {journal} {Annals of Physics}\ }\textbf {\bibinfo {volume} {310}},\ \bibinfo {pages} {493} (\bibinfo {year} {2004})}\BibitemShut {NoStop}%
\bibitem [{\citenamefont {Papanikolaou}\ \emph {et~al.}(2007)\citenamefont {Papanikolaou}, \citenamefont {Raman},\ and\ \citenamefont {Fradkin}}]{PapanikolaouRamanFradkin}%
  \BibitemOpen
  \bibfield  {author} {\bibinfo {author} {\bibfnamefont {S.}~\bibnamefont {Papanikolaou}}, \bibinfo {author} {\bibfnamefont {K.~S.}\ \bibnamefont {Raman}},\ and\ \bibinfo {author} {\bibfnamefont {E.}~\bibnamefont {Fradkin}},\ }\bibfield  {title} {\bibinfo {title} {{Topological phases and topological entropy of two-dimensional systems with finite correlation length}},\ }\href {https://doi.org/10.1103/PhysRevB.76.224421} {\bibfield  {journal} {\bibinfo  {journal} {Phys. Rev. B}\ }\textbf {\bibinfo {volume} {76}},\ \bibinfo {pages} {224421} (\bibinfo {year} {2007})}\BibitemShut {NoStop}%
\bibitem [{\citenamefont {Isakov}\ \emph {et~al.}(2011)\citenamefont {Isakov}, \citenamefont {Fendley}, \citenamefont {Ludwig}, \citenamefont {Trebst},\ and\ \citenamefont {Troyer}}]{IsakovFendleyLudwigTrebstTroyer}%
  \BibitemOpen
  \bibfield  {author} {\bibinfo {author} {\bibfnamefont {S.~V.}\ \bibnamefont {Isakov}}, \bibinfo {author} {\bibfnamefont {P.}~\bibnamefont {Fendley}}, \bibinfo {author} {\bibfnamefont {A.~W.~W.}\ \bibnamefont {Ludwig}}, \bibinfo {author} {\bibfnamefont {S.}~\bibnamefont {Trebst}},\ and\ \bibinfo {author} {\bibfnamefont {M.}~\bibnamefont {Troyer}},\ }\bibfield  {title} {\bibinfo {title} {{Dynamics at and near conformal quantum critical points}},\ }\href {https://doi.org/10.1103/PhysRevB.83.125114} {\bibfield  {journal} {\bibinfo  {journal} {Phys. Rev. B}\ }\textbf {\bibinfo {volume} {83}},\ \bibinfo {pages} {125114} (\bibinfo {year} {2011})}\BibitemShut {NoStop}%
\bibitem [{\citenamefont {Zhu}\ and\ \citenamefont {Zhang}(2019)}]{Zhu19deform}%
  \BibitemOpen
  \bibfield  {author} {\bibinfo {author} {\bibfnamefont {G.-Y.}\ \bibnamefont {Zhu}}\ and\ \bibinfo {author} {\bibfnamefont {G.-M.}\ \bibnamefont {Zhang}},\ }\bibfield  {title} {\bibinfo {title} {{Gapless Coulomb State Emerging from a Self-Dual Topological Tensor-Network State}},\ }\href {https://doi.org/10.1103/PhysRevLett.122.176401} {\bibfield  {journal} {\bibinfo  {journal} {Phys. Rev. Lett.}\ }\textbf {\bibinfo {volume} {122}},\ \bibinfo {pages} {176401} (\bibinfo {year} {2019})}\BibitemShut {NoStop}%
\bibitem [{\citenamefont {Sahay}\ \emph {et~al.}(2025)\citenamefont {Sahay}, \citenamefont {von Keyserlingk}, \citenamefont {Verresen},\ and\ \citenamefont {Zhang}}]{Verresen25deform}%
  \BibitemOpen
  \bibfield  {author} {\bibinfo {author} {\bibfnamefont {R.}~\bibnamefont {Sahay}}, \bibinfo {author} {\bibfnamefont {C.}~\bibnamefont {von Keyserlingk}}, \bibinfo {author} {\bibfnamefont {R.}~\bibnamefont {Verresen}},\ and\ \bibinfo {author} {\bibfnamefont {C.}~\bibnamefont {Zhang}},\ }\href {https://arxiv.org/abs/2503.01977} {\bibinfo {title} {{Enforced Gaplessness from States with Exponentially Decaying Correlations}}} (\bibinfo {year} {2025}),\ \Eprint {https://arxiv.org/abs/2503.01977} {arXiv:2503.01977 [cond-mat.str-el]} \BibitemShut {NoStop}%
\bibitem [{\citenamefont {Wiese}\ \emph {et~al.}(2026)\citenamefont {Wiese}, \citenamefont {Das},\ and\ \citenamefont {Nahum}}]{WieseDasNahum}%
  \BibitemOpen
  \bibfield  {author} {\bibinfo {author} {\bibfnamefont {K.~J.}\ \bibnamefont {Wiese}}, \bibinfo {author} {\bibfnamefont {A.}~\bibnamefont {Das}},\ and\ \bibinfo {author} {\bibfnamefont {A.}~\bibnamefont {Nahum}},\ }\href {https://arxiv.org/abs/2604.23346} {\bibinfo {title} {{Bayesian phase transition for the critical Ising model: Enlarged replica symmetry in the epsilon expansion and in 2D}}} (\bibinfo {year} {2026}),\ \Eprint {https://arxiv.org/abs/2604.23346} {arXiv:2604.23346 [cond-mat.stat-mech]} \BibitemShut {NoStop}%
\bibitem [{\citenamefont {Le~Doussal}\ and\ \citenamefont {Harris}(1989)}]{LeDoussalHarrisII}%
  \BibitemOpen
  \bibfield  {author} {\bibinfo {author} {\bibfnamefont {P.}~\bibnamefont {Le~Doussal}}\ and\ \bibinfo {author} {\bibfnamefont {A.~B.}\ \bibnamefont {Harris}},\ }\bibfield  {title} {\bibinfo {title} {{\ensuremath{\epsilon} expansion for the Nishimori multicritical point of spin glasses}},\ }\href {https://doi.org/10.1103/PhysRevB.40.9249} {\bibfield  {journal} {\bibinfo  {journal} {Phys. Rev. B}\ }\textbf {\bibinfo {volume} {40}},\ \bibinfo {pages} {9249} (\bibinfo {year} {1989})}\BibitemShut {NoStop}%
\bibitem [{\citenamefont {Georges}\ \emph {et~al.}(1987)\citenamefont {Georges}, \citenamefont {Hansel}, \citenamefont {Le~Doussal},\ and\ \citenamefont {Maillard}}]{GeorgesHanselLeDoussalMaillard}%
  \BibitemOpen
  \bibfield  {author} {\bibinfo {author} {\bibfnamefont {A.}~\bibnamefont {Georges}}, \bibinfo {author} {\bibfnamefont {D.}~\bibnamefont {Hansel}}, \bibinfo {author} {\bibfnamefont {P.}~\bibnamefont {Le~Doussal}},\ and\ \bibinfo {author} {\bibfnamefont {J.}~\bibnamefont {Maillard}},\ }\bibfield  {title} {\bibinfo {title} {{The replica momenta of a spin-glass and the phase diagram of n-colour Ashkin-Teller models}},\ }\href {https://doi.org/10.1051/jphys:019870048010100} {\bibfield  {journal} {\bibinfo  {journal} {J. Phys. France}\ }\textbf {\bibinfo {volume} {48}},\ \bibinfo {pages} {1} (\bibinfo {year} {1987})}\BibitemShut {NoStop}%
\bibitem [{\citenamefont {Gruzberg}\ \emph {et~al.}(2001)\citenamefont {Gruzberg}, \citenamefont {Read},\ and\ \citenamefont {Ludwig}}]{GRL2001}%
  \BibitemOpen
  \bibfield  {author} {\bibinfo {author} {\bibfnamefont {I.~A.}\ \bibnamefont {Gruzberg}}, \bibinfo {author} {\bibfnamefont {N.}~\bibnamefont {Read}},\ and\ \bibinfo {author} {\bibfnamefont {A.~W.~W.}\ \bibnamefont {Ludwig}},\ }\bibfield  {title} {\bibinfo {title} {{Random-bond Ising model in two dimensions: The Nishimori line and supersymmetry}},\ }\href {https://doi.org/10.1103/PhysRevB.63.104422} {\bibfield  {journal} {\bibinfo  {journal} {Phys. Rev. B}\ }\textbf {\bibinfo {volume} {63}},\ \bibinfo {pages} {104422} (\bibinfo {year} {2001})}\BibitemShut {NoStop}%
\bibitem [{\citenamefont {Jacobsen}\ and\ \citenamefont {Picco}(2002)}]{JacobsenPicco}%
  \BibitemOpen
  \bibfield  {author} {\bibinfo {author} {\bibfnamefont {J.~L.}\ \bibnamefont {Jacobsen}}\ and\ \bibinfo {author} {\bibfnamefont {M.}~\bibnamefont {Picco}},\ }\bibfield  {title} {\bibinfo {title} {{Phase diagram and critical exponents of a Potts gauge glass}},\ }\href {https://doi.org/10.1103/PhysRevE.65.026113} {\bibfield  {journal} {\bibinfo  {journal} {Phys. Rev. E}\ }\textbf {\bibinfo {volume} {65}},\ \bibinfo {pages} {026113} (\bibinfo {year} {2002})}\BibitemShut {NoStop}%
\bibitem [{\citenamefont {Dotsenko}(1984)}]{DOTSENKO1984Potts}%
  \BibitemOpen
  \bibfield  {author} {\bibinfo {author} {\bibfnamefont {V.}~\bibnamefont {Dotsenko}},\ }\bibfield  {title} {\bibinfo {title} {{Critical behaviour and associated conformal algebra of the Z$_3$ Potts model}},\ }\href {https://doi.org/https://doi.org/10.1016/0550-3213(84)90148-2} {\bibfield  {journal} {\bibinfo  {journal} {Nuclear Physics B}\ }\textbf {\bibinfo {volume} {235}},\ \bibinfo {pages} {54} (\bibinfo {year} {1984})}\BibitemShut {NoStop}%
\bibitem [{\citenamefont {Ludwig}\ and\ \citenamefont {Cardy}(1987)}]{LUDWIGCardy}%
  \BibitemOpen
  \bibfield  {author} {\bibinfo {author} {\bibfnamefont {A.~W.~W.}\ \bibnamefont {Ludwig}}\ and\ \bibinfo {author} {\bibfnamefont {J.~L.}\ \bibnamefont {Cardy}},\ }\bibfield  {title} {\bibinfo {title} {{Perturbative evaluation of the conformal anomaly at new critical points with applications to random systems}},\ }\href {https://doi.org/https://doi.org/10.1016/0550-3213(87)90362-2} {\bibfield  {journal} {\bibinfo  {journal} {Nuclear Physics B}\ }\textbf {\bibinfo {volume} {285}},\ \bibinfo {pages} {687} (\bibinfo {year} {1987})}\BibitemShut {NoStop}%
\bibitem [{\citenamefont {Ludwig}(1987)}]{Ludwig-Potts-OPE-RG-NPB285-1987-97}%
  \BibitemOpen
  \bibfield  {author} {\bibinfo {author} {\bibfnamefont {A.~W.~W.}\ \bibnamefont {Ludwig}},\ }\bibfield  {title} {\bibinfo {title} {{Critical behavior of the two-dimensional random q-state Potts model by expansion in (q - 2)}},\ }\href {https://doi.org/https://doi.org/10.1016/0550-3213(87)90330-0} {\bibfield  {journal} {\bibinfo  {journal} {Nuclear Physics B}\ }\textbf {\bibinfo {volume} {285}},\ \bibinfo {pages} {97 } (\bibinfo {year} {1987})}\BibitemShut {NoStop}%
\bibitem [{\citenamefont {Zhu}\ \emph {et~al.}(2023)\citenamefont {Zhu}, \citenamefont {Tantivasadakarn}, \citenamefont {Vishwanath}, \citenamefont {Trebst},\ and\ \citenamefont {Verresen}}]{ZhuTantivasadakarnVishwanthTrebstVerresen}%
  \BibitemOpen
  \bibfield  {author} {\bibinfo {author} {\bibfnamefont {G.-Y.}\ \bibnamefont {Zhu}}, \bibinfo {author} {\bibfnamefont {N.}~\bibnamefont {Tantivasadakarn}}, \bibinfo {author} {\bibfnamefont {A.}~\bibnamefont {Vishwanath}}, \bibinfo {author} {\bibfnamefont {S.}~\bibnamefont {Trebst}},\ and\ \bibinfo {author} {\bibfnamefont {R.}~\bibnamefont {Verresen}},\ }\bibfield  {title} {\bibinfo {title} {{Nishimori's Cat: Stable Long-Range Entanglement from Finite-Depth Unitaries and Weak Measurements}},\ }\href {https://doi.org/10.1103/PhysRevLett.131.200201} {\bibfield  {journal} {\bibinfo  {journal} {Phys. Rev. Lett.}\ }\textbf {\bibinfo {volume} {131}},\ \bibinfo {pages} {200201} (\bibinfo {year} {2023})}\BibitemShut {NoStop}%
\bibitem [{\citenamefont {Elitzur}(1975)}]{Elitzur1975}%
  \BibitemOpen
  \bibfield  {author} {\bibinfo {author} {\bibfnamefont {S.}~\bibnamefont {Elitzur}},\ }\bibfield  {title} {\bibinfo {title} {{Impossibility of spontaneously breaking local symmetries}},\ }\href {https://doi.org/10.1103/PhysRevD.12.3978} {\bibfield  {journal} {\bibinfo  {journal} {Phys. Rev. D}\ }\textbf {\bibinfo {volume} {12}},\ \bibinfo {pages} {3978} (\bibinfo {year} {1975})}\BibitemShut {NoStop}%
\bibitem [{\citenamefont {Wegner}(1971)}]{Wegner1971}%
  \BibitemOpen
  \bibfield  {author} {\bibinfo {author} {\bibfnamefont {F.~J.}\ \bibnamefont {Wegner}},\ }\bibfield  {title} {\bibinfo {title} {{Duality in Generalized Ising Models and Phase Transitions without Local Order Parameters}},\ }\href {https://doi.org/10.1063/1.1665530} {\bibfield  {journal} {\bibinfo  {journal} {Journal of Mathematical Physics}\ }\textbf {\bibinfo {volume} {12}},\ \bibinfo {pages} {2259} (\bibinfo {year} {1971})}\BibitemShut {NoStop}%
\bibitem [{\citenamefont {Fradkin}\ and\ \citenamefont {Shenker}(1979)}]{FradkinShenker1979}%
  \BibitemOpen
  \bibfield  {author} {\bibinfo {author} {\bibfnamefont {E.}~\bibnamefont {Fradkin}}\ and\ \bibinfo {author} {\bibfnamefont {S.~H.}\ \bibnamefont {Shenker}},\ }\bibfield  {title} {\bibinfo {title} {{Phase diagrams of lattice gauge theories with Higgs fields}},\ }\href {https://doi.org/10.1103/PhysRevD.19.3682} {\bibfield  {journal} {\bibinfo  {journal} {Phys. Rev. D}\ }\textbf {\bibinfo {volume} {19}},\ \bibinfo {pages} {3682} (\bibinfo {year} {1979})}\BibitemShut {NoStop}%
\bibitem [{\citenamefont {Patil}\ and\ \citenamefont {Ludwig}(2025)}]{PatilLudwig20251}%
  \BibitemOpen
  \bibfield  {author} {\bibinfo {author} {\bibfnamefont {R.~A.}\ \bibnamefont {Patil}}\ and\ \bibinfo {author} {\bibfnamefont {A.~W.~W.}\ \bibnamefont {Ludwig}},\ }\href {https://arxiv.org/abs/2507.07959} {\bibinfo {title} {{Shannon entropy of the measurement record at measurement-dominated criticality and RG flow: A c-theorem for effective central charge and a g-theorem for effective boundary entropy}}} (\bibinfo {year} {2025}),\ \Eprint {https://arxiv.org/abs/2507.07959} {arXiv:2507.07959 [cond-mat.stat-mech]} \BibitemShut {NoStop}%
\bibitem [{\citenamefont {Patil}\ and\ \citenamefont {Ludwig}(2026)}]{PatilLudwig20262}%
  \BibitemOpen
  \bibfield  {author} {\bibinfo {author} {\bibfnamefont {R.~A.}\ \bibnamefont {Patil}}\ and\ \bibinfo {author} {\bibfnamefont {A.~W.~W.}\ \bibnamefont {Ludwig}},\ }\href@noop {} {} (\bibinfo {year} {2026}),\ \bibinfo {note} {in preparation}\BibitemShut {NoStop}%
\bibitem [{\citenamefont {Dotsenko}\ \emph {et~al.}(1999)\citenamefont {Dotsenko}, \citenamefont {Jacobsen}, \citenamefont {Lewis},\ and\ \citenamefont {Picco}}]{DotsenkoJacobsenLewisPicco}%
  \BibitemOpen
  \bibfield  {author} {\bibinfo {author} {\bibfnamefont {V.}~\bibnamefont {Dotsenko}}, \bibinfo {author} {\bibfnamefont {J.~L.}\ \bibnamefont {Jacobsen}}, \bibinfo {author} {\bibfnamefont {M.-A.}\ \bibnamefont {Lewis}},\ and\ \bibinfo {author} {\bibfnamefont {M.}~\bibnamefont {Picco}},\ }\bibfield  {title} {\bibinfo {title} {{Coupled Potts models: Self-duality and fixed point structure}},\ }\href {https://doi.org/https://doi.org/10.1016/S0550-3213(99)00097-8} {\bibfield  {journal} {\bibinfo  {journal} {Nuclear Physics B}\ }\textbf {\bibinfo {volume} {546}},\ \bibinfo {pages} {505} (\bibinfo {year} {1999})}\BibitemShut {NoStop}%
\bibitem [{\citenamefont {Mittag}\ and\ \citenamefont {Stephen}(1974)}]{MittagStephen}%
  \BibitemOpen
  \bibfield  {author} {\bibinfo {author} {\bibfnamefont {L.}~\bibnamefont {Mittag}}\ and\ \bibinfo {author} {\bibfnamefont {M.~J.}\ \bibnamefont {Stephen}},\ }\bibfield  {title} {\bibinfo {title} {{Mean-field theory of the many component Potts model}},\ }\href {https://doi.org/10.1088/0305-4470/7/9/003} {\bibfield  {journal} {\bibinfo  {journal} {Journal of Physics A: Mathematical, Nuclear and General}\ }\textbf {\bibinfo {volume} {7}},\ \bibinfo {pages} {L109} (\bibinfo {year} {1974})}\BibitemShut {NoStop}%
\bibitem [{Note1()}]{Note1}%
  \BibitemOpen
  \bibinfo {note} {Equivalently, ${\gamma }=(e^{q\protect \tilde {\gamma }}-1)/(e^{q\protect \tilde {\gamma }}+(q-1))$}\BibitemShut {NoStop}%
\bibitem [{\citenamefont {Haegeman}\ \emph {et~al.}(2015)\citenamefont {Haegeman}, \citenamefont {Van~Acoleyen}, \citenamefont {Schuch}, \citenamefont {Cirac},\ and\ \citenamefont {Verstraete}}]{Haegeman_2015}%
  \BibitemOpen
  \bibfield  {author} {\bibinfo {author} {\bibfnamefont {J.}~\bibnamefont {Haegeman}}, \bibinfo {author} {\bibfnamefont {K.}~\bibnamefont {Van~Acoleyen}}, \bibinfo {author} {\bibfnamefont {N.}~\bibnamefont {Schuch}}, \bibinfo {author} {\bibfnamefont {J.~I.}\ \bibnamefont {Cirac}},\ and\ \bibinfo {author} {\bibfnamefont {F.}~\bibnamefont {Verstraete}},\ }\bibfield  {title} {\bibinfo {title} {{Gauging Quantum States: From Global to Local Symmetries in Many-Body Systems}},\ }\href {https://doi.org/10.1103/PhysRevX.5.011024} {\bibfield  {journal} {\bibinfo  {journal} {Phys. Rev. X}\ }\textbf {\bibinfo {volume} {5}},\ \bibinfo {pages} {011024} (\bibinfo {year} {2015})}\BibitemShut {NoStop}%
\bibitem [{Note2()}]{Note2}%
  \BibitemOpen
  \bibinfo {note} {The undeformed $\protect \mathbb {Z}_q$ toric code state is dual to the RK state in Eq.~\protect \eqref {EqRKState} at infinite temperature $\beta =0$, which is the product state $|+\rangle ^{\otimes N}$ of eigenstates of unit eigenvalue of the ${\protect \hat X}_{i}$ operators at each site $i$ lying on the face of the square lattice [Fig.~\ref {FigSchematic}]. The RK state at $\beta =0$ is an equal-weight superposition of eigenstates of operators ${\protect \hat Z}_{i}$ located on the faces of the square lattice. Eigenstates of operators ${\protect \hat Z}_{ij}$ with non-unit eigenvalues $\omega _{ij}$ correspond to domain wall configurations on the links of the lattice: Each configuration of eigenvalues $\omega _i$ on the faces of the square lattice uniquely determines a domain wall configuration (up to boundary terms) on links via $\omega _{ij}=$ $\omega _i {\omega _j}^{-1}$ (going from left to right and top to bottom). Compare Fig.~\ref {FigSchematic}. The equal-weight superposition of
  the latter domain-wall configurations, which satisfy a local Gauss-law constraint, is the ground state of the (undeformed) $\protect \mathbb {Z}_q$ toric code $\vert TC\rangle $ with a $q$-dimensional Hilbert space and operators $\protect \hat {Z}_{ij}$ defined on the links of the square lattice. When acting on the undeformed $\protect \mathbb {Z}_q$ toric code state, the factor \vskip -3mm \protect \[ \hskip 2em\relax \exp \{ {\beta \protect \primfrac {over}2} \DOTSB \sum@ \slimits@ _{\langle i j \rangle } \DOTSB \sum@ \slimits@ _{r=1}^{q-1} [ ({\protect \hat Z}_{ij})^r + ({{\protect \hat Z}_{ij}}^{-1})^r ] \} \propto \exp \{\beta q \DOTSB \sum@ \slimits@ _{\langle i j \rangle } \delta _{\omega _{ij}, 1} \} \protect \,, \protect \] \vskip -1mm where $\omega _{ij}$ is the eigenvalue of the operator ${\protect \hat Z}_{ij}$, assigns a factor $\exp \{ - q\beta \times ({\protect \rm total \ number \ domain \ wall \ segments})\}$ to each domain wall configuration. Thus, the RK state \protect \eqref {EqRKState}
  for the finite-temperature $q$-state Potts model is exactly dual to the deformed $Z_q$ toric code state, which, following the above logic, reads $\exp \{ {1\protect \primfrac {over}2} {\beta \protect \primfrac {over}2} \DOTSB \sum@ \slimits@ _{\langle i j \rangle } \DOTSB \sum@ \slimits@ _{r=1}^{q-1} [ ({\protect \hat Z}_{ij})^r + ({{\protect \hat Z}_{ij}}^{-1})^r ] \} \ \vert TC\rangle $}\BibitemShut {NoStop}%
\bibitem [{Note3()}]{Note3}%
  \BibitemOpen
  \bibinfo {note} {That is, on each link $\langle i j \rangle $, $g=g_{ij}\in \protect \mathbb {Z}_q$ is the state variable on this link, and $m = m_{ij}\in \protect \mathbb {Z}_q$ is the discrete measurement outcome on the same link.}\BibitemShut {Stop}%
\bibitem [{Note4()}]{Note4}%
  \BibitemOpen
  \bibinfo {note} {A continuous complex Gaussian pointer}\BibitemShut {NoStop}%
\bibitem [{\citenamefont {P\"utz}\ \emph {et~al.}(2025)\citenamefont {P\"utz}, \citenamefont {Vasseur}, \citenamefont {Ludwig}, \citenamefont {Trebst},\ and\ \citenamefont {Zhu}}]{Putz25nishimori}%
  \BibitemOpen
  \bibfield  {author} {\bibinfo {author} {\bibfnamefont {M.}~\bibnamefont {P\"utz}}, \bibinfo {author} {\bibfnamefont {R.}~\bibnamefont {Vasseur}}, \bibinfo {author} {\bibfnamefont {A.~W.~W.}\ \bibnamefont {Ludwig}}, \bibinfo {author} {\bibfnamefont {S.}~\bibnamefont {Trebst}},\ and\ \bibinfo {author} {\bibfnamefont {G.-Y.}\ \bibnamefont {Zhu}},\ }\bibfield  {title} {\bibinfo {title} {{Flow to Nishimori Universality in Weakly Monitored Quantum Circuits with Qubit Loss}},\ }\href {https://doi.org/10.1103/ygfz-crvp} {\bibfield  {journal} {\bibinfo  {journal} {PRX Quantum}\ }\textbf {\bibinfo {volume} {6}},\ \bibinfo {pages} {040372} (\bibinfo {year} {2025})}\BibitemShut {NoStop}%
\bibitem [{\citenamefont {Schumacher}\ and\ \citenamefont {Nielsen}(1996)}]{SchumacherNielsen1996}%
  \BibitemOpen
  \bibfield  {author} {\bibinfo {author} {\bibfnamefont {B.}~\bibnamefont {Schumacher}}\ and\ \bibinfo {author} {\bibfnamefont {M.~A.}\ \bibnamefont {Nielsen}},\ }\bibfield  {title} {\bibinfo {title} {{Quantum data processing and error correction}},\ }\href {https://doi.org/10.1103/PhysRevA.54.2629} {\bibfield  {journal} {\bibinfo  {journal} {Phys. Rev. A}\ }\textbf {\bibinfo {volume} {54}},\ \bibinfo {pages} {2629} (\bibinfo {year} {1996})}\BibitemShut {NoStop}%
\bibitem [{\citenamefont {Wilde}(2017)}]{WildeQuantumInformation}%
  \BibitemOpen
  \bibfield  {author} {\bibinfo {author} {\bibfnamefont {M.~M.}\ \bibnamefont {Wilde}},\ }\href {https://doi.org/10.1017/9781316809976} {\emph {\bibinfo {title} {{Quantum Information Theory}}}},\ \bibinfo {edition} {2nd}\ ed.\ (\bibinfo  {publisher} {Cambridge University Press},\ \bibinfo {year} {2017})\BibitemShut {NoStop}%
\bibitem [{\citenamefont {Andrist}\ \emph {et~al.}(2015)\citenamefont {Andrist}, \citenamefont {Wootton},\ and\ \citenamefont {Katzgraber}}]{PhysRevA.91.042331}%
  \BibitemOpen
  \bibfield  {author} {\bibinfo {author} {\bibfnamefont {R.~S.}\ \bibnamefont {Andrist}}, \bibinfo {author} {\bibfnamefont {J.~R.}\ \bibnamefont {Wootton}},\ and\ \bibinfo {author} {\bibfnamefont {H.~G.}\ \bibnamefont {Katzgraber}},\ }\bibfield  {title} {\bibinfo {title} {{Error thresholds for Abelian quantum double models: Increasing the bit-flip stability of topological quantum memory}},\ }\href {https://doi.org/10.1103/PhysRevA.91.042331} {\bibfield  {journal} {\bibinfo  {journal} {Phys. Rev. A}\ }\textbf {\bibinfo {volume} {91}},\ \bibinfo {pages} {042331} (\bibinfo {year} {2015})}\BibitemShut {NoStop}%
\bibitem [{\citenamefont {Mukherjee}\ and\ \citenamefont {Trebst}(2026)}]{mukherjee2026nishimorithresholdestimationbayesian}%
  \BibitemOpen
  \bibfield  {author} {\bibinfo {author} {\bibfnamefont {R.}~\bibnamefont {Mukherjee}}\ and\ \bibinfo {author} {\bibfnamefont {S.}~\bibnamefont {Trebst}},\ }\href {https://arxiv.org/abs/2607.18374} {\bibinfo {title} {{Nishimori Threshold Estimation for Bayesian Inference and $\mathbb{Z}_q$ Surface Code Decoding}}} (\bibinfo {year} {2026}),\ \Eprint {https://arxiv.org/abs/2607.18374} {arXiv:2607.18374 [quant-ph]} \BibitemShut {NoStop}%
\bibitem [{\citenamefont {Nishimori}\ and\ \citenamefont {Stephen}(1983)}]{NishimoriStephen}%
  \BibitemOpen
  \bibfield  {author} {\bibinfo {author} {\bibfnamefont {H.}~\bibnamefont {Nishimori}}\ and\ \bibinfo {author} {\bibfnamefont {M.~J.}\ \bibnamefont {Stephen}},\ }\bibfield  {title} {\bibinfo {title} {{Gauge-invariant frustrated Potts spin-glass}},\ }\href {https://doi.org/10.1103/PhysRevB.27.5644} {\bibfield  {journal} {\bibinfo  {journal} {Phys. Rev. B}\ }\textbf {\bibinfo {volume} {27}},\ \bibinfo {pages} {5644} (\bibinfo {year} {1983})}\BibitemShut {NoStop}%
\bibitem [{Note5()}]{Note5}%
  \BibitemOpen
  \bibinfo {note} {As discussed below, the replica theory in Eq.~\protect \eqref {EqReplicatedPottsHamiltonianUngauged} is an exact description of the learning phase diagram for a specialized Gaussian measurement protocol. The latter protocol is discussed in detail in Sec.~\ref {SecSpecializedMeasurementProtocol}.}\BibitemShut {Stop}%
\bibitem [{\citenamefont {Honecker}\ \emph {et~al.}(2002)\citenamefont {Honecker}, \citenamefont {Jacobsen}, \citenamefont {Picco},\ and\ \citenamefont {Pujol}}]{HoneckerJacobsenPiccoPujol}%
  \BibitemOpen
  \bibfield  {author} {\bibinfo {author} {\bibfnamefont {A.}~\bibnamefont {Honecker}}, \bibinfo {author} {\bibfnamefont {J.~L.}\ \bibnamefont {Jacobsen}}, \bibinfo {author} {\bibfnamefont {M.}~\bibnamefont {Picco}},\ and\ \bibinfo {author} {\bibfnamefont {P.}~\bibnamefont {Pujol}},\ }\bibinfo {title} {{Nishimori Point in Random-Bond Ising and Potts Models in 2D}},\ in\ \href {https://doi.org/10.1007/978-94-010-0514-2_23} {\emph {\bibinfo {booktitle} {Statistical Field Theories}}},\ \bibinfo {editor} {edited by\ \bibinfo {editor} {\bibfnamefont {A.}~\bibnamefont {Cappelli}}\ and\ \bibinfo {editor} {\bibfnamefont {G.}~\bibnamefont {Mussardo}}}\ (\bibinfo  {publisher} {Springer Netherlands},\ \bibinfo {address} {Dordrecht},\ \bibinfo {year} {2002})\ pp.\ \bibinfo {pages} {251--261}\BibitemShut {NoStop}%
\bibitem [{\citenamefont {Potts}(1952)}]{Potts_1952}%
  \BibitemOpen
  \bibfield  {author} {\bibinfo {author} {\bibfnamefont {R.~B.}\ \bibnamefont {Potts}},\ }\bibfield  {title} {\bibinfo {title} {{Some generalized order-disorder transformations}},\ }\href {https://doi.org/10.1017/S0305004100027419} {\bibfield  {journal} {\bibinfo  {journal} {Mathematical Proceedings of the Cambridge Philosophical Society}\ }\textbf {\bibinfo {volume} {48}},\ \bibinfo {pages} {106} (\bibinfo {year} {1952})}\BibitemShut {NoStop}%
\bibitem [{Note6()}]{Note6}%
  \BibitemOpen
  \bibinfo {note} {Modulo the specific form of the measurement operation in Eq.~\protect \eqref {eq:PmsigmaGaussianPotts}, in general, one might expect a Gaussian distribution with `{\protect \sl blurred}' measurement outcomes to more closely model an experiment than the sharp Kronecker-delta measurement outcomes in Eq.~\protect \eqref {EqMeasurementProtocolA}. For a quantum measurement setup, a protocol to perform exact Gaussian quantum weak measurements was discussed in Ref.~\cite {GarrattWeinsteinAltman2022} by weakly coupling the measurement operator to a quantum harmonic oscillator.}\BibitemShut {Stop}%
\bibitem [{Note7()}]{Note7}%
  \BibitemOpen
  \bibinfo {note} {As discussed in Sec.~\ref {SecQuantumProtocol}, the Bayesian inference problem can be formulated as a quantum Born-rule measurement problem on the Rokhsar-Kivelson wavefunction for the stat-mech model~\cite {PutzGarrattNishimoriTrebstZhu,PatilLudwig20251}. In the quantum formulation, the second equality in Eq.~\protect \eqref {EqPottsEqualityLongDistance} can be understood as a simple consequence of the POVM condition, $\DOTSB \sum@ \slimits@ _{\protect \vec {m}}\protect \hat {K}_{\protect \vec {m}}^{\dagger }\protect \hat {K}_{\protect \vec {m}}=1$. Note that \begin {align*} &\protect \overline {\langle \protect \hat {\protect \mathcal {O}}_1\rangle _{\protect \vec {m}}}=\DOTSB \sum@ \slimits@ _{\protect \vec {m}}P(\protect \vec {m})\langle \Psi _{\protect \vec {m}}\vert \protect \hat {\protect \mathcal {O}}_1\vert \Psi _{\protect \vec {m}}\rangle \\&=\DOTSB \sum@ \slimits@ _{\protect \vec {m}}\langle RK\vert \protect \hat {K}^{\dagger }_{\protect \vec {m}}\protect \hat {K}_{\protect \vec
  {m}}\vert RK\rangle \protect \frac {\langle RK\vert \protect \hat {K}^{\dagger }_{\protect \vec {m}}\protect \hat {\protect \mathcal {O}}_1\protect \hat {K}_{\protect \vec {m}}\vert RK\rangle }{\langle RK\vert \protect \hat {K}^{\dagger }_{\protect \vec {m}}\protect \hat {K}_{\protect \vec {m}}\vert RK\rangle }\\ &=\DOTSB \sum@ \slimits@ _{\protect \vec {m}}{\langle RK\vert \protect \hat {K}^{\dagger }_{\protect \vec {m}}\protect \hat {\protect \mathcal {O}}_1\protect \hat {K}_{\protect \vec {m}}\vert RK\rangle }\\&={\langle RK\vert \protect \hat {\protect \mathcal {O}}_1\DOTSB \sum@ \slimits@ _{\protect \vec {m}}\protect \hat {K}^{\dagger }_{\protect \vec {m}}\protect \hat {K}_{\protect \vec {m}}\vert RK\rangle }={\langle RK\vert \protect \hat {\protect \mathcal {O}}_1\vert RK\rangle }=\langle \protect \hat {\protect \mathcal {O}}_1\rangle \protect \,, \end {align*} where we have commuted $\protect \hat {\protect \mathcal {O}}_1$ through the Kraus operator $\protect \hat {K}_{\protect \vec {m}}$, since
  both are diagonal in the basis defined by stat-mech configurations, and {then} used the POVM condition.}\BibitemShut {Stop}%
\bibitem [{\citenamefont {Le~Doussal}\ and\ \citenamefont {Harris}(1988)}]{LeDoussalHarrisI}%
  \BibitemOpen
  \bibfield  {author} {\bibinfo {author} {\bibfnamefont {P.}~\bibnamefont {Le~Doussal}}\ and\ \bibinfo {author} {\bibfnamefont {A.~B.}\ \bibnamefont {Harris}},\ }\bibfield  {title} {\bibinfo {title} {{Location of the Ising Spin-Glass Multicritical Point on Nishimori's Line}},\ }\href {https://doi.org/10.1103/PhysRevLett.61.625} {\bibfield  {journal} {\bibinfo  {journal} {Phys. Rev. Lett.}\ }\textbf {\bibinfo {volume} {61}},\ \bibinfo {pages} {625} (\bibinfo {year} {1988})}\BibitemShut {NoStop}%
\bibitem [{Note8()}]{Note8}%
  \BibitemOpen
  \bibinfo {note} {The `spin-glass' phase is characterized by long-range order in the measurement-averaged second moment of the spin-spin correlator while its measurement-averaged first moment remains short-ranged.}\BibitemShut {Stop}%
\bibitem [{Note9()}]{Note9}%
  \BibitemOpen
  \bibinfo {note} {Note that the second equality in the above equation remains exact because measurement-averaged first moments are unaffected by conditioning on measurement outcomes~\cite {Note7}.}\BibitemShut {Stop}%
\bibitem [{Note10()}]{Note10}%
  \BibitemOpen
  \bibinfo {note} {Most of the terms in the Taylor expansion about the unperturbed ordinary Nishimori transition at $\beta =0$ [cf. Eq.~\protect \eqref {EqReplicatedPottsHamiltonianProtocolI}], in terms of the correlation functions of the perturbation, vanish identically due to the local symmetry, by Elitzur's theorem~\cite {Elitzur1975}. This is except for `contact-terms' where terms in a multi-bond correlator (calculated with the unperturbed theory) appear in combinations invariant under the above local symmetry at any given bond $\langle ij\rangle $. The latter terms, which are invariant under the above local symmetry as well as (by construction) under the global $\protect \mathbb {Z}_q$ symmetry, are the same as the (local and global) symmetry allowed terms that are already present on the ordinary Nishimori line $\beta =0$ [Eq.~\protect \eqref {EqReplicatedPottsHamiltonianProtocolI}]. These terms, therefore, only shift the (non-universal) location of the transition along the ordinary Nishimori line, and
  do not introduce additional relevant scaling fields at the critical point.}\BibitemShut {Stop}%
\bibitem [{\citenamefont {Trebst}\ \emph {et~al.}(2007)\citenamefont {Trebst}, \citenamefont {Werner}, \citenamefont {Troyer}, \citenamefont {Shtengel},\ and\ \citenamefont {Nayak}}]{TrebstWernerTroyerShtengelNayak}%
  \BibitemOpen
  \bibfield  {author} {\bibinfo {author} {\bibfnamefont {S.}~\bibnamefont {Trebst}}, \bibinfo {author} {\bibfnamefont {P.}~\bibnamefont {Werner}}, \bibinfo {author} {\bibfnamefont {M.}~\bibnamefont {Troyer}}, \bibinfo {author} {\bibfnamefont {K.}~\bibnamefont {Shtengel}},\ and\ \bibinfo {author} {\bibfnamefont {C.}~\bibnamefont {Nayak}},\ }\bibfield  {title} {\bibinfo {title} {{Breakdown of a Topological Phase: Quantum Phase Transition in a Loop Gas Model with Tension}},\ }\href {https://doi.org/10.1103/PhysRevLett.98.070602} {\bibfield  {journal} {\bibinfo  {journal} {Phys. Rev. Lett.}\ }\textbf {\bibinfo {volume} {98}},\ \bibinfo {pages} {070602} (\bibinfo {year} {2007})}\BibitemShut {NoStop}%
\bibitem [{\citenamefont {Tupitsyn}\ \emph {et~al.}(2010)\citenamefont {Tupitsyn}, \citenamefont {Kitaev}, \citenamefont {Prokof'ev},\ and\ \citenamefont {Stamp}}]{TupitsynKitaevProkofevStamp}%
  \BibitemOpen
  \bibfield  {author} {\bibinfo {author} {\bibfnamefont {I.~S.}\ \bibnamefont {Tupitsyn}}, \bibinfo {author} {\bibfnamefont {A.}~\bibnamefont {Kitaev}}, \bibinfo {author} {\bibfnamefont {N.~V.}\ \bibnamefont {Prokof'ev}},\ and\ \bibinfo {author} {\bibfnamefont {P.~C.~E.}\ \bibnamefont {Stamp}},\ }\bibfield  {title} {\bibinfo {title} {{Topological multicritical point in the phase diagram of the toric code model and three-dimensional lattice gauge Higgs model}},\ }\href {https://doi.org/10.1103/PhysRevB.82.085114} {\bibfield  {journal} {\bibinfo  {journal} {Phys. Rev. B}\ }\textbf {\bibinfo {volume} {82}},\ \bibinfo {pages} {085114} (\bibinfo {year} {2010})}\BibitemShut {NoStop}%
\bibitem [{\citenamefont {Weinstein}\ and\ \citenamefont {Garratt}(2026)}]{WeinsteinGarratt2026}%
  \BibitemOpen
  \bibfield  {author} {\bibinfo {author} {\bibfnamefont {Z.}~\bibnamefont {Weinstein}}\ and\ \bibinfo {author} {\bibfnamefont {S.~J.}\ \bibnamefont {Garratt}},\ }\href {https://arxiv.org/abs/2603.14098} {\bibinfo {title} {{Intrinsic Error Thresholds in Nearly Critical Toric Codes}}} (\bibinfo {year} {2026}),\ \Eprint {https://arxiv.org/abs/2603.14098} {arXiv:2603.14098 [cond-mat.stat-mech]} \BibitemShut {NoStop}%
\bibitem [{\citenamefont {Dotsenko}\ \emph {et~al.}(1995)\citenamefont {Dotsenko}, \citenamefont {Picco},\ and\ \citenamefont {Pujol}}]{DotsenkoPiccoPujol}%
  \BibitemOpen
  \bibfield  {author} {\bibinfo {author} {\bibfnamefont {V.}~\bibnamefont {Dotsenko}}, \bibinfo {author} {\bibfnamefont {M.}~\bibnamefont {Picco}},\ and\ \bibinfo {author} {\bibfnamefont {P.}~\bibnamefont {Pujol}},\ }\bibfield  {title} {\bibinfo {title} {{Renormalisation-group calculation of correlation functions for the 2D random bond Ising and Potts models}},\ }\href {https://doi.org/https://doi.org/10.1016/0550-3213(95)00534-Y} {\bibfield  {journal} {\bibinfo  {journal} {Nuclear Physics B}\ }\textbf {\bibinfo {volume} {455}},\ \bibinfo {pages} {701} (\bibinfo {year} {1995})}\BibitemShut {NoStop}%
\bibitem [{\citenamefont {Binder}\ and\ \citenamefont {Young}(1986)}]{BinderYoungSpinGlass}%
  \BibitemOpen
  \bibfield  {author} {\bibinfo {author} {\bibfnamefont {K.}~\bibnamefont {Binder}}\ and\ \bibinfo {author} {\bibfnamefont {A.~P.}\ \bibnamefont {Young}},\ }\bibfield  {title} {\bibinfo {title} {{Spin glasses: Experimental facts, theoretical concepts, and open questions}},\ }\href {https://doi.org/10.1103/RevModPhys.58.801} {\bibfield  {journal} {\bibinfo  {journal} {Rev. Mod. Phys.}\ }\textbf {\bibinfo {volume} {58}},\ \bibinfo {pages} {801} (\bibinfo {year} {1986})}\BibitemShut {NoStop}%
\bibitem [{Note11()}]{Note11}%
  \BibitemOpen
  \bibinfo {note} {In the projective limit, the configuration of the underlying Potts spins is completely pinned (up to a $\protect \mathbb {Z}_q$ flip) by the measurement outcomes. Then the measurement-averaged moments of the correlation functions are straightforwardly determined using the known correlation functions at the unmeasured critical point. This is in contrast with the $L^{(1)}$ and $N^{(2)}$ critical points [Fig.~\ref {FigPottsLearningPhaseDiagram}], where moments of correlation functions (except for one or two lowest moments) are expected to display completely distinct critical behavior from the unmeasured critical point.}\BibitemShut {Stop}%
\bibitem [{\citenamefont {Pandey}\ \emph {et~al.}(2026)\citenamefont {Pandey}, \citenamefont {Mahadevan}, \citenamefont {Middleton},\ and\ \citenamefont {Fisher}}]{PandeyMahadevanMiddletonFisher}%
  \BibitemOpen
  \bibfield  {author} {\bibinfo {author} {\bibfnamefont {A.}~\bibnamefont {Pandey}}, \bibinfo {author} {\bibfnamefont {A.}~\bibnamefont {Mahadevan}}, \bibinfo {author} {\bibfnamefont {A.~A.}\ \bibnamefont {Middleton}},\ and\ \bibinfo {author} {\bibfnamefont {D.~S.}\ \bibnamefont {Fisher}},\ }\href {https://arxiv.org/abs/2603.02308} {\bibinfo {title} {{Low-temperature transition of 2d random-bond Ising model and quantum infinite randomness}}} (\bibinfo {year} {2026}),\ \Eprint {https://arxiv.org/abs/2603.02308} {arXiv:2603.02308 [cond-mat.stat-mech]} \BibitemShut {NoStop}%
\bibitem [{Note12()}]{Note12}%
  \BibitemOpen
  \bibinfo {note} {We note that the strong-disorder RG for the Ising case in~\cite {PandeyMahadevanMiddletonFisher} proceeds via a mapping to a $2D$ quantum free-fermion Hamiltonian (see also Ref.~\cite {MotrunichDamleHuse}). For the $q>2$ state Potts model, there is no obvious free-fermion mapping, and it might be difficult to find a controlled strong-disorder RG analysis for the $S^{(0)}$ critical point in the RBPM [Fig.~\ref {FigPottsLearningPhaseDiagram}(b)].}\BibitemShut {Stop}%
\bibitem [{\citenamefont {Aharony}(1978)}]{Aharony_1978}%
  \BibitemOpen
  \bibfield  {author} {\bibinfo {author} {\bibfnamefont {A.}~\bibnamefont {Aharony}},\ }\bibfield  {title} {\bibinfo {title} {{Low-temperature phase diagram and critical properties of a dilute spin glass}},\ }\href {https://doi.org/10.1088/0022-3719/11/11/004} {\bibfield  {journal} {\bibinfo  {journal} {Journal of Physics C: Solid State Physics}\ }\textbf {\bibinfo {volume} {11}},\ \bibinfo {pages} {L457} (\bibinfo {year} {1978})}\BibitemShut {NoStop}%
\bibitem [{\citenamefont {Domany}(1979)}]{Domany_1979}%
  \BibitemOpen
  \bibfield  {author} {\bibinfo {author} {\bibfnamefont {E.}~\bibnamefont {Domany}},\ }\bibfield  {title} {\bibinfo {title} {{Some results for the two-dimensional Ising model with competing interactions}},\ }\href {https://doi.org/10.1088/0022-3719/12/3/007} {\bibfield  {journal} {\bibinfo  {journal} {Journal of Physics C: Solid State Physics}\ }\textbf {\bibinfo {volume} {12}},\ \bibinfo {pages} {L119} (\bibinfo {year} {1979})}\BibitemShut {NoStop}%
\bibitem [{\citenamefont {Dotsenko}\ and\ \citenamefont {Fateev}(1984)}]{DotsenkoFateev}%
  \BibitemOpen
  \bibfield  {author} {\bibinfo {author} {\bibfnamefont {V.}~\bibnamefont {Dotsenko}}\ and\ \bibinfo {author} {\bibfnamefont {V.}~\bibnamefont {Fateev}},\ }\bibfield  {title} {\bibinfo {title} {{Conformal algebra and multipoint correlation functions in 2D statistical models}},\ }\href {https://doi.org/https://doi.org/10.1016/0550-3213(84)90269-4} {\bibfield  {journal} {\bibinfo  {journal} {Nuclear Physics B}\ }\textbf {\bibinfo {volume} {240}},\ \bibinfo {pages} {312} (\bibinfo {year} {1984})}\BibitemShut {NoStop}%
\bibitem [{\citenamefont {Belavin}\ \emph {et~al.}(1984)\citenamefont {Belavin}, \citenamefont {Polyakov},\ and\ \citenamefont {Zamolodchikov}}]{BelavinPolyakovZamolodchikov}%
  \BibitemOpen
  \bibfield  {author} {\bibinfo {author} {\bibfnamefont {A.}~\bibnamefont {Belavin}}, \bibinfo {author} {\bibfnamefont {A.}~\bibnamefont {Polyakov}},\ and\ \bibinfo {author} {\bibfnamefont {A.}~\bibnamefont {Zamolodchikov}},\ }\bibfield  {title} {\bibinfo {title} {{Infinite conformal symmetry in two-dimensional quantum field theory}},\ }\href {https://doi.org/https://doi.org/10.1016/0550-3213(84)90052-X} {\bibfield  {journal} {\bibinfo  {journal} {Nuclear Physics B}\ }\textbf {\bibinfo {volume} {241}},\ \bibinfo {pages} {333} (\bibinfo {year} {1984})}\BibitemShut {NoStop}%
\bibitem [{\citenamefont {Ludwig}(1990)}]{LUDWIG1990infinitehierarchy}%
  \BibitemOpen
  \bibfield  {author} {\bibinfo {author} {\bibfnamefont {A.~W.~W.}\ \bibnamefont {Ludwig}},\ }\bibfield  {title} {\bibinfo {title} {{Infinite hierarchies of exponents in a diluted ferromagnet and their interpretation}},\ }\href {https://doi.org/https://doi.org/10.1016/0550-3213(90)90126-X} {\bibfield  {journal} {\bibinfo  {journal} {Nuclear Physics B}\ }\textbf {\bibinfo {volume} {330}},\ \bibinfo {pages} {639} (\bibinfo {year} {1990})}\BibitemShut {NoStop}%
\bibitem [{\citenamefont {{Lykke Jacobsen}}\ and\ \citenamefont {Cardy}(1998)}]{JacobsenCardy}%
  \BibitemOpen
  \bibfield  {author} {\bibinfo {author} {\bibfnamefont {J.}~\bibnamefont {{Lykke Jacobsen}}}\ and\ \bibinfo {author} {\bibfnamefont {J.}~\bibnamefont {Cardy}},\ }\bibfield  {title} {\bibinfo {title} {{Critical behaviour of random-bond Potts models: a transfer matrix study}},\ }\href {https://doi.org/https://doi.org/10.1016/S0550-3213(98)00024-8} {\bibfield  {journal} {\bibinfo  {journal} {Nuclear Physics B}\ }\textbf {\bibinfo {volume} {515}},\ \bibinfo {pages} {701} (\bibinfo {year} {1998})}\BibitemShut {NoStop}%
\bibitem [{\citenamefont {Wang}\ \emph {et~al.}(2025)\citenamefont {Wang}, \citenamefont {Vasseur}, \citenamefont {Trebst}, \citenamefont {Ludwig},\ and\ \citenamefont {Zhu}}]{Wang2025}%
  \BibitemOpen
  \bibfield  {author} {\bibinfo {author} {\bibfnamefont {Q.}~\bibnamefont {Wang}}, \bibinfo {author} {\bibfnamefont {R.}~\bibnamefont {Vasseur}}, \bibinfo {author} {\bibfnamefont {S.}~\bibnamefont {Trebst}}, \bibinfo {author} {\bibfnamefont {A.~W.~W.}\ \bibnamefont {Ludwig}},\ and\ \bibinfo {author} {\bibfnamefont {G.-Y.}\ \bibnamefont {Zhu}},\ }\href {https://arxiv.org/abs/2502.14034} {\bibinfo {title} {{Decoherence-induced self-dual criticality in topological states of matter}}} (\bibinfo {year} {2025}),\ \Eprint {https://arxiv.org/abs/2502.14034} {arXiv:2502.14034 [quant-ph]} \BibitemShut {NoStop}%
\bibitem [{\citenamefont {Feller}(1966)}]{fellerintroduction}%
  \BibitemOpen
  \bibfield  {author} {\bibinfo {author} {\bibfnamefont {W.}~\bibnamefont {Feller}},\ }\href@noop {} {\emph {\bibinfo {title} {{An Introduction to Probability Theory and Its Applications, Volume 2}}}}\ (\bibinfo  {publisher} {J. Wiley and Sons, New York},\ \bibinfo {year} {1966})\BibitemShut {NoStop}%
\bibitem [{\citenamefont {Aizenman}\ and\ \citenamefont {Wehr}(1990)}]{AizenmanWehr}%
  \BibitemOpen
  \bibfield  {author} {\bibinfo {author} {\bibfnamefont {M.}~\bibnamefont {Aizenman}}\ and\ \bibinfo {author} {\bibfnamefont {J.}~\bibnamefont {Wehr}},\ }\bibfield  {title} {\bibinfo {title} {{Rounding effects of quenched randomness on first-order phase transitions}},\ }\href {https://doi.org/10.1007/BF02096933} {\bibfield  {journal} {\bibinfo  {journal} {Communications in Mathematical Physics}\ }\textbf {\bibinfo {volume} {130}},\ \bibinfo {pages} {489} (\bibinfo {year} {1990})}\BibitemShut {NoStop}%
\bibitem [{\citenamefont {Jacobsen}\ and\ \citenamefont {Picco}(2000)}]{JacobsenPiccoLargeq}%
  \BibitemOpen
  \bibfield  {author} {\bibinfo {author} {\bibfnamefont {J.~L.}\ \bibnamefont {Jacobsen}}\ and\ \bibinfo {author} {\bibfnamefont {M.}~\bibnamefont {Picco}},\ }\bibfield  {title} {\bibinfo {title} {{Large-q asymptotics of the random-bond Potts model}},\ }\href {https://doi.org/10.1103/PhysRevE.61.R13} {\bibfield  {journal} {\bibinfo  {journal} {Phys. Rev. E}\ }\textbf {\bibinfo {volume} {61}},\ \bibinfo {pages} {R13(R)} (\bibinfo {year} {2000})}\BibitemShut {NoStop}%
\bibitem [{\citenamefont {Pal{\'a}gyi}\ \emph {et~al.}(2000)\citenamefont {Pal{\'a}gyi}, \citenamefont {Chatelain}, \citenamefont {Berche},\ and\ \citenamefont {Igl{\'o}i}}]{Palagyi2000}%
  \BibitemOpen
  \bibfield  {author} {\bibinfo {author} {\bibfnamefont {G.}~\bibnamefont {Pal{\'a}gyi}}, \bibinfo {author} {\bibfnamefont {C.}~\bibnamefont {Chatelain}}, \bibinfo {author} {\bibfnamefont {B.}~\bibnamefont {Berche}},\ and\ \bibinfo {author} {\bibfnamefont {F.}~\bibnamefont {Igl{\'o}i}},\ }\bibfield  {title} {\bibinfo {title} {{Boundary critical behaviour of two-dimensional random Potts models}},\ }\href {https://doi.org/10.1007/s100510050042} {\bibfield  {journal} {\bibinfo  {journal} {The European Physical Journal B - Condensed Matter and Complex Systems}\ }\textbf {\bibinfo {volume} {13}},\ \bibinfo {pages} {357} (\bibinfo {year} {2000})}\BibitemShut {NoStop}%
\bibitem [{\citenamefont {Chatelain}\ \emph {et~al.}(2001)\citenamefont {Chatelain}, \citenamefont {Berche},\ and\ \citenamefont {Shchur}}]{Chatelain_2001}%
  \BibitemOpen
  \bibfield  {author} {\bibinfo {author} {\bibfnamefont {C.}~\bibnamefont {Chatelain}}, \bibinfo {author} {\bibfnamefont {B.}~\bibnamefont {Berche}},\ and\ \bibinfo {author} {\bibfnamefont {L.~N.}\ \bibnamefont {Shchur}},\ }\bibfield  {title} {\bibinfo {title} {{Quenched bond dilution in two-dimensional Potts models}},\ }\href {https://doi.org/10.1088/0305-4470/34/45/301} {\bibfield  {journal} {\bibinfo  {journal} {Journal of Physics A: Mathematical and General}\ }\textbf {\bibinfo {volume} {34}},\ \bibinfo {pages} {9593} (\bibinfo {year} {2001})}\BibitemShut {NoStop}%
\bibitem [{\citenamefont {Jos\'e}\ \emph {et~al.}(1977)\citenamefont {Jos\'e}, \citenamefont {Kadanoff}, \citenamefont {Kirkpatrick},\ and\ \citenamefont {Nelson}}]{Jose1977}%
  \BibitemOpen
  \bibfield  {author} {\bibinfo {author} {\bibfnamefont {J.~V.}\ \bibnamefont {Jos\'e}}, \bibinfo {author} {\bibfnamefont {L.~P.}\ \bibnamefont {Kadanoff}}, \bibinfo {author} {\bibfnamefont {S.}~\bibnamefont {Kirkpatrick}},\ and\ \bibinfo {author} {\bibfnamefont {D.~R.}\ \bibnamefont {Nelson}},\ }\bibfield  {title} {\bibinfo {title} {{Renormalization, vortices, and symmetry-breaking perturbations in the two-dimensional planar model}},\ }\href {https://doi.org/10.1103/PhysRevB.16.1217} {\bibfield  {journal} {\bibinfo  {journal} {Phys. Rev. B}\ }\textbf {\bibinfo {volume} {16}},\ \bibinfo {pages} {1217} (\bibinfo {year} {1977})}\BibitemShut {NoStop}%
\bibitem [{\citenamefont {Vijay}\ and\ \citenamefont {Lee}(2025)}]{vijay2025}%
  \BibitemOpen
  \bibfield  {author} {\bibinfo {author} {\bibfnamefont {A.}~\bibnamefont {Vijay}}\ and\ \bibinfo {author} {\bibfnamefont {J.~Y.}\ \bibnamefont {Lee}},\ }\href {https://arxiv.org/abs/2512.22121} {\bibinfo {title} {{Information Critical Phases under Decoherence}}} (\bibinfo {year} {2025}),\ \Eprint {https://arxiv.org/abs/2512.22121} {arXiv:2512.22121 [quant-ph]} \BibitemShut {NoStop}%
\bibitem [{\citenamefont {Patil}\ \emph {et~al.}(2026{\natexlab{b}})\citenamefont {Patil}, \citenamefont {P\"utz}, \citenamefont {Mukherjee}, \citenamefont {Zhu}, \citenamefont {Trebst},\ and\ \citenamefont {Ludwig}}]{zenodo_potts_nishimori}%
  \BibitemOpen
  \bibfield  {author} {\bibinfo {author} {\bibfnamefont {R.~A.}\ \bibnamefont {Patil}}, \bibinfo {author} {\bibfnamefont {M.}~\bibnamefont {P\"utz}}, \bibinfo {author} {\bibfnamefont {R.}~\bibnamefont {Mukherjee}}, \bibinfo {author} {\bibfnamefont {G.-Y.}\ \bibnamefont {Zhu}}, \bibinfo {author} {\bibfnamefont {S.}~\bibnamefont {Trebst}},\ and\ \bibinfo {author} {\bibfnamefont {A.~W.~W.}\ \bibnamefont {Ludwig}},\ }\bibfield  {title} {\bibinfo {title} {{Data for ``Learning Potts Models and $\mathbb{Z}_3$ Toric Codes: Higher and Ordinary Nishimori Criticality''}},\ }\bibfield  {journal} {\bibinfo  {journal} {Zenodo}\ }\href {https://doi.org/10.5281/zenodo.21918134} {10.5281/zenodo.21918134} (\bibinfo {year} {2026}{\natexlab{b}})\BibitemShut {NoStop}%
\bibitem [{\citenamefont {Garratt}\ \emph {et~al.}(2023)\citenamefont {Garratt}, \citenamefont {Weinstein},\ and\ \citenamefont {Altman}}]{GarrattWeinsteinAltman2022}%
  \BibitemOpen
  \bibfield  {author} {\bibinfo {author} {\bibfnamefont {S.~J.}\ \bibnamefont {Garratt}}, \bibinfo {author} {\bibfnamefont {Z.}~\bibnamefont {Weinstein}},\ and\ \bibinfo {author} {\bibfnamefont {E.}~\bibnamefont {Altman}},\ }\bibfield  {title} {\bibinfo {title} {{Measurements Conspire Nonlocally to Restructure Critical Quantum States}},\ }\href {https://doi.org/10.1103/PhysRevX.13.021026} {\bibfield  {journal} {\bibinfo  {journal} {Phys. Rev. X}\ }\textbf {\bibinfo {volume} {13}},\ \bibinfo {pages} {021026} (\bibinfo {year} {2023})}\BibitemShut {NoStop}%
\bibitem [{\citenamefont {Motrunich}\ \emph {et~al.}(2002)\citenamefont {Motrunich}, \citenamefont {Damle},\ and\ \citenamefont {Huse}}]{MotrunichDamleHuse}%
  \BibitemOpen
  \bibfield  {author} {\bibinfo {author} {\bibfnamefont {O.}~\bibnamefont {Motrunich}}, \bibinfo {author} {\bibfnamefont {K.}~\bibnamefont {Damle}},\ and\ \bibinfo {author} {\bibfnamefont {D.~A.}\ \bibnamefont {Huse}},\ }\bibfield  {title} {\bibinfo {title} {{Particle-hole symmetric localization in two dimensions}},\ }\href {https://doi.org/10.1103/PhysRevB.65.064206} {\bibfield  {journal} {\bibinfo  {journal} {Phys. Rev. B}\ }\textbf {\bibinfo {volume} {65}},\ \bibinfo {pages} {064206} (\bibinfo {year} {2002})}\BibitemShut {NoStop}%
\end{thebibliography}%

\end{document}